\documentclass[
pagesize=pdftex,   
paper=a4,          
fontsize=10 pt,     
twoside=false,      
openany,           
liststotoc,        
idxtotoc,          
headsepline
]{scrbook}
 
\usepackage[
paper=a4paper,
inner=3.4cm,
outer=2.3cm,
top=32mm,
bottom=40mm
]{geometry}
\usepackage[usenames,dvipsnames,svgnames,table]{xcolor}
\usepackage[utf8]{inputenc}
\usepackage[ngerman,english]{babel} 
\usepackage[T1]{fontenc}
\usepackage{float}
\usepackage{mathrsfs}
\usepackage[pdftex]{graphicx}
\usepackage[right]{eurosym}
\usepackage{fancyhdr}
\usepackage{helvet}
\usepackage[protrusion=true, expansion=false]{microtype}
\usepackage{./package/chngcntr}
\usepackage{./package/pifont}
\usepackage{makeidx}
\usepackage{hyperref}
\usepackage[all]{hypcap}
\usepackage{tikz}
\usepackage{lscape}
\usepackage{multirow}
\usepackage{tabularx}
\usepackage{xcolor}
\usepackage{graphicx}
\usepackage{setspace}
\usepackage{amsmath}
\usepackage{amssymb}
\usepackage{titlesec}
\usepackage{adjustbox}
\usepackage{color}
\usepackage{framed}

\definecolor{shadecolor}{gray}{0.95}

\usepackage{listings}
\usepackage{color} 
\definecolor{mygreen}{RGB}{28,172,0} 
\definecolor{mylilas}{RGB}{170,55,241}

\usepackage[labelfont=bf,labelsep=endash]{caption} 

\usepackage[
backend=biber,
style=numeric, 
citestyle=numeric-comp,
sortlocale=de_DE,
natbib=true,
url=false, 
doi=true,
eprint=false,
firstinits=true,
sorting=none
]{biblatex}
\fancypagestyle{plain}{

\fancyfoot[OR]{\thepage}
\fancyfoot[EL]{\thepage}
\fancyfoot[OL]{\textsc{Stefano De Blasi, UAS Aschaffenburg, SS 2018}}
}
\fancypagestyle{frontback}{
\fancyhf{}

\fancyfoot[OR]{\thepage}
\fancyfoot[EL]{\thepage}
\fancyfoot[OL]{\textsc{Stefano De Blasi, UAS Aschaffenburg, SS 2018}}
}
\fancypagestyle{main}{
\fancyhf{}

\fancyfoot[OR]{\thepage}
\fancyfoot[EL]{\thepage}
\fancyhead[L]{\textsc{\nouppercase{\leftmark}}}
\fancyfoot[OL]{\textsc{Stefano De Blasi, UAS Aschaffenburg, SS 2018}}
}
\fancypagestyle{em}{

	\fancyfoot[OR]{}
	\fancyfoot[EL]{}
	\fancyfoot[OL]{}
}
\fancypagestyle{plain}{%
\fancyhf{}

\fancyfoot[OR]{\thepage}
\fancyfoot[EL]{\thepage}
\fancyhead[L]{}
\fancyfoot[OL]{\textsc{Stefano De Blasi, UAS Aschaffenburg, SS 2018}}
}

\renewcommand{\footnoterule}{%
	\kern -3pt
	\hrule width \textwidth height 1.0pt
	\kern 2pt%
}
\usepackage{minitoc}
\KOMAoption{chapterprefix}{true}

\addtokomafont{chapter}{\raggedleft\Huge\bfseries}

\counterwithout{footnote}{chapter}

\usepackage{textcomp} 
\usepackage{acronym} 
\usepackage[toc,page]{appendix}

\deffootnote{1em}{1em}{\textsuperscript{\thefootnotemark\ }}

\newcommand{\ARBEITART}{Masterthesis}
\newcommand{\TITEL}{Connectivity estimation of high dimensional data recorded from neuronal cells}
\newcommand{\AUTORNAME}{Stefano De Blasi}
\newcommand{\AUTORSTR}{Prozelterner Weg 6}
\newcommand{\AUTORPLZ}{63933}
\newcommand{\AUTORORT}{Mönchberg}

\newcommand{\ERSTPRUEFER}{Prof. Dr.-Ing. Christiane Thielemann}
\newcommand{\ZWEITPRUEFER}{Prof. Dr. Hans-Georg Stark}
\newcommand{\BETREUER}{M.Eng. Manuel Ciba}

\newcommand{\HOCHSCHULE}{University of Applied Science Aschaffenburg}
\newcommand{\FAKULTAT}{Engineering Sciences}
\newcommand{\HOCHSCHULSTRASSE}{Würzburger Straße 45}
\newcommand{\HOCHSCHULEORT}{63743 Aschaffenburg}

\newsavebox\foobox
\newlength{\foodim}

\hypersetup{
	pdfauthor 	= {\AUTORNAME},
	pdftitle	= {\TITEL},
	pdfsubject	= {\ARBEITART \HOCHSCHULE},
	pdfkeywords	= {\AUTORNAME, \TITEL, \ARBEITART, \HOCHSCHULE},
	pdfcreator	= {LaTex},
	pdfproducer	= {pdflatex},
	colorlinks      = {false},
	}

\makeindex

\usepackage{pdfpages}

\begin{document}

\lstset{language=Matlab,%
	basicstyle=\small,
	breaklines=true,%
	morekeywords={matlab2tikz},
	keywordstyle=\color{blue},%
	morekeywords=[2]{1}, keywordstyle=[2]{\color{black}},
	identifierstyle=\color{black},%
	stringstyle=\color{mylilas},
	commentstyle=\color{mygreen},%
	showstringspaces=false,
	numbers=left,%
	numberstyle={\tiny \color{black}},
	numbersep=9pt, 
	emph=[1]{for,end,break},emphstyle=[1]\color{red}, 
}

\frontmatter
\pagenumbering{Roman}
\pagestyle{empty}

\includepdf[pages=-]{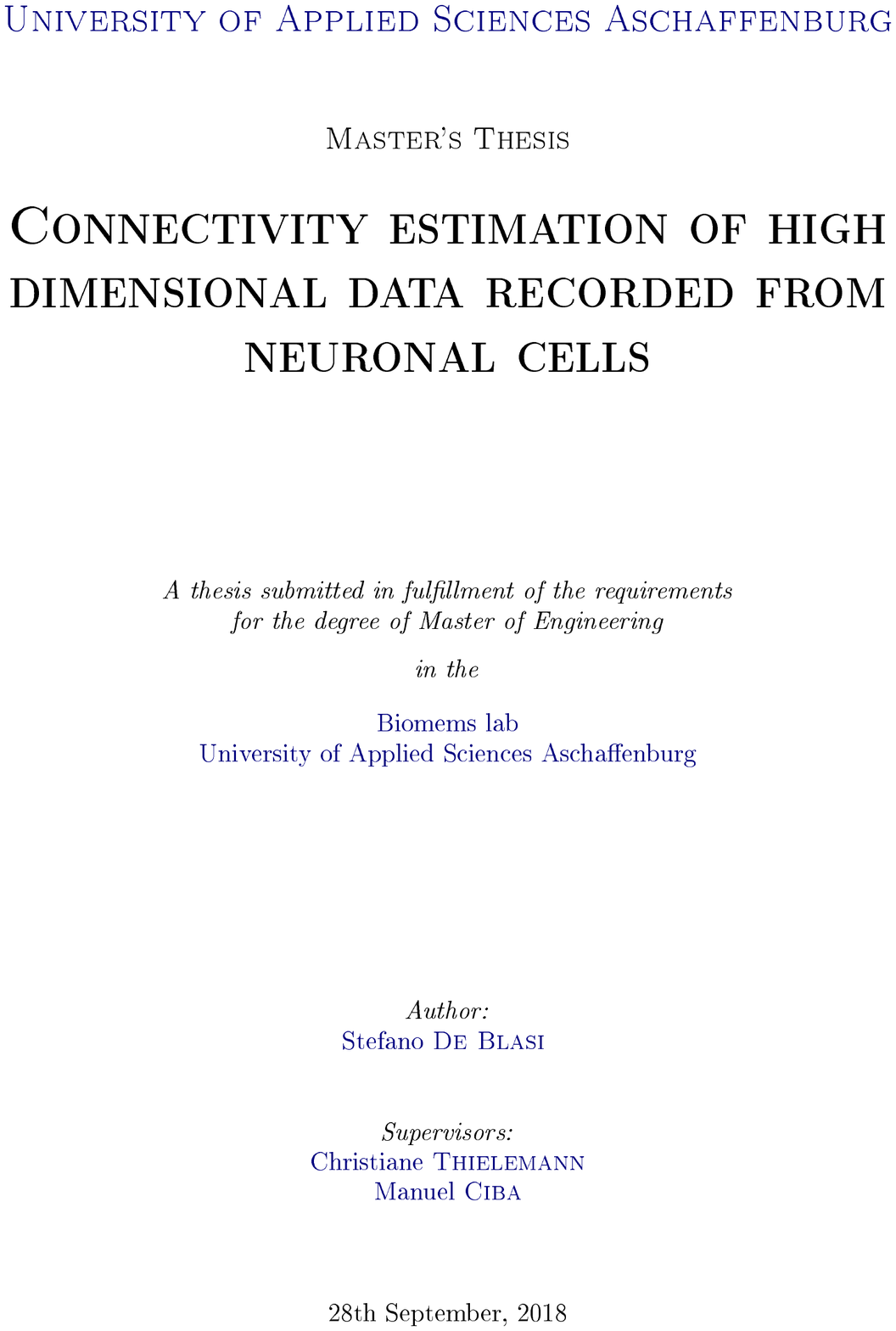}

\cleardoubleemptypage
\chapter*{Declaration of Authorship}\thispagestyle{empty}
	\noindent I, Stefano De Blasi, declare that this thesis titled, \textit{\TITEL} and the work presented in it are my own. I confirm that:
	\begin{itemize} 
		\item This work was done wholly or mainly while in candidature for a research degree at the \ac{UAS} Aschaffenburg.
		\item Where any part of this report has previously been submitted for a degree at the \ac{UAS} Aschaffenburg, this has been clearly stated.
		\item Where I have consulted the published work of others, this is always clearly attributed.
		\item Where I have quoted from the work of others, the source is always given. With the exception of such quotations, this thesis is entirely my own work.
		\item Some parts of this work are the pre-print versions of the publications I submitted during my candidature for a research degree at the \ac{UAS} Aschaffenburg. These publications have not been submitted for a degree.
		\item I have acknowledged all main sources of help.
		\item Where the report is based on work done by myself jointly with others, I have made clear exactly what was done by others and what I have contributed myself.\\
	\end{itemize}
	\vspace{5cm}
	\noindent Signed:\\
	\rule[0.5em]{25em}{0.5pt} 
	
	\noindent Date: \hspace{2cm} 28th September, 2018\\
	\rule[0.5em]{25em}{0.5pt} 

\cleardoubleemptypage

\vspace*{\fill}{
\vfill{\large
\noindent
Author:\newline
\AUTORNAME \newline
\AUTORSTR  \newline
\AUTORPLZ~\AUTORORT
}
\vfill{
\noindent
\large {First examiner and supervisor:}
\newline
\large \ERSTPRUEFER
\newline
\newline
\newline
\large {Second examiner:}
\newline
\large \ZWEITPRUEFER
\newline
\newline
\newline
\newline
\large {Supervisor:}
\newline
\large \BETREUER
}
\bigskip
\noindent
\bigskip
\newline
\bigskip
\noindent
\bigskip
\newline
\begin{minipage}[t]{0.68\textwidth}
	\vfill{\large
		\ac{UAS} Aschaffenburg \newline
		Faculty \FAKULTAT \newline
		\HOCHSCHULSTRASSE \newline
		\HOCHSCHULEORT
	}
\end{minipage}
\bigskip
\begin{minipage}[t]{0.28\textwidth}
	\centering
	\vfill{ \includegraphics[height=2.5cm]{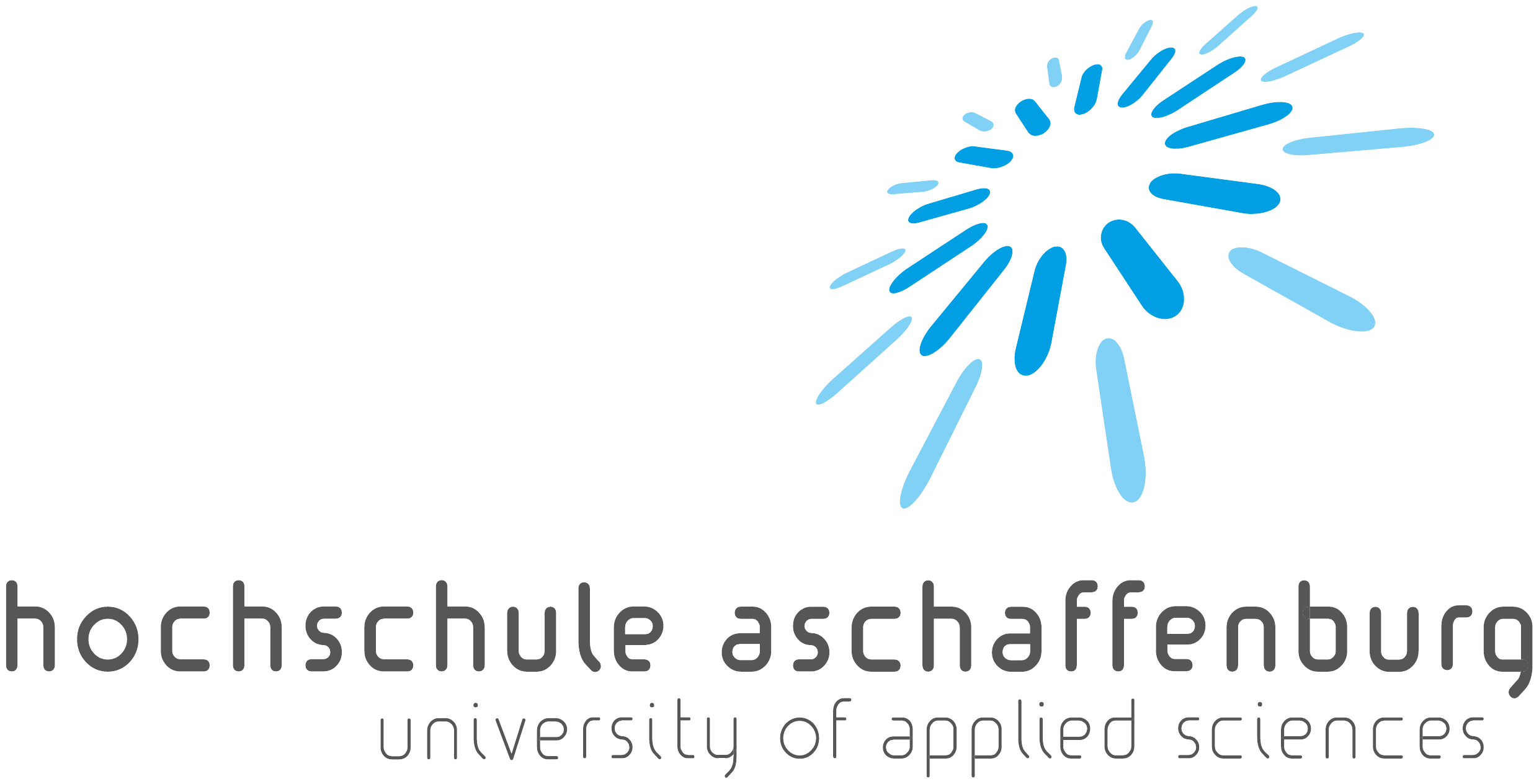}}
\end{minipage}
\vspace{4cm}
\bigskip
\noindent


}
\clearpage

\pagestyle{frontback}
\tableofcontents


\listoffigures
\clearpage

\chapter*{Acronyms}
\addcontentsline{toc}{chapter}{Acronyms}
\begin{acronym}[ticktocktick]
	\acro{APS}{Active Pixel Sensor}
	\acro{BA}{Barabasi--Albert}
	\acro{BaCaTeC}{Bavaria California Technology Center}
	\acro{CA1}{Cornu Ammonis 1}
	\acro{CA3}{Cornu Ammonis 3}
	\acro{CC}{Cross Correlation}
	\acro{CCW}{Counter Clockwise}	
	\acro{CDHOTE}{Combined Delayed Higher Order Transfer Entropy}
	\acro{CH}{Chattering}	
	\acro{CI}{Coincidence Index}
	\acro{CM}[CM]{Connectivity Matrix}	
	\acro{CMOS}{Complementary Metal-Oxide Semiconductor}	
	\acro{COBA}{Conductance-Based}
	\acro{CTU}{Czech Technical University}
	\acro{CUBA}{Current-Based}
	\acro{CW}{Clockwise}
	\acro{D1TE}{Delay One Transfer Entropy}	
	\acro{DG}{Dentate Gyrus}
	\acro{DHOTE}{Delayed Higher Order Transfer Entropy}
	\acro{DHOTECI}{Delayed Higher Order Transfer Entropy Coincidence Index}
	\acro{DIV}{Days \textit{in vitro}}
	\acro{DM}{Delay Matrix}	
	\acro{DTE}{Delayed Transfer Entropy}	
	\acro{DTECI}{Delayed Transfer Entropy Coincidence Index}
	\acro{EC}{Entorhinal Cortex}	
	\acro{ER}{Erdos--Renyi}		
	\acro{FIR}{Finite Impulse Response}
	\acro{FLOP}{Float Operation}
	\acro{FN}{False Negative}
	\acro{FP}{False Positive}
	\acro{FPR}{False Positive Rate}
	\acro{FS}{Fast Spiking}
	\acro{HDMEA}{High Density Microelectrode Array} 
	\acro{HF}{High Frequency}
	\acro{HH}{Hodgkin–Huxley}
	\acro{HOTE}{Higher Order Transfer Entropy}
	\acro{HT}{Hard Threshold}
	\acro{IB}{Intrinsically Bursting}		
	\acro{IC}{Implementation of Catanzaro}		
	\acro{IF}{Integrate-and-Fire}	
	\acro{K-S test}{Kolmogorov-Smirnov test}
	\acro{LR}{Long Recording}
	\acro{LTS}{Low-Threshold Spiking}
	\acro{MCS}{Multi Channel Systems} 
	\acro{MEA}{Microelectrode Array} 
	\acro{MEX}{MATLAB executable}
	\acro{MFR}{Mean Firing Rate}	
	\acro{MI}{Mutual Information}
	\acro{MIND}{Institute for Memory Impairments and Neurological Disorders}
	\acro{MPL}{Mean Path Length}
	\acro{NCC}{Normalized Cross Correlation}
	\acro{NCCCI}{Normalized Cross Correlation Coincidence Index}
	\acro{NCCH}{Normalized Cross Correlation Histogram}
	\acro{PDMS}{Polydimethylsiloxane}
	\acro{PP}{Paired Pulse}
	\acro{PTSD}{Precise Timing Spike Detection}
	\acro{QBD}{Quantile Based Detection}
	\acro{ROC}{Receiver Operating Characteristic}
	\acro{RS}{Regular Spiking}
	\acro{SD}{Standard Deviation}
	\acro{SII}{Standard Implementation of Izhikevich}
	\acro{SNR}{Signal-to-Noise Ratio}
	\acro{SPE}{Spiking Probability Edges}
	\acro{SS}{Summer Semester}	
	\acro{STDP}{Spike-Timing Dependent Plasticity}
	\acro{SWDT}{Sliding Window Differential Threshold}
	\acro{SWM}{Synaptic Weight Matrix}
	\acro{TC}{Thalamo-Cortical}
	\acro{TCM}{Thresholded Connectivity Matrix}
	\acro{TE}{Transfer Entropy}
	\acro{TETRA}{Terrestrial Trunked Radio}
	\acro{TN}{True Negative}
	\acro{TP}{True Positive}
	\acro{TPR}{True Positive Rate}
	\acro{TSPE}{Total Spiking Probability Edges}
	
	\acro{UAS}{University of Applied Sciences}  
	\acro{UCI}{University of California Irvine}
\end{acronym}

\cleardoubleemptypage

\newcounter{verz}
\setcounter{verz}{\value{page}}

\mainmatter
 
\pagestyle{main}

\chapter{Motivation}
The human brain consists of billions of nerve cells, so-called neurons, and is one of the most exciting mysteries of our time.
Many neuroscientists study nervous systems in order to understand the brain, which is essential for human consciousness, motor control or memory and learning. Since brain diseases have devastating effects on our lives~\cite{Olesen.2003}, the need for an improvement of our knowledge about these diseases and their causes is critical in order to develop more effective treatment. Fundamental understanding of the brain could even help to cure neural diseases like epilepsy, Parkinson's disease or Alzheimer's disease~\cite{Bhatti.2017}. Moreover, developments such as bionic prostheses can restore the patient's lost senses. These developments are highly desirable and possible with acquired knowledge about our brain. For example, cochlear implants can be used to make a deaf person hear~\cite{Pinyon.2014}. In addition, retinal prostheses, so-called bionic eyes, are currently attracting a lot of attention and are able to grant blind people visual perception~\cite{Luo.2014}.

To understand neural networks like the brain it is fundamentally necessary to know the topology of them first. How are single neurons connected with each other? Is there some kind of a connection map similar to power grids or motorways? Next, the information flow has to be analysed. How is an information designed and how is it processed by the brain? Are there similarities with already known information networks such as the Internet?
Finally, by observing the information flow, we would be able to detect functions of certain neurons, what provides the basis for precise treatments or bionic interfaces.

Many issues are directly or indirectly connected to the international task understanding the human brain. There are many different approaches to solving these problems.
For example, the biological technology of optogenetics has recently attracted international attention. 
It allows to use genetically modified cells in order to understand functions of rat brains by literally switching parts of a living rat brain on or off to observe effects on the behaviour of the subject~\cite{Deisseroth.2011}. 
An optical fiber goes directly though skin into the brain of living animals to stimulate certain areas or functions. This kind of experiments is called \textit{in vivo} and is ethically questionable not only from the point of view of animal welfare organizations~\cite{Svendsen.2005}. 
While countless researchers already work on mapping brains of humans or animals by using methods like optogenetics or reconstructions of dead rat brain slices~\cite{Shepherd.1998}
there are still many unsolved questions even at small networks with only thousands of neurons. Wouldn't it be more effective to understand small neural networks complexity first?

\acp{MEA} can be used to measure electrical signals of \textit{in vitro} cultures, which means these experiments do not involve living animals. 
By detecting action potentials of a neuronal network there exist methods to make statements about connections between neurons. This task is called estimation of neuronal connectivity and it is part of cutting-edge fundamental research. 

At the biomems lab of \ac{UAS} Aschaffenburg, such \textit{in vitro} experiments are carried out in order to learn more about external influences on neural systems by using cell-based biosensors in form of cultures on \acp{MEA}.
The portfolio of the biomems labs is represented by three projects which could gain in significance in the future through connectivity estimation. Electrophysiological effects on neural network communication of human embryonic stem cell derived neurospheres are investigated~\cite{Mayer.2016}. In Aschaffenburg cell-based sensor chips are also used for neurotoxicity measurements in drinking water~\cite{Flachs.2016}.
In addition, impacts of irradiation like \ac{TETRA} on neural \textit{in vitro} networks cells are explored also in long-terms~\cite{Kohler.2016}.
Observing a significant change in connectivity during these experiments would lead to important biological statements. Furthermore, with an ability to estimate connectivity there are several methods of graph theory to get even more knowledge about the neural network like node degree, path length or efficiency, connection density or cost, hubs, centrality or robustness~\cite{Bullmore.2009}.
These parameters would improve the meaningfulness of future experiments. Moreover, network dynamics could be observed in experimental environment.

The main result of this thesis is the development of a novel connectivity estimation method, called \textit{Total Spiking Probability Edges} (TSPE). Based on cross-correlation and edge filtering at different time scales this method is proposed and the theoretical framework is outlined in this work. TSPE enables the classification between inhibitory and excitatory connections by using recorded action potentials.

To compare this method learning about state of the art algorithms to estimate 
connectivity is necessary. After a research, promising algorithms are implemented and evaluated for further research topics, among others in the biomems lab of \ac{UAS} Aschaffenburg.
To evaluate these algorithms \textit{in silico} networks are used, because of their known connectivity. This makes it possible to validate the correctness of our algorithm results. Therefore, a biophysically representative neuronal network simulation is needed first. Datasets were simulated in different ways and analysed in order to develop an evaluation framework.  
After a successful evaluation with \textit{in silico} networks, \textit{in vitro} experiments and their analyses complete this project. 
\chapter{Fundamentals}
This chapter covers relevant fundamentals which are necessary to understand the project beginning with some basics of neuronal systems. Section~\ref{Sec:Typesofconnectivity} provides knowledge about connectivity in neuronal networks. In Section~\ref{Sec:SpikeTrain}, the extraction and preprocessing of neuronal data from real cultures \textit{in vitro} is explained.

\section{Elements of neuronal systems}
\begin{figure}[!htbp]
	\begin{center}
		\includegraphics[width=0.85\textwidth]{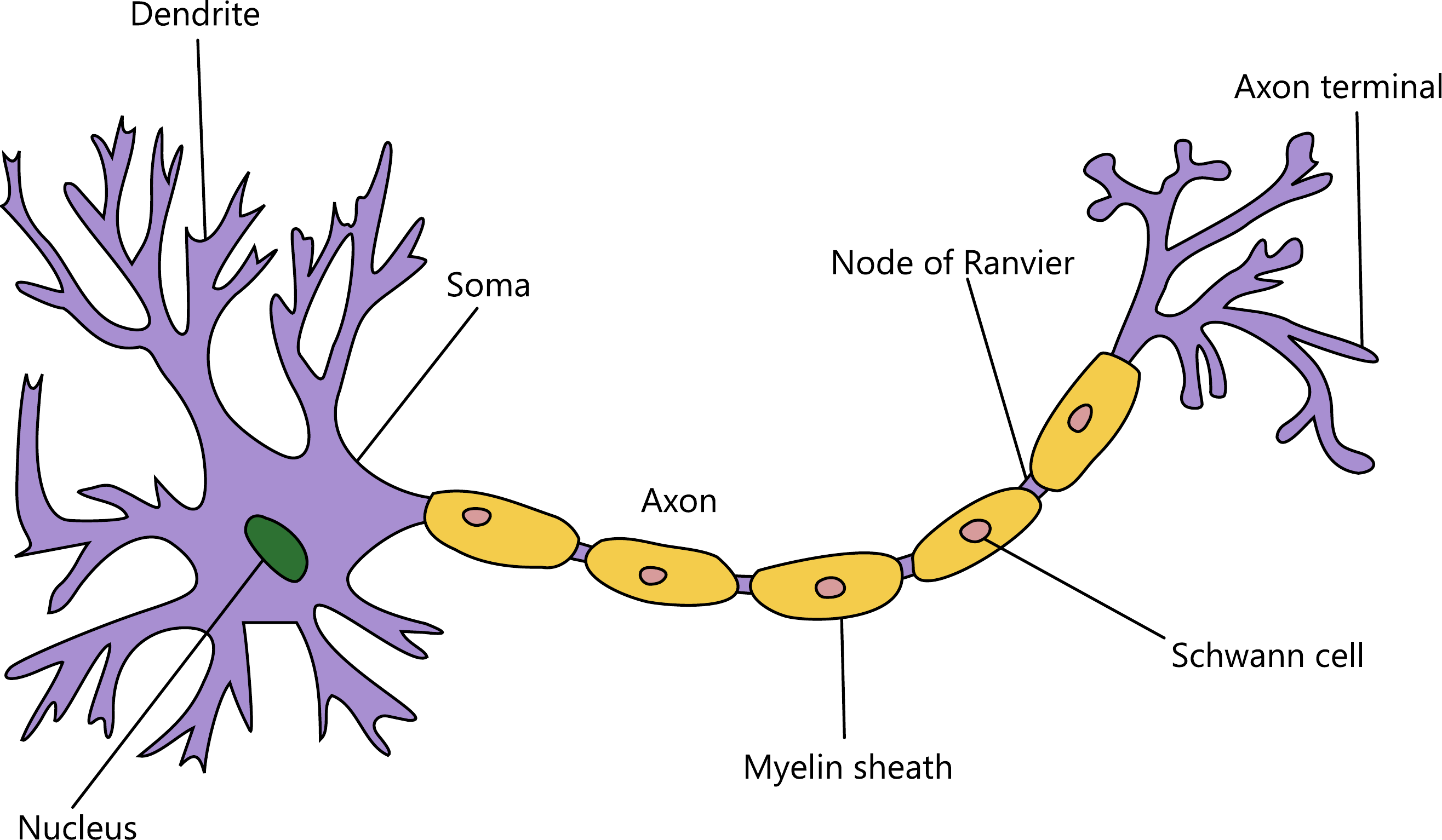}
	\end{center}
	\caption[Structure of a typical neuron]{\textbf{Structure of a typical neuron:}  
	On the left-hand side dendrites are the input side of a neuron. The body of a neuron is called soma. Through an axon an action potential will be emitted to the axon terminals, which are the output side. The connection between a dendrite and terminal is a so-called synapse.	
	Figure by~\cite{Jarosz.2009}.}\label{fig:neuron}
\end{figure}

\begin{figure}[!htbp]
	\centering
	\includegraphics[width=0.85\textwidth]{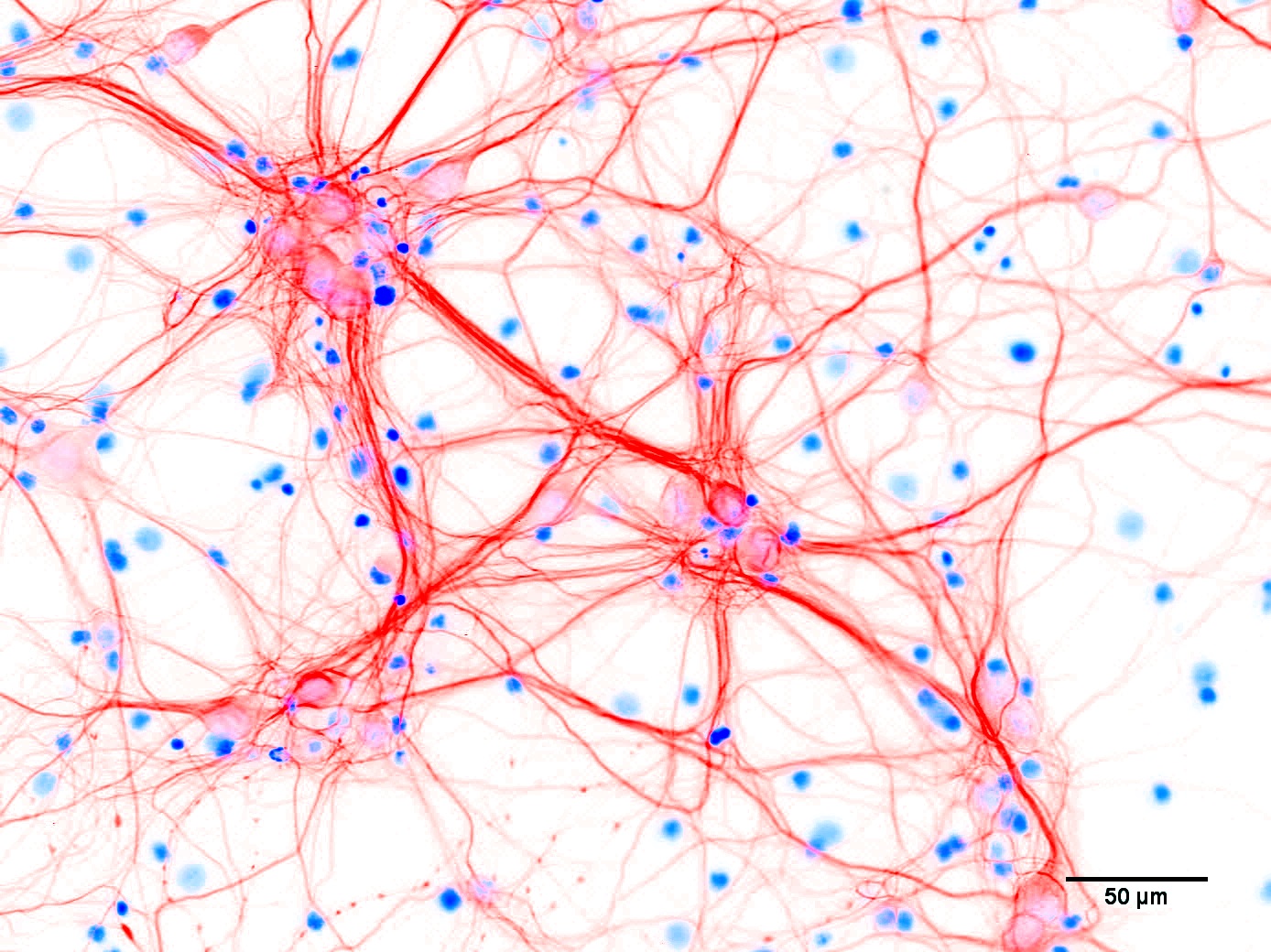}
	\caption[Biological neural network \textit{in vitro}]{\textbf{Biological neural network  \textit{in vitro}: \\}A fluorescence image of a neuronal \textit{in vitro} culture was taken, with blue coloured cell bodies and red coloured dendrites. Origin picture taken by Margot Mayer at \ac{UAS} Aschaffenburg. \\}\label{fig:nn}
\end{figure}

A neuronal network consists of many neurons connected which each other. Each neuron has the ability to change its membrane potential, that is given by different ion concentrations within (intracellular) and outside (extracellular) the nerve cell. Changes are enabled by ion channels and pumps transferring mainly sodium ions $Na^{+}$ and potassium ions $K^{+}$ across the cell membrane. The flow of sodium ions produces an increase in membrane potential, leading to the rapid opening of even more channels (reaction series). This immense influx inactivates the sodium ion channels by polarization and activates the potassium ion channels (membrane potential decreases again).
Between two neurons, information is transmitted using these described electrical pulses, so-called action potentials~\cite{Dayan.2005}. These pulses are generated by each individual neuron and sent to adjacent nerve cells along the connections of a network. In Figure~\ref{fig:neuron}, the structure of a typical neuron is illustrated. 
These neurons are connected to each other in a neuronal network (see Figure~\ref{fig:nn}).
The axon can be seen as the output of a neuron while dendrites are the inputs received from other neurons. Therefore, the way of an action potential consists basically of three parts:
\begin{itemize}
	\item Axon of the emitting, also presynaptic, neuron
	\item Synapse connecting an axon and a dendrite
	\item Dendrite ending at the soma of a receiving, also postsynaptic, neuron
\end{itemize}
While the soma can be seen as a processing unit, it emits new action potentials depending on its input. Basically there are two types of neurons: Excitatory and inhibitory. Very simplified you may imagine a threshold for each soma, excitatory inputs add a value while inhibitory subtract another value. If the sum of these inputs is big enough to pass the postsynaptic threshold its neuron will emit an action potential, this process is also known as firing~\cite{Gerstner.2014}.

Neural networks are able to toughen a synapse by using it, which is basically the ability to learn something. This hypothesis for learning processes in the brain is called Hebbian theory~\cite{Hebb.1949}.
Since neuronal coding theories take only action potentials into account, the amplitude is not of interest. Thus, the important information of recorded neuronal data is the history of detected peaks, which are then called spikes. Temporal sequences of all spikes are referred to spike trains. In this work discussed connectivity estimation algorithms use such spike trains. Multiple spikes of a same emitting neuron in a short time are called bursts. Furthermore, in neuronal networks neurons may fire almost synchronously in short time intervals, which is known as network bursts.

Further information of neuronal systems can be found in literature like the books \textit{Theoretical Neuroscience} by Dayan or \textit{Neuronal Dynamics: From Single Neurons to Networks and Models of Cognition} by Gerstner~\cite{Dayan.2005,Gerstner.2014}.

\section{Types of connectivity}\label{Sec:Typesofconnectivity}
\begin{figure}[htbp]
	
	\begin{center}
		\includegraphics[width=1\textwidth]{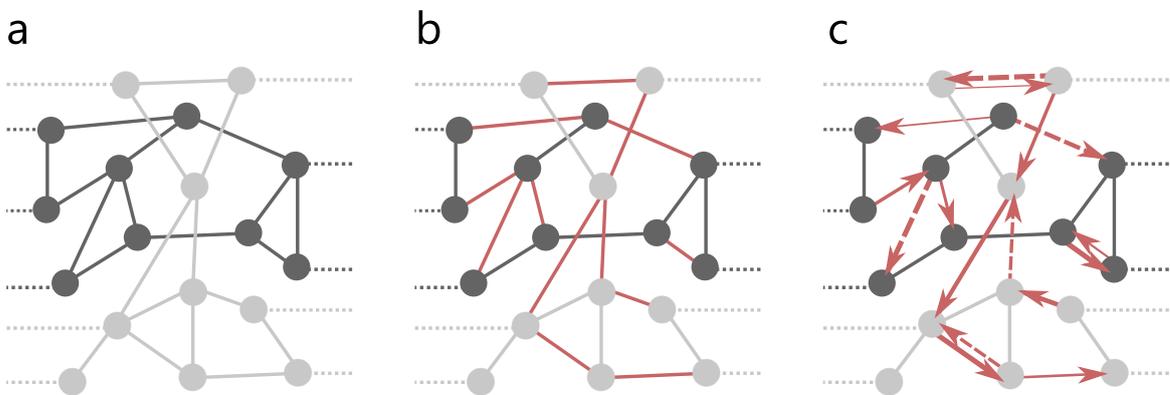}
	\end{center}
	\caption[Types of connectivity]{\textbf{Types of connectivity:} (a) Structural connectivity, connections and neurons are marked with same color. Two unconnected networks have no anatomical connection to each other. (b) Functional connectivity, used connections are marked with red. Information was transferred over these connections. (c) Effective connectivity, used connections are marked with red, bigger connection widths denotes stronger influences and arrows label causality. Dashed connections indicates inhibitory behaviour, while normal ones stays for excitatory behaviour. Detailed knowledge about transferred information is possible. (a), (b) and (c) illustrate the same network in different perspectives in the form of connection types. Figure inspired by~\cite{Poli.2015}.}\label{fig:contypes}
\end{figure}

Connectivity provides knowledge of how a network is connected. Like every kind of network also neuronal networks are based on nodes and connections. In this context, the nodes are nerve cells and, like mentioned before, the connections consist basically of axons, synapses and dendrites. 
There are three different types of connectivity~\cite{Friston.1994, Poli.2015} which are often used to describe the topology of neuronal networks:
\begin{itemize}
	\item Structural connectivity, the pure existence of connections (see Figure \ref{fig:contypes}.a)
	\item Functional connectivity, the knowledge about used connections (see Figure \ref{fig:contypes}.b)
	\item Effective connectivity, the detailed knowledge about used connections (see Figure \ref{fig:contypes}.c)
\end{itemize}
In addition to the illustration, their similarities, differences and characteristics are explained in the following.

\subsection{Structural connectivity}
A structural or anatomical connection refers to physical interactions. In this particular case, it is an electrical or biochemical link between two neurons known as synapse. Thus, structural connectivity is the ability to communicate which does not mean connections are necessarily used for observed network activity. For example, the connectome can be seen as a structural network map of the human brain~\cite{Sporns.2005}.
All synaptic connections are time- or activity-dependent~\cite{Saneyoshi.2010}.
Since short timescale investigations showed static morphological connections, the structural connectivity is manifest over a short time. Over longer terms anatomical connections are able to grow or fully degenerate~\cite{Buchs.1996}.
Knowledge about this type of connectivity does not offer us to understand real functions of a network but about the ability of a function which could be possible. It is similar to the knowledge about existing streets of a motorway network without knowing anything about the traffic.

\subsection{Functional connectivity}
As the name indicates functional connectivity allows statements about functions of a neuronal network. Such a functional connection exists if the activity of two neurons is correlated somehow. 
This interaction is not described in more detail, since each functional connection is just defined as an activity correlation.
In contrast to structural connectivity, the functional one can not be detected by using optical methods without genetic modifications (e.g. sequences of fluorescent calcium indicator proteins are incorporated into the genome for so-called calcium imaging methods)~\cite{Renault.2015}.
On the contrary, there would still be attained knowledge using normal imaging, since functional connections are a subset of structural connections~\cite{Sporns.2002}. 
The ability of synapses to strengthen or weaken over time is called synaptic plasticity and leads to an even greater time-dependency of functional connectivity, due to more direct influences of activity and the fact that structural connectivity is just manifest for a specific time~\cite{Abbott.1997}.

\subsection{Effective connectivity}
By knowing the third and final type of connectivity we are able to describe effects of each interaction in detail. 
The amount of influence between neurons offers to distinguish weak connections from strong connections. Investigations can also differ between inhibitory or excitatory behaviour of a synaptic connection. 
Effective connectivity makes it even possible to answer causality questions of synapses~\cite{Poli.2015}. For instance, which neuron is source and which is target? 
By knowing all these details, a neuronal network can be fully reconstructed in its characteristics. Since the observed effects are defined by the used model, effective connectivity is not model free like the other types of connectivity~\cite{Friston.2011}.

As one may already noticed, effective connectivity is a subset of functional connectivity and in order of that also a subset of structural connectivity. 
Like functional connectivity, the same impact of synaptic plasticity appeals effective connectivity. Thus, it changes its properties with time fast.

\section{Spike train data \textit{in vitro}}\label{Sec:SpikeTrain}
Since information transfer is decisive, estimation of functional or effective connectivity is only possible by observing the communication events between neurons, which are action potentials as mentioned. Therefore, these communication events are required to be recorded. The neuronal raw data then has to be preprocessed for a demanded format, which is based on detected action potential peak times, so-called spikes. The history of these spikes is called spike train data. In this section the way of signals from neuronal cultures \textit{in vitro} to connectivity estimation algorithms will be explained.
\subsection{Measurement of neuronal raw data in vitro}\label{sec:MEA}
\begin{figure}[!htbp]
	\begin{center}
		\includegraphics[width=0.96\textwidth]{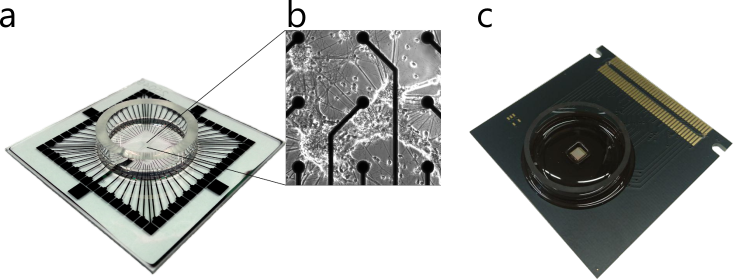} 
	\end{center}
	\caption[MEA and HDMEA chip]{\textbf{\ac{MEA} and \ac{HDMEA} chip:} (a) 60 electrode \ac{MEA} chip by \ac{MCS} with an electrode diameter of 30\,\textmu m and spacing of 200\,\textmu m. (b) An enlargement of some electrodes (black dots) and a neuronal culture whose signals can be measured at the electrodes and transmitted via trace (black lines). (c) 4096 electrode \ac{HDMEA} chip by 3brain with an electrode length of 21\,\textmu m and spacing of 21\,\textmu m.}\label{fig:meachips}
\end{figure}

To measure signals of \textit{in vitro} cultures various methods exist. These techniques are divided into two subgroups: Intracellular and extracellular. While intracellular measurements require at least one electrode inside the neuron or axon and destroy them afterwards, extracellular methods try to measure at the surface of cells without damaging them~\cite{Bhatti.2017}.
In this project non-implantable \acp{MEA} are used. \ac{MEA} chips are capable of simultaneously recording multiple electrical signals from investigated objects such as cardiac muscle cells, neuronal cultures, hippocampal slices or stem cells extracellularly with a dense array of biocompatible electrodes. Common \ac{MEA} chips (see Figure \ref{fig:meachips}.a) of \ac{MCS} (Reutlingen, Germany, \url{http://www.multichannelsystems.com}) consist of 60 electrodes distributed on an area size of about 1.4x1.4\,mm$^2$, but there are also chips with 120 electrodes.

New technologies like \acfp{HDMEA} make it possible to improve our measurements by recording a culture with more electrodes which are closer to each other. Instead of 60 channels an \ac{HDMEA} chip (see Figure \ref{fig:meachips}.c) by 3brain (Wädenswil, Switzerland, \url{http://www.3brain.com/}) is able to record signals from up to 4096 (64x64) channels on an area size of about 2.67x2.67\,mm$^2$. 
This also means a reduction of inter-electrode-distance from 200 to 21\,\textmu m. Electrodes are sampled simultaneously at a frequency up to a maximum of 18\,kHz, that is enabled by using a special \ac{APS} based on \ac{CMOS} technology~\cite{Berdondini.2009}. In Figure \ref{fig:meas} layouts of \ac{MEA} and \ac{HDMEA} are comparably visualized. 

\begin{figure}[!htbp]
	\begin{center}
		\includegraphics[width=0.88\textwidth]{./pic/meacompare.pdf}
	\end{center}
	\caption[Layout of MEA and HDMEA]{\textbf{Layout of \ac{MEA} and \ac{HDMEA}:} (a) On the left a drawn to scale 60 electrodes \ac{MEA}: Electrodes have a diameter of 30\,\textmu m, the electrode distance is 200\,\textmu m and working area is about 1.4x1.4\,mm$^2$. (b) On the right a drawn to scale 4096 electrode \ac{HDMEA} by 3brain: Electrodes have a size of 21x21\,\textmu m$^2$, each electrode distance is 21\,\textmu m and the working area is about 2.67x2.67\,mm$^2$. Figures taken at \ac{UAS} Aschaffenburg}\label{fig:meas}
\end{figure}

The improvement of electrode density is obvious. Consequently, \ac{HDMEA} chips are able to help us understanding how neuronal networks work even better by increasing the percentage of recorded neurons in a network~\cite{Bhatti.2017}. However, with smaller spaces between electrodes also some disadvantages in signal processing are produced, which will be explained in the next step.

\subsection{Signal preprocessing}\label{sec:preprocess}
\begin{figure}[!htbp]
	\begin{center}
		\includegraphics[width=1\textwidth]{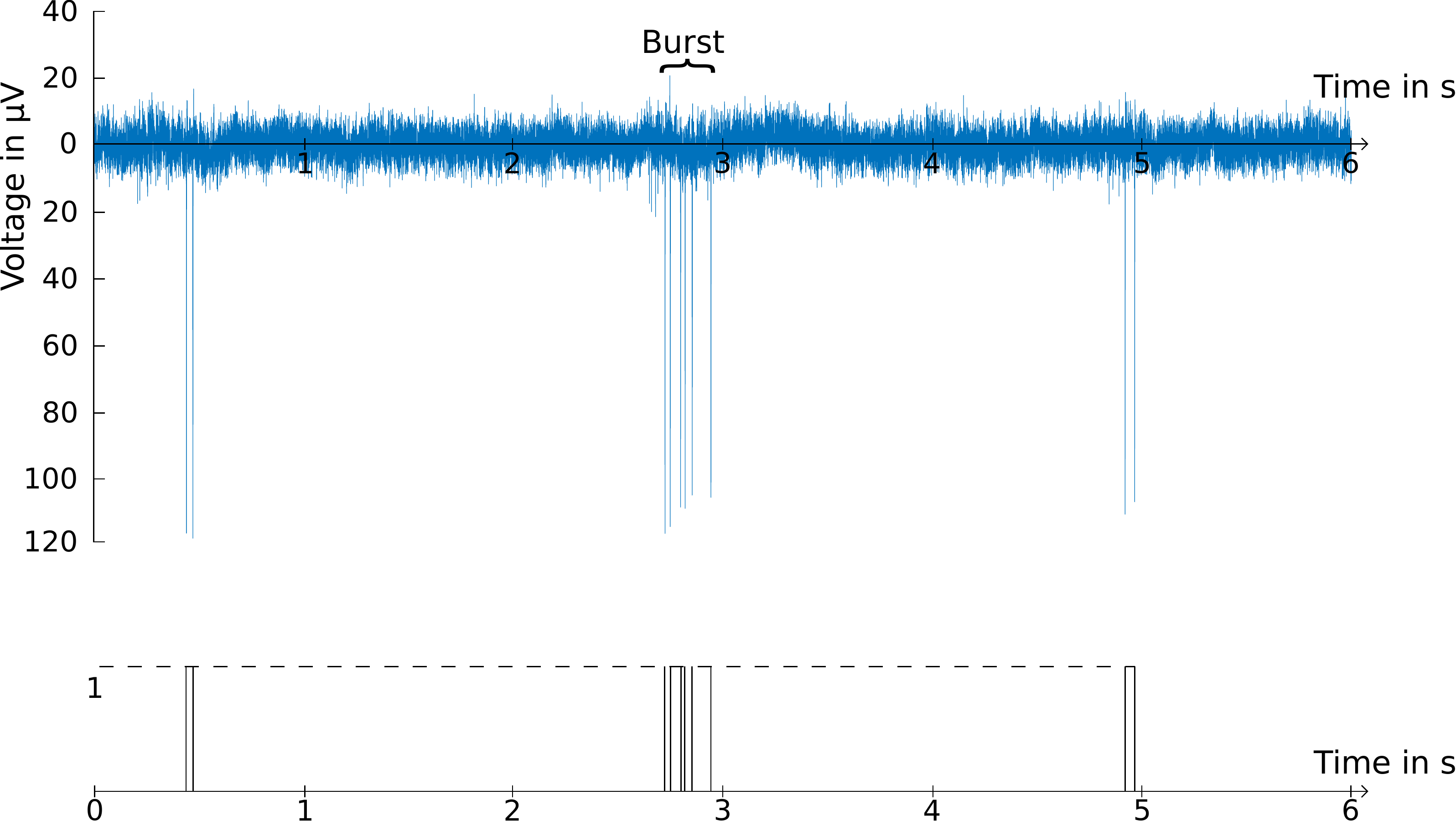}
	\end{center}
	\caption[Spike detection]{\textbf{Spike detection:} Recorded raw data of neuronal cells is converted into a binary spike train via filtering and a spike detection algorithm. Multiple spikes in a short time window define a burst.}\label{fig:spikedetection}
\end{figure}

After a successful measurement neuronal raw data is noisy because of externally coupled signals and neurons, which are too far away from the electrode to be measured but near enough to take influence in form of background noise. Furthermore, recorded neuronal signals have broad frequency spectra. Since spikes are short voltage pulses (see Figure \ref{fig:spikedetection}), digital filters can be used to improve the detection of spikes. At \ac{UAS} Aschaffenburg the standard procedure is a reduction of low-frequency portions by using a high-pass \ac{FIR} filter. 

The next crucial step is spike detection, which determines the peaks of network activity. Especially in context with \acp{HDMEA} it is still an up-to-date topic~\cite{Muthmann.2015}. The easiest and fastest spike detection is simple \ac{HT}, which is widely used~\cite{Obeid.2004, Maccione.2009b}. Since a low signal-to-noise ratio can be problematic, there are several different methods to detect spikes with adaptive thresholds~\cite{Chan.2008}. By using stationary wavelet transform or time-frequency based algorithms a better performance is also possible~\cite{Lieb.2017}. In Figure \ref{fig:spikedetection} the basic function of spike detection is illustrated. The preprocessing software for \ac{HDMEA} chips of 3brain \textit{BrainWave} provides some spike detection methods like \ac{PTSD}, \ac{SWDT}, \ac{HT} or \ac{QBD}. In Aschaffenburg, for standard 60 channel \ac{MEA} chips the \textit{MATLAB} based software \textit{DrCell} is used and comes also with tools like filtering and spike detection~\cite{Nick.2013}. The result of spike detection for a signal is then the spike train, which is the binary history of spike times: One for spikes or zero for no spiking at a sampling time.

Since the signals from more than just one cell are able to be measured with an electrode, some researchers try to separate and assign the detected action potential to a specific cell. This step is called Spike Sorting~\cite{Bestel.2012}.

\begin{figure}[H]
	\begin{center}
		\includegraphics[width=1\textwidth]{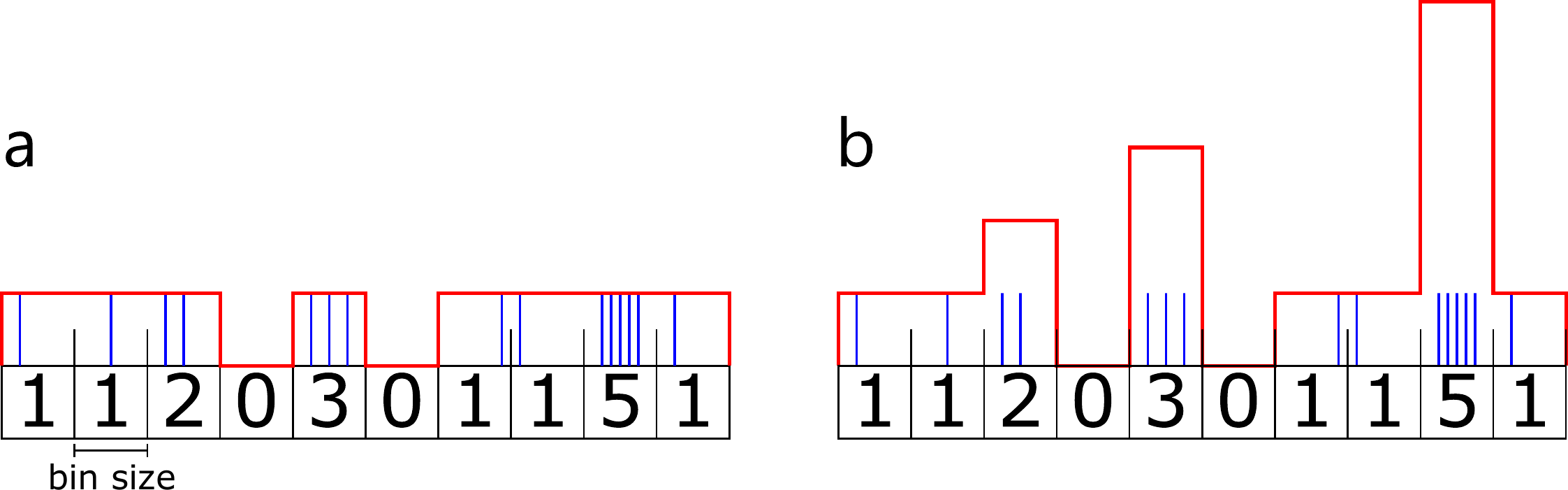}
	\end{center}
	\caption[Binning types]{\textbf{Binning types:} A spike train (blue-colored) is binned (red-colored) by select a bin size and sum up all underlaying spikes (numbers in boxes). 	
	Binary binning (a) compromises sum values to binary states: One for each sum value bigger than zero else zero.	
	The second, also widely used method is multi-stage binning (b), which directly uses the sum values for each bin.}\label{fig:binning}
\end{figure}

However, many connectivity estimation algorithms require to use binned spike trains for improving their performance in terms of quality and speed. Selecting a bin size the spike train is divided into equal pieces of that size. For each bin, the spikes are added together in this time window. Basically there are two types of binning:
\begin{itemize}
	\item Binary binning, in which the exact amount of spikes in a bin is irrelevant (see Figure \ref{fig:binning}.a)
	\item Multistage binning, in which the exact amount of spikes in a bin is relevant (see Figure \ref{fig:binning}.b)
\end{itemize}

In addition to the choice of a binning type, the choice of a bin size also influences the performance of an algorithm.
Moreover, spike sorting algorithms are used to assign spikes to specific neurons. Since it is possible for an electrode to have contact with more than one cell, these algorithms try to discriminate between shapes of spikes in order to identify which spike is emitted by which neuron~\cite{Lewicki.1998}. Especially for \ac{HDMEA} chips, it can also be a problem if a neuron is recorded multiple times by adjacent electrodes.

\chapter{Simulation framework for evaluation applications}\label{cha:sim}
In contrast to real neuronal data (see Section~\ref{Sec:SpikeTrain}), simulations are able to produce datasets with a known network connectivity. 
These datasets play a major role to quantify how accurate the results of connectivity estimation methods are.
As biological neural networks - even \textit{in vitro} - are far too complex (see Figure.~\ref{fig:nn}) to reveal connections between single neurons, \textit{in silico} neural networks with known topology are typically modelled.
These simulated networks should be biophysically representative for a meaningful evaluation of neurocomputational algorithms, like connectivity estimation.

As the topology of an \textit{in silico} network can affect the results and accuracy of algorithms~\cite{Kadirvelu.2017}, it is essential to evaluate these methods with realistic topologies.
In the following, known \textit{in silico} models are compared with respect to their applicability in large-scale simulations for biological neural networks. Implementations for different topologies are analysed in terms of biological plausibility. Based on these findings, an evaluation framework for the evaluation of connectivity estimation algorithms is developed. Parts of this section have already been published during the work on the project~\cite{DeBlasi.2018}.

\section{Selection of a neuron model}
Like mentioned in Section \ref{Sec:Typesofconnectivity} a neuronal network consists of somata (nodes) and axons, synapses and dendrites (connections). There are many types of neuron models with different strengths and weaknesses. We considered three well known types and rated them for the sensuous usage for testing connectivity algorithms. 
First, the \ac{HH} model takes experimentally measured ion channels and pumps into account but is too complex for large network simulations, due to its high number of needed differential equations~\cite{Hodgkin.1952} (see Appendix \ref{A:HH} for further information).
Second, the \ac{IF} model handles just essential functions of a neuron to be simulated in large quantities~\cite{Vogels.2005}. Since many observed processes are not captured by the \ac{IF} model, it is too simplified for a meaningful evaluation (see Appendix \ref{A:IF} for further information).  Several models, also \ac{HH} and \ac{IF}, were compared in computing complexity and biological plausibility by investigate their possible features~\cite{Izhikevich.2004}. The Izhikevich model is a simplification of the \ac{HH} model by using two-differential equations. This model was chosen because of its good trade-off and its good reputation for biological simulations. In Appendix \ref{A:Izh_plot} further information can be found.

\subsection*{Izhikevich model}
The Izhikevich model is especially designed for so-called large-scale simulations, which means up to tens of thousands of neurons. By combining the biological plausibility of \ac{HH} model with the computational efficiency of \ac{IF} model the Izhikevich model is able to reproduce spiking and bursting of cortical neurons in real time~\cite{Izhikevich.2003}. 
The operating principle is based on the variable $v$, which represents the membrane potential while $u$ handles a slower recovery. For this reason $u$ is also called recovery variable and takes the inactivation of sodium $Na^{+}$ and activation of potassium $K^{+}$ channels into account. Both variables obey the dynamics (\ref{equ:Izh_v}) and (\ref{equ:Izh_u}). Impacts driven by synaptic currents are realized with variable $I$.
\begin{equation}\label{equ:Izh_v}
\dot{v} = 0.04 v^2 + 5 v + 140 - u + I
\end{equation}
\begin{equation}\label{equ:Izh_u}
\dot{u} = a \cdot (b \cdot v - u)
\end{equation}
By reaching the threshold of $v \ge 30  \mathrm{~mV} $, the neuron emits an action potential and the auxiliary after-spike resetting is activated. In that case variables $v$ and $u$ are changed obeying the rule, 
\begin{equation}\label{equ:Izh_aux}
\mathrm{~if~}  v \ge 30  \mathrm{~mV, then} 
\begin{cases}
v = c \\
u = u + d .
\end{cases}
\end{equation}
Parameters adapt the model for different types of neurons. $a$ is a time scale parameter for recovery variable $u$. Small values will lead to a slow recovery. For excitatory neurons $0.02$ is chosen for $a$, while inhibitory ones use $0.1$ normally. Thus, excitatory neurons recover faster than inhibitory cells.
$b$ describes the sensitivity of $u$ to the subthreshold fluctuations of $v$. Great values of $b$ will enable a strong coupling of $v$ and $u$. This could end up in low-threshold spiking behaviour or subthreshold oscillations. A typical value for $b$ is $0.2$.
$c$ is the reset potential after each spike. Typically $-65$\,mV is used for both, excitatory and inhibitory neurons.
Finally parameter $d$ describes the reset value of the recovery variable $u$ after each spike. While inhibitory neurons use $2$ for $d$, excitatory ones are enabled with $8$ usually.
Users of the Izhikevich model are able to select between several neuron types by configure the parameters $a$, $b$, $c$ and $d$. The published combinations~\cite{Izhikevich.2003} are mentioned in Table \ref{tab:Izh}. 
\vspace{-1mm}
\begin{table}[h]
	\caption[Configuration of neuron types]{\textbf{Configuration of neuron types:} Options of the Izhikevich model to define different types of neurons only by using parameter combinations. Three excitatory (exc.) and two inhibitory (inh.) types are provided as well as a thalamo-cortical type. Information by Izhikevich~\cite{Izhikevich.2003}.}\label{tab:Izh}
	
	\begin{center}		
		\begin{tabular}{c | l | l | l | l | l }
			\hline
			\multicolumn{1}{c}{}& \multicolumn{1}{c}{Neuron type} & \multicolumn{1}{c}{$a$} & \multicolumn{1}{c}{$b$ } & \multicolumn{1}{c}{$c$} & \multicolumn{1}{c}{$d$}\\
			\hline
			\parbox[t]{2mm}{\multirow{3}{*}{\rotatebox[origin=c]{90}{Exc.}}} 
			
			& \ac{RS} & 0.02 & 0.20 & -65 & 8.00 \\ 
			
			& \ac{IB} & 0.02 & 0.20 & -55 & 4.00 \\ 
			
			& \ac{CH} & 0.02 & 0.20 & -50 & 2.00 \\ 
			
			\hline
			
			\parbox[t]{2mm}{\multirow{3}{*}{\rotatebox[origin=c]{90}{Inh.}}} & \ac{FS} & 0.10 & 0.20 & -65 & 2.00 \\ 
			
			& \ac{LTS} & 0.02 & 0.25 & -65 & 2.00 \\ 
			&&&&&\\
			\hline
			& \ac{TC} & 0.02 & 0.25 & -65 & 0.05 \\ 
			\hline
			
		\end{tabular}
	\end{center}
\end{table}\vspace{-5mm}

\section{Modeling of networks}
Besides defining neuron types by using parameter combinations (see Table \ref{tab:Izh}), there are also synaptic properties to adjust for a neuronal network \textit{in silico}. Therefore, two matrices will be introduced which handle the construction of networks: \ac{SWM} and \ac{DM}. 
\subsection{Synaptic Weight Matrix}
Strengths of connections are stored in a square symmetric matrix with size $N$, which is the number of neurons. This parameterizable matrix is known as \ac{SWM}. Each row represents targets of one neuron. The column index of the \ac{SWM} then indicates source neurons of these targets. Thus, for instance, element $(4,8)$ of the \ac{SWM} stays for the synaptic weight of a connection from neuron 4 to neuron 8. Value ranges depends on the used model while the sign indicates types of connections and value 0 means no connection. Weights of the Izhikevich model are chosen for a maximum of 10 for excitatory and a minimum of -5 for inhibitory synapses. Adjusting the \ac{SWM} we are able to design any topologies.
\subsection{Delay Matrix}\label{DM}
Structured like the \ac{SWM} a second matrix stores durations for each signal transfer taken place. In a realistic picture long axons could lead to longer transmission time. Times are randomly distributed over a value range of 1 to 20\,ms. Since reported, these range is realistic for monosynaptic delay times in mammalian cortex~\cite{Mason.1991, Swadlow.1994}. By knowing \ac{DM} and \ac{SWM} it would be possible to reconstruct the network topology in an geometric formation: Connection existences depend on the \ac{SWM} and their lengths depend on the \ac{DM}.

\section{Network types}
All kinds of networks are able to manage certain goods in very different ways. Therefore, the topology depends on its application. For example, a motorway network is simpler constructed than the World Wide Web. In particular, sums of all incoming and outgoing connections of nodes can be very different. Such a sum is also called degree in graph theory. By investigating these degrees in the whole network and using a histogram, the so-called degree distribution is a common tool to characterize the network.
Basically in network theory there are four subgroups of networks to distinguish. This section will briefly introduce these network types. For further information the book \textit{Network Science} by Barabasi is recommended~\cite{Barabasi.2016}.

\subsection{Regular networks}
\textit{Regular networks} are defined with a fixed number of connections for each node, which means a constant degree $k$ for the whole network and a standard deviation for degree distribution of zero~\cite{Poli.2015}. The connection probability is one for degree $k$ and else zero, 
\begin{equation}\label{equ:reg}
P{(deg = k )} = 1 .
\end{equation}
Mostly, the regularity is ensured with connections between neighbours. Since such networks are not found in nature, uniform grid structure can only be artificially created for biological neuronal networks.

\subsection{Random networks}
A \textit{random network} is constructed by using a constant connection probability following a Poisson distribution,
\begin{equation}\label{equ:rand_d}
P{(deg = k )} = e ^ {-Np} \cdot \frac{(N \cdot p)^k}{k!} .
\end{equation}
Nodes with significantly higher or lower numbers of connections are very rare, but unlike the \textit{regular network}, they exist~\cite{Barabasi.2003}. A clearly identifiable mean degree can be recognized for $N \cdot p$, where $N$ is the number of neurons and $p$ is the connection probability. All nodes are connected to the same number of other nodes on average, but there is a standard deviation.
For the illustrated \textit{random network} in Figure~\ref{fig:scalefree_vs_rand}, mean degree would be around five, which is the maximum of degree distribution.

\subsection{Small-world networks}
The combination of \textit{random} and \textit{regular networks} is called \textit{small-world network}. By slowly decreasing regularity of a \textit{regular network} small-worlds will arise: Most nodes are connected in their own neighbourhood, but also some long range connections are existing~\cite{Watts.1998}. In addition, the standard deviation of connection distribution increases and some outliers in form of high degree nodes emerge.

\subsection{Scale-free networks}
\begin{figure}[!htb]
	\centering
	\includegraphics[width=1\textwidth]{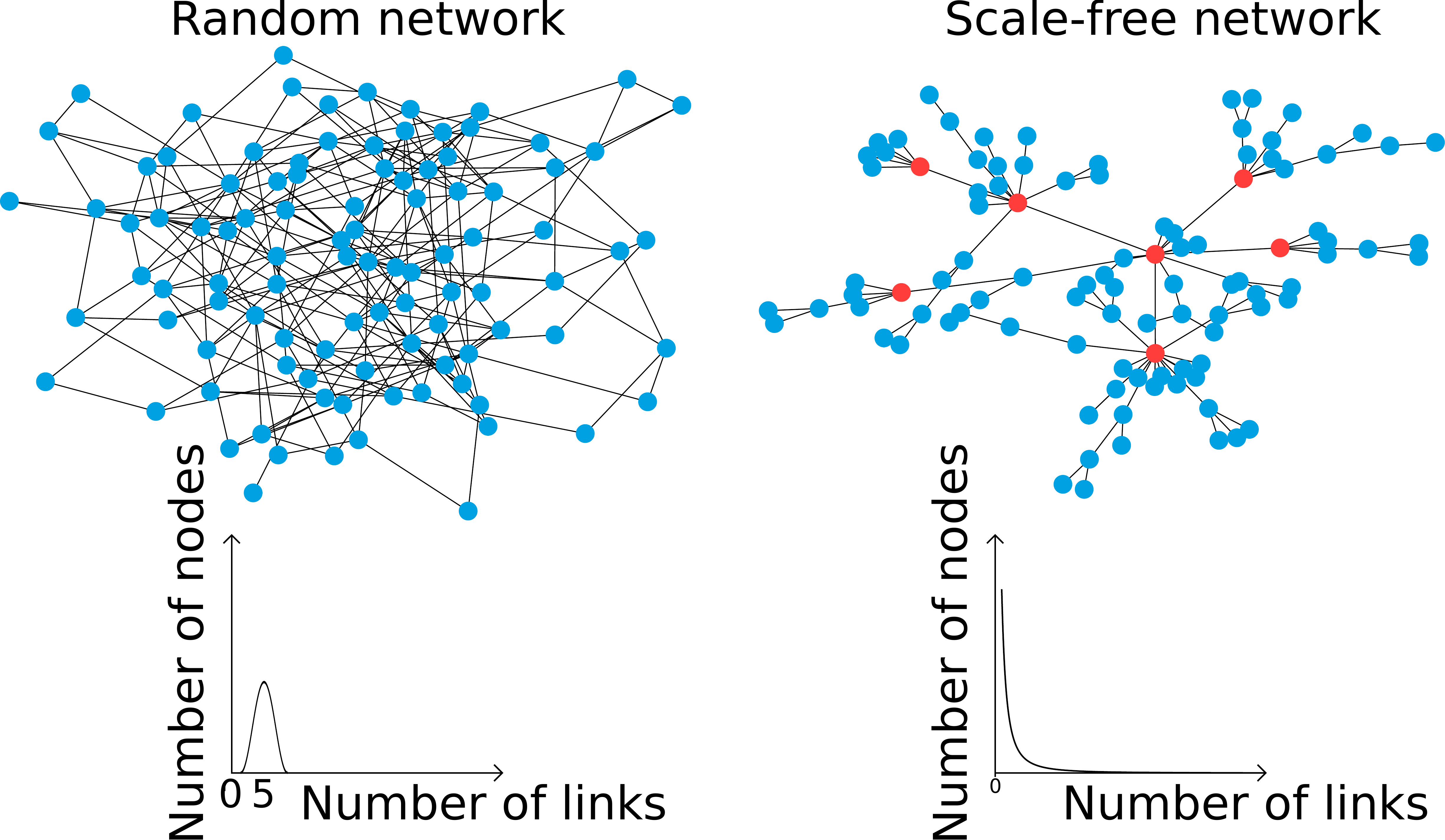} 
	\caption[Difference between random and scale-free neuronal networks]{\textbf{Difference between \textit{random} and \textit{scale-free} neuronal networks:} A \textit{random network} is constructed by a constant connection probability. All neurons have approximately a same number of connections. The distribution of node linkages for \textit{random networks} is bell shaped. The most neurons for the illustrated \textit{random network} would be connected with about five synapses, which is the number of links at the maximum of degree distribution. A \textit{scale-free network} is constructed by special methods~\cite{Catanzaro.2005}. Hub neurons are red marked and have an immense number of connections to average. The distribution of node degrees is formed like a power law function. Thus, there are many nodes which are sparsely connected while some hubs are able to have lots of links.}\label{fig:scalefree_vs_rand}
\end{figure}

In \textit{scale-free networks} some nodes, so-called hubs, have an immense number of connections to other nodes~\cite{Barabasi.2003}. 
As some nodes are barely connected while hubs are able to have 100 times more connections, the \textit{scale-free networks} could also be called ultrasmall-world networks: Some nodes are almost isolated of other groups~\cite{Cohen.2003}.
\textit{Scale-free} and also \textit{small-world networks} find usage in many applications of nature like the brain. Furthermore, hub neurons were already detected in regions of the brain and characterised~\cite{Sporns.2007, Bonifazi.2009}.
In Figure~\ref{fig:scalefree_vs_rand} an exemplary \textit{scale-free network} is illustrated with seven red marked hub neurons, resulting in a fundamentally different distribution of node degrees. 
This kind of log normal distribution can be described by a power low function with free parameter $\gamma$,
\begin{equation}\label{equ:sf_d}
P{(deg = k )} \propto  k^{-\gamma} .
\end{equation}

\newpage
\section{Neuronal network simulations}
\begin{figure}[!htbp]
	
	\centering
	\includegraphics[width=0.86\textwidth]{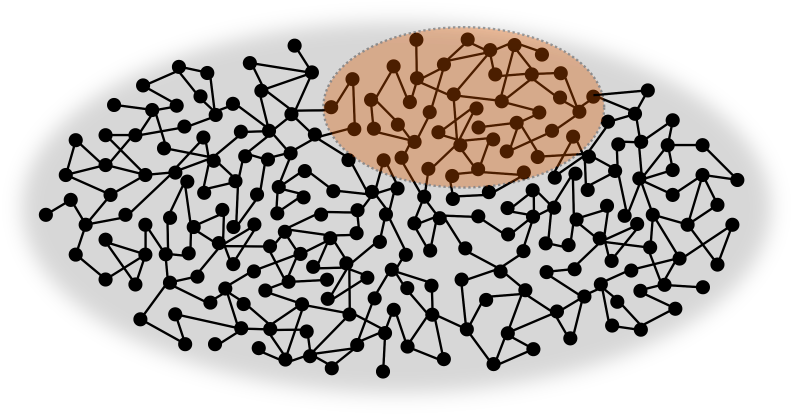}
	\caption[Neuronal network \textit{in silico}]{\textbf{Neuronal network \textit{in silico}:} 1000 Izhikevich neurons will be simulated and a subset (orange marked) is randomly chosen. It contains 80 excitatory and 20 inhibitory neurons.}\label{fig:sim}
\end{figure}

Simulated networks will contain 1000 neurons (like the original model~\cite{Izhikevich.2003}), which will be connected randomly for specific network types. Thus, each simulation will deliver different datasets. As software \textit{MATLAB} of MathWorks is chosen because of its modifiability (for details see Appendix \ref{A:SimSoft}). Since \ac{MEA} chips usually do not measure signals from each neuron only a randomly chosen subset of 100 neurons is recorded (see Figure \ref{fig:sim}). For all subsets the ratio of excitatory to inhibitory neurons is just like for the whole network chosen to be 4 to 1. In these networks model, parameters for \ac{RS} ($a=0.02; b=0.2; c=-65; d=8$) are selected for excitatory neurons and \ac{FS} ($a=0.1; b=0.2; c=-65; d=2$) for inhibitory respectively (see Table \ref{tab:Izh}). While excitatory synapses contribute to the membrane potential of the receiving neuron, inhibitory ones counteract.

\begin{figure}[!htb]
	\centering
	\includegraphics[width=1\textwidth]{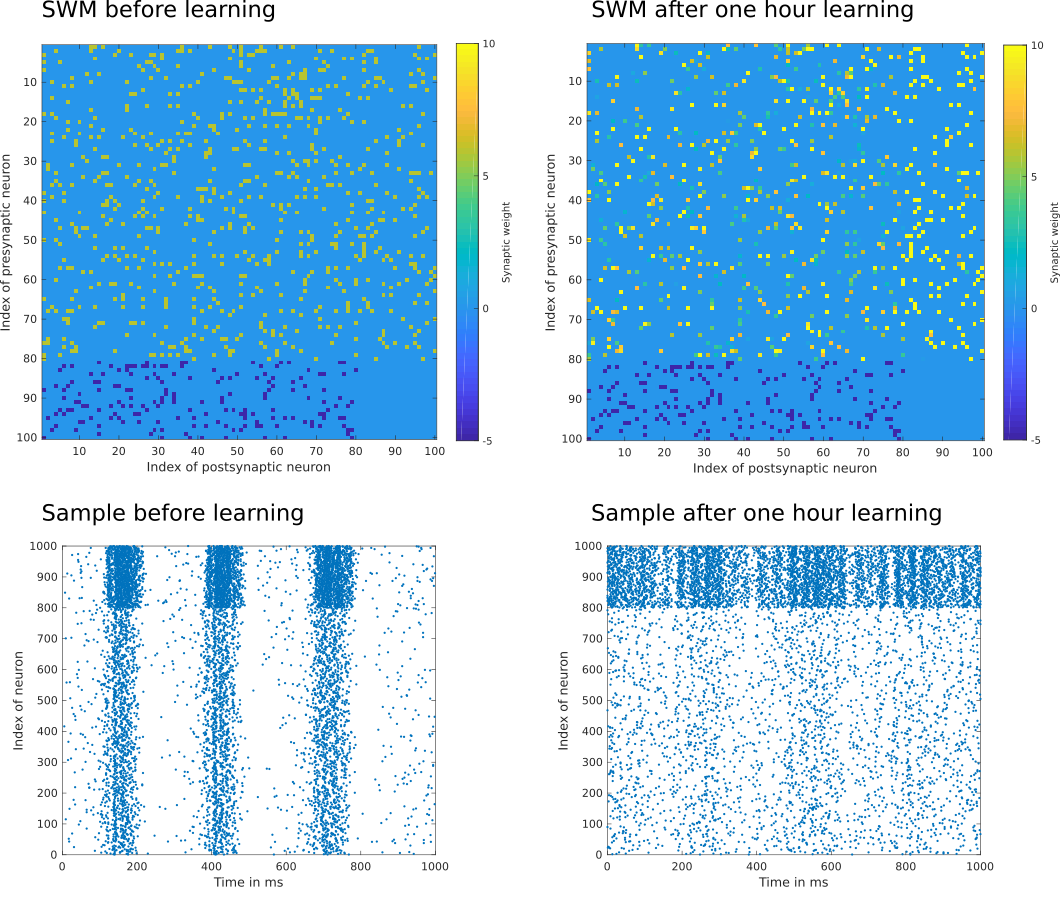} 
	\caption[Effects of STDP]{\textbf{Effects of \ac{STDP}:} At the beginning all excitatory synapses have a weight of 6 and all inhibitory ones -5. After one hour learning with \ac{STDP} (only allowed for excitatory connections) some excitatory synapses expand to their maximum of 10 while others shrink to nearly 0. For the purposes of illustration both \acp{SWM} only show a subset for 100 neurons (representatively 80 excitatory and 20 inhibitory neurons). These changes of the \ac{SWM} have a massive impact on spike trains, which can be seen in both samples. Before learning there is a much higher synchrony. }\label{fig:learningeffect}
\end{figure}

As reported in~\cite{Swadlow.1994}, synaptic transmission times of 1 to 20\,ms are realistic in mammalian cortex. However, the original Izhikevich model simulation of~\cite{Izhikevich.2003} uses uniform delay times of 1\,ms for all connections.
To increase the biological plausibility, here implemented simulations are based on the code by Izhikevich in 2006~\cite{Izhikevich.2006}, which is an advanced version with plasticity obeying the Hebbian theory and axonal conduction delays. The ability of plasticity is called \ac{STDP}, where synchrony dramatically decreases after seconds and realistic network bursts do not appear any more (see Figure \ref{fig:learningeffect}). 
In this approach a static topology is more desirable. Therefore \ac{STDP} is not used, but transmission times randomly distributed with values between 1 and 20\,ms. 

For comparison of possible network topologies, their characteristics are described in Table~\ref{tab:charac}. The \textit{regular network} was not used for simulation because of its unrealistic construction in nature. The \ac{SII}~\cite{Izhikevich.2006} for \textit{random networks} was constructed with randomly chosen 100 of outgoing synapses for each neuron, where no inhibitory-to-inhibitory connections are allowed. In this way, for a network of 1000 neurons, the resulting connection probability of outgoing synapses is set to $\frac{100}{1000}=0.1$ without standard deviation. For incoming synapses, the probability is also 0.1 with standard deviation.
Furthermore, a second \textit{random network} will be evaluated in form of an implemented \ac{ER} network topology~\cite{Erdos.1959} with a connection probability of 0.1, following a Poisson distribution for input- and output-degrees. Another difference to the \ac{SII} \textit{random network} is the possibility of connections between inhibitory neurons.
Thus, both \textit{random networks} have a theoretical mean degree of 200 synapses: 100 inputs and 100 outputs.

\begin{table}[!htbp]
	\caption[Summery of network type characteristics]{\textbf{Summery of network type characteristics:\\} Different network types are characterized by the existence of hubs, the distribution of input- and output-degree.} \label{tab:charac}	\vspace{0.0cm}
	\begin{center}	
		\begin{tabular}{ l | c | c | c}
			\hline
			& In-degrees & Out-degrees  & Hubs\\
			\hline
			\textit{Regular network} & constant & constant   & no\\
			SII \textit{random network} ~\cite{Izhikevich.2006} & Poisson & constant   & very unlikely\\
			ER \textit{random network}~\cite{Erdos.1959} & Poisson & Poisson  & unlikely\\
			IC \textit{scale-free network}~\cite{Catanzaro.2005}& power-law & power-law   & yes\\
			BA \textit{scale-free network}~\cite{Barabasi.2003}& power-law & power-law  & yes\\
			\hline
		\end{tabular}
	\end{center}
\end{table}

Moreover, two ways of \textit{scale-free} construction are investigated. First, the \ac{IC}~\cite{Catanzaro.2005} for uncorrelated random \textit{scale-free networks} uses a connection probability function in form of formula~(\ref{equ:sf_d}). The parameters applied are a minimum degree of 10 and $\gamma = 2.0$. Second, the \ac{BA}~\cite{Barabasi.2003} network, which is also a form of \textit{scale-free network}, with 24 connections, 12 input- and 12 output-synapses, per each growing step were constructed. Thus, the minimum degree of \ac{BA} networks is 24. Inhibitory to inhibitory connections are permitted for both \textit{scale-free network} types.

For all networks, self-connections and parallel synapses are prohibited, while antiparallel synapses are allowed.
Binary masks for respective \acp{SWM} were constructed by modified implementations of the \textit{Python} complex network package \textit{NetworkX}~\cite{Hagberg.2008}.
\acp{SWM} were filled up with log-normal distributed synaptic weights with a maximum of 10. The mean synaptic weight is chosen for each simulation in such a way that network bursts appear. It was found that a higher density of connections required a lower average synaptic weight for regular network bursts.


\section{Results and discussion of simulation}
\begin{figure}[H]
	\centering
	\includegraphics[width=0.49\textwidth]{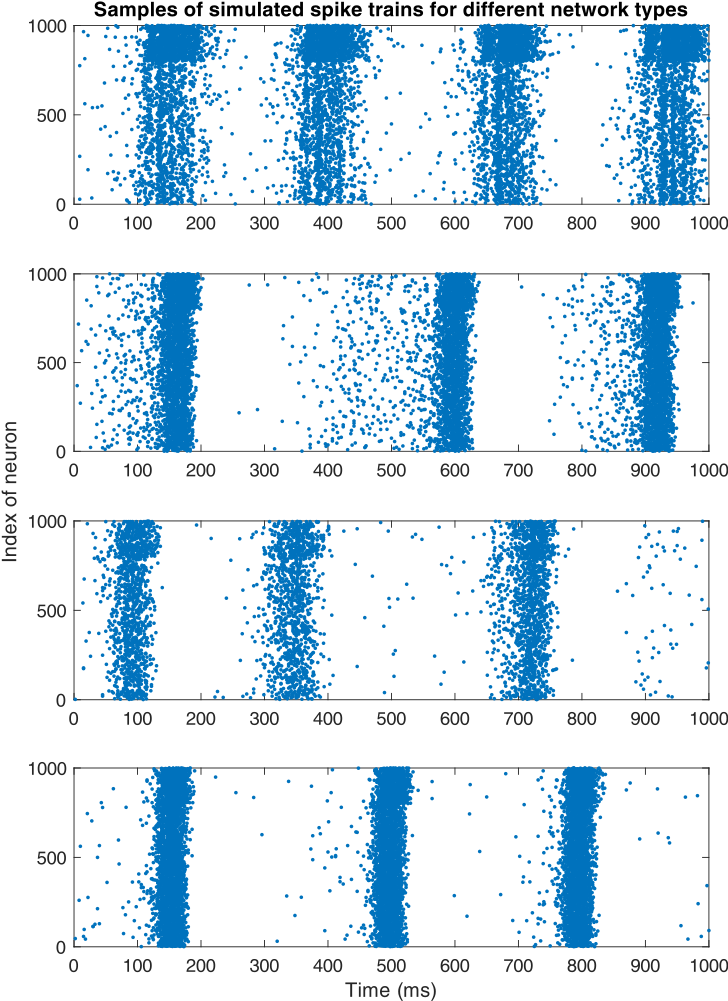} 
	\caption[Samples of simulated spike trains]{\textbf{Samples of simulated spike trains:\\}  Each dot represents an emitted action potential of the neuron with index of respective y-coordinate. The networks with overall 1000 neurons each produced three or four network bursts in the shown second of simulation. The samples are sorted for network types from top to bottom: \ac{SII} \textit{random network}, \ac{ER} \textit{random network}, \ac{IC} \textit{scale-free network} and \ac{BA} network. \\}\label{fig:sample}
\end{figure}

Samples of the simulated spike train data are shown in Figure~\ref{fig:sample}. Network bursts were ensured for all simulations with manual regulation of synaptic weights. In this way, all samples of the explored network types show similar patterns, even if the spiking density varies in network bursts.

The resulting distributions for input-, output- and total-degree (sum of input- and output-degree), separated into excitatory and inhibitory neurons, are analysed. Furthermore, respective \acp{MFR} are measured by counting all spikes of neuron $i$, whose history is called spike train $S_{i}(t)$, and dividing the number by the considered time range,
\begin{equation}\label{equ:MFR}
MFR_{i} =  \frac{\sum_{t=t_{start}}^{t_{end}} S_{i}(t)}{t_{end} - t_{start}} .
\end{equation}

\begin{figure}[!htb]
	\centering
	\includegraphics[width=1\textwidth]{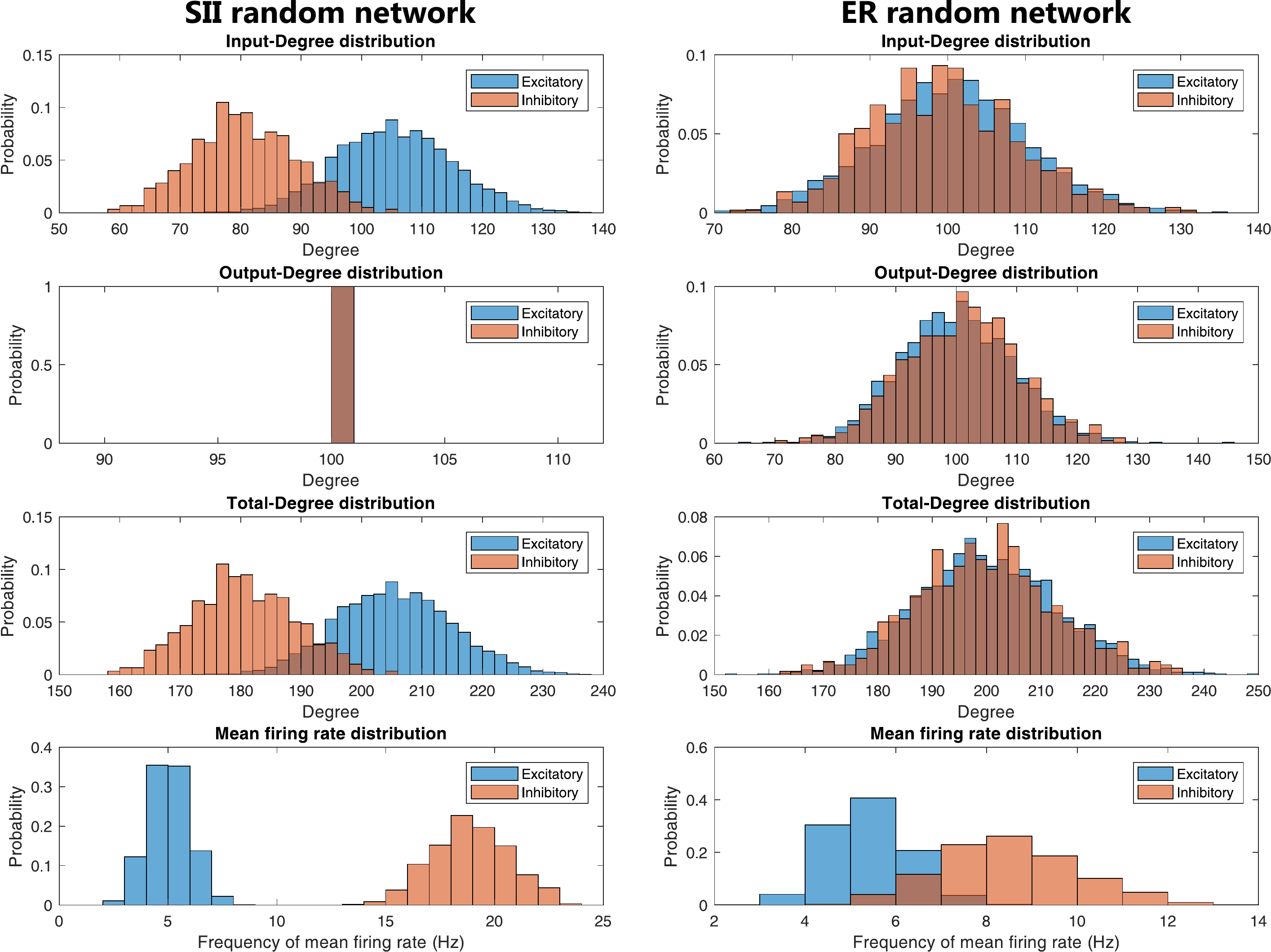}
	\caption[Analysis of SII and ER random network]{\textbf{Analysis of \ac{SII} and \ac{ER} \textit{random network}:\\} A lower mean degree of inhibitory neurons is the result of prohibited inhibitory-to-inhibitory connections. The inhibitory effects lead to lower \acp{MFR} of excitatory neurons, while inhibitory neurons will not be inhibited. In contrast to \ac{SII} \textit{random network}, the distribution of \acp{MFR} for both neuron types of \ac{ER} \textit{random network} have a superimposed area.}\label{fig:Rand_nets}
\end{figure}

Due to the constant output-degrees of 100 for \ac{SII} \textit{random networks}, the difference between its normal distributions of total-degree and input-degree is an offset by 100. Since every fifth neuron is an inhibitory neuron and inhibitory-to-inhibitory connections are not allowed for \ac{SII}, the offset between input-degrees of both neuron types is theoretically $\frac{100}{5}=20$, which is same for the total-degrees. The theoretical assumption can be confirmed by the results, see Figure~\ref{fig:Rand_nets}, as the mean values of the total degree distributions differ by about 22. It is also the reason for a separation of normal distributed \acp{MFR} for both neuron types. The inhibitory effects lead to inhibited \acp{MFR} of excitatory neurons, while \acp{MFR} of inhibitory neurons are not inhibited. Knowing the effects, in the upper plot of Figure~\ref{fig:sample} the inhibitory neurons can be identified easily at indexes 801 to 1000.

The \ac{ER} model reduces this separation with a non-constant output-degree and the permission for inhibitory-to-inhibitory connections (see Figure~\ref{fig:Rand_nets}). Its normal distributions of input-degree and output-degree are similar to each other with a mean value of 100. In this way the normal distribution of total-degrees has a $\sqrt{2}$ higher standard deviation. Despite superimposed area for \acp{MFR} of both neuron types, the distribution of \acp{MFR} of inhibitory neurons still has a higher mean value with higher standard deviation.

\begin{figure}[!htb]
	\centering
	\includegraphics[width=1\textwidth]{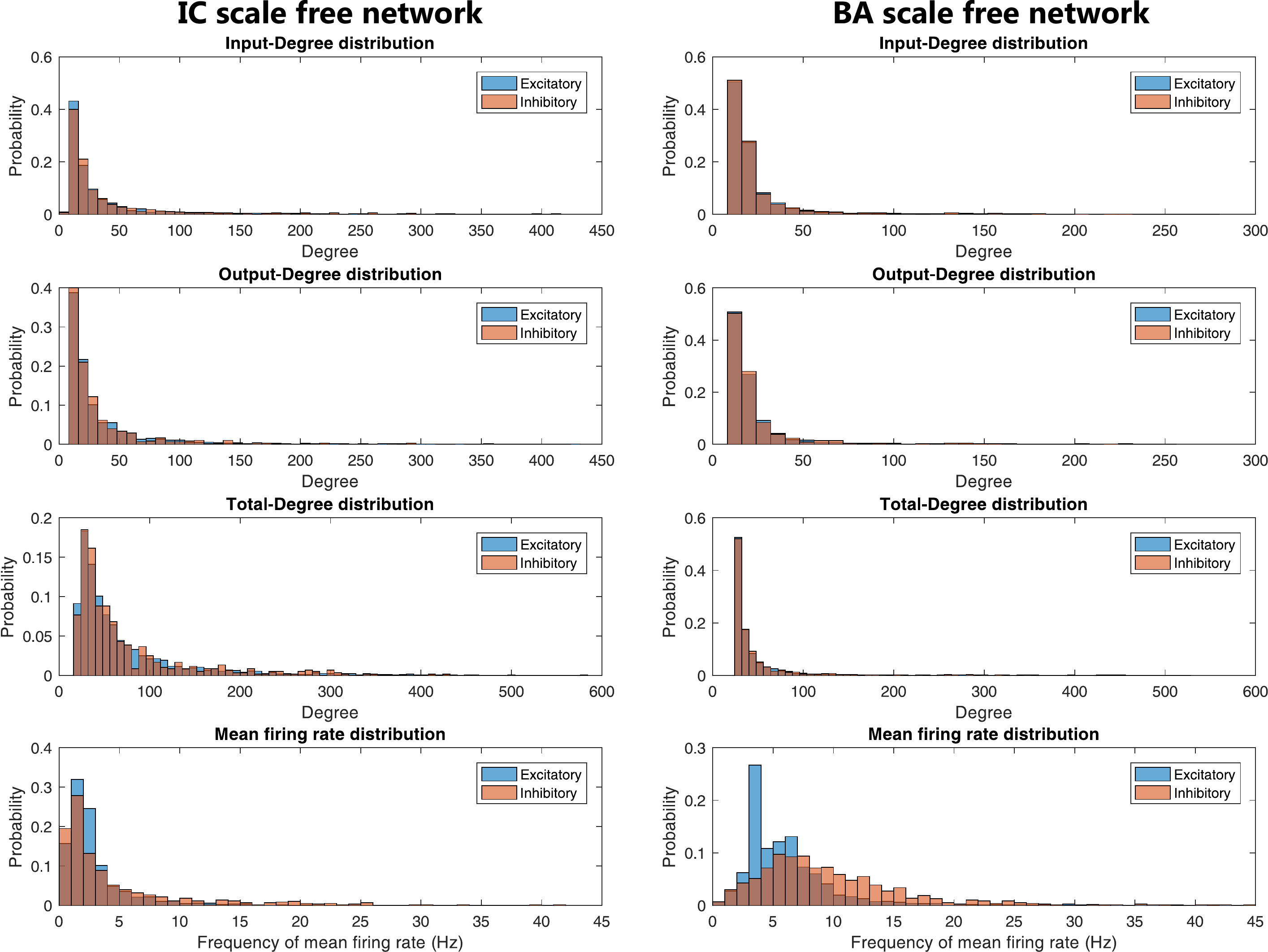}
	\caption[Analysis of generated IC and BA scale-free network]{\textbf{Analysis of generated \ac{IC} and \ac{BA} \textit{scale-free network}:\\} Resulting \ac{MFR} distributions of excitatory and inhibitory neurons are similar and can be fitted by a power law function. For \ac{BA} \textit{scale-free networks}, log normal distributed \acp{MFR} have different standard deviations for both neuron types.}\label{fig:SF_nets}
\end{figure}

Both \textit{scale-free} implementations lead to intersections of inhibitory and excitatory \acp{MFR}. For the \ac{IC} networks, log normal distributions of input- and output-degree are similar with a maximal probability at low degrees and outliers can be found at up to 400 connections (see Figure~\ref{fig:SF_nets}). Thus, total-degrees are also log-normal distributed. Resulting \ac{MFR} distributions of excitatory and inhibitory neurons are similar and can be fitted by a power law function. For the \ac{BA} networks, log normal distributions of input-, output- and total-degree are almost identical for inhibitory and excitatory neurons. The minimum input- and output-degrees are added up to minimum total-degrees. The \ac{BA} network leads to a larger deviation in the distribution of inhibitory \acp{MFR} than that of excitatory \acp{MFR} (see Figure~\ref{fig:SF_nets}). Respective outliers can be identified at up to 300 inputs or outputs. The total-degree for hub neurons are up to 600. Even with a big superimposed area, the log normal distribution of \acp{MFR} for inhibitory neurons has a higher standard deviation than the one for excitatory neurons.

\newpage
Summarizing, the distribution of \acp{MFR} can be immensely influenced by modelling different network topologies.
Since a log normal distribution for intracortical spontaneous \acp{MFR} was already demonstrated~\cite{Song.2005}, many neurons fire far above network average in biological \textit{in vitro} cultures. One can conclude that more realistic \acp{MFR} can be measured by using \textit{scale-free} topologies for large scale neural network simulations.

\section{Conclusion of simulation}
For a meaningful evaluation of neurocomputational algorithms analysing large neuro datasets, \textit{in silico} neural networks shall be employed with good biological relevance.
However, the complexity and topology of commonly used \textit{in silico} neural networks and thus the evaluation results can vary.

The result is indeed that the implemented neural network topology immensely influences the distribution of the \acp{MFR} and thus the meaningfulness of evaluation results. 
Whereas \ac{SII} topologies are used in many evaluations, e.g. without delay times~\cite{Garofalo.2009}, with delay times, and with different use of \ac{STDP}~\cite{Pastore.2017,Ito.2011}, more realistic \textit{scale-free} topologies are rarely used for connectivity estimation evaluations~\cite{Kadirvelu.2017}. 
One can concluded that searching for the best algorithms can be confusing, since apparently better results are not synonymous with physiologic relevance for real biological neural networks. 

For future evaluations, a standardised method improves an effective research of neurocomputational algorithms. Widely used and uniform benchmarking also makes it easier to compare newly developed methods with previous methods.
Moreover, the further development and improvement of intergroup research is possible in a simpler way.
To ensure topology independent algorithms, a multiple model evaluation is used. Usage of at least one \textit{scale-free network} implementation is necessary for good, sufficient biological plausibility. Furthermore, to strengthen the significance repeated simulations are required. Since electrophysiological recording methods do not measure all signals of a neuronal culture, it is recommended to use spike train data of a small subset, e.g. 100 spike trains of a simulated network with 1000 neurons.

\chapter{Methods of connectivity estimation}\label{cha:alg}
\begin{figure}[!htb]
	\centering
	\includegraphics[width=1\textwidth]{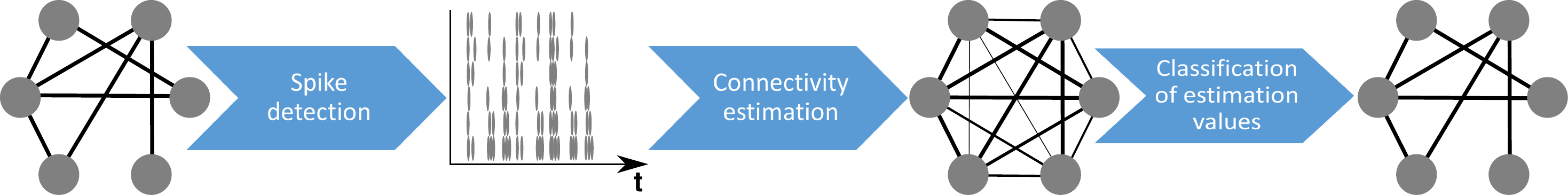}
	\caption[Workflow of connectivity estimation]{\textbf{Workflow of connectivity estimation:}
		Signals of a considered neuronal network (left side) are measured. After applying a spike detection algorithm spike trains are obtained. The result of connectivity estimation algorithms is the \ac{CM}, which describes a graph with values for each possible connection. A classification of these estimation values leads to a \ac{TCM}, which describes the final graph (right side). Ideally this graph should be identical to the considered neuronal network.
	}\label{fig:WF}
\end{figure}

The goal of effective and functional connectivity estimation algorithms is gaining knowledge about the \ac{SWM}. Thus, the result of these methods is also a matrix of same size, a so-called \ac{CM}.
The definition of \ac{CM} is very similar to \ac{SWM}: While \ac{SWM} stores real synaptic weights the \ac{CM} is filled up with estimated values, which describe connection strengths. Since \acp{CM} are just results of algorithms, their value ranges strongly depend on the used method: 
\begin{itemize}
	\item Functional connectivity estimation algorithms do not indicate connection types and causality. However, some algorithms are able to extract a few characteristics of effective connectivity. 
	\item Effective connectivity estimation algorithms indicate connection types (e.g. inhibitory synapses with negative and excitatory ones with positive values) and causality. 
\end{itemize}
Because value ranges of \ac{CM} also strongly depend on the measured signal, there is no information to know what connection strength corresponds to a real synapse. Since values of \ac{CM} should ideally be proportional to real synaptic weights of \ac{SWM}, the need of a threshold selection is given to distinguish between a true connection and just statistical correlation. The result of this step is then called \ac{TCM}. The general workflow is illustrated in Figure~\ref{fig:WF}.

To estimate the functional or effective connectivity different methods based on information theory~\cite{Gourevitch.2007, Garofalo.2009}, pattern recognition~\cite{Perkel.1967, Masud.2017}, model fitting~\cite{Isomura.2014, Isomura.2015, Friston.2011c} or data mining~\cite{Diekman.2014} have been developed and applied.
The quality of the results is strongly dependent on datasets gained from \textit{in vivo} or \textit{in vitro} neural networks. 

Some methods are not developed for large and complex networks, for example estimating the connectivity by modeling \textit{in silico} neuronal networks~\cite{Isomura.2014, Isomura.2015, Friston.2011c} or by using data mining methods~\cite{Diekman.2014}. Another method is Partial Correlation, which is able to distinguish between direct and indirect connections by considering linear contributions. Partial Correlation estimates connectivity precisely for \textit{in silico} with a small amount of neurons, e.g. 130 neurons~\cite{Kadirvelu.2017}. However, Partial Correlation is inaccurate for large scale network models~\cite{Poli.2016} because of the complex amount of possibilities for considered indirect contributions. Therefore, in this work we focus on a scenario measuring signals of a small subset of a large scale neuronal network, which is more realistic for most \textit{in vitro} and \textit{in vivo} applications. 

As the number of electrodes in electrophysiological recordings have increased up to several thousands accompanied with huge datasets, used algorithms for connectivity analysis have to be highly computationally effective.

All chosen algorithms investigate pairwise a source spike train $X$ for effects on a target spike train $Y$. Thus, statements about causality are possible, which is a characteristic of effective connectivity. 



Furthermore, only algorithms with a higher potential of good performances are considered. For example, Granger Causality~\cite{Ding.2006} was often used to estimate connectivity~\cite{Chavez.2003, Saalmann.2012, Nakhnikian.2014} and is available as \textit{MATLAB} based toolbox~\cite{Seth.2010}. Since Granger Causality is insensitive to nonlinear correlations~\cite{Timme.2014}, more promising methods were chosen, which will be introduced in the following.

\section{Cross Correlation}
\ac{CC} is similar to normal convolution and a classic tool to recognize a pattern. In 1967 \ac{CC} was already used to measure relations between spike trains~\cite{Perkel.1967}. By multiplication of a time-shifted signal $x(i-d)$ element-wise with another non-shifted signal $y{(i)}$, similarity can be recognized as a function of time shift $d$. Ideally, the greatest similarity should be at the delay, where the result of \ac{CC} has its largest peak staying for the greatest correlation. The essential idea of \ac{CC} can be expressed as
\begin{equation}\label{equ:CC}
CC_{XY}(d) =\sum_{i=-\infty}^{\infty} y{(i)} \cdot x{(i-d)} .
\end{equation}
Since the binary spike trains $X$ and $Y$ only add up values in formula (\ref{equ:CC}) and \ac{MFR} of a spike train can vary, this form has problems with its value range. Thus, to solve this problem and enable comparability between results normalizations are used. This means a variable division of formula (\ref{equ:CC}) depending on $X$ and $Y$. Using different types of normalization can strongly influence results.
While some implementations have trouble detecting inhibitory connections~\cite{Masud.2017}, other methods of \ac{CC} are even able to distinguish between inhibitory (negative peak of \ac{CC}) and excitatory (positive peak of \ac{CC}) connections~\cite{Bartho.2004}.

\subsection{Normalized Cross Correlation Histogram}
The most common method is \ac{NCCH}, which normalizes results with the geometric mean of total spiking times $n_x$ and $n_y$ of both spike trains~\cite{Pasquale.2008, Berdondini.2009b, Maccione.2012, Poli.2015, Pastore.2016, Ito.2011, Brosch.1999, Kiemel.1998, Eytan.2004},
\begin{equation}\label{equ:NCCH}
NCCH_{XY}(d) = \frac{1}{\sqrt{n_x \cdot n_y}} \sum_{i=-\infty}^{\infty} y{(i)} \cdot x{(i-d)}
\end{equation}

\subsection{Normalized Cross Correlation}
\begin{figure}[!htb]
	\centering
	\includegraphics[width=1\textwidth]{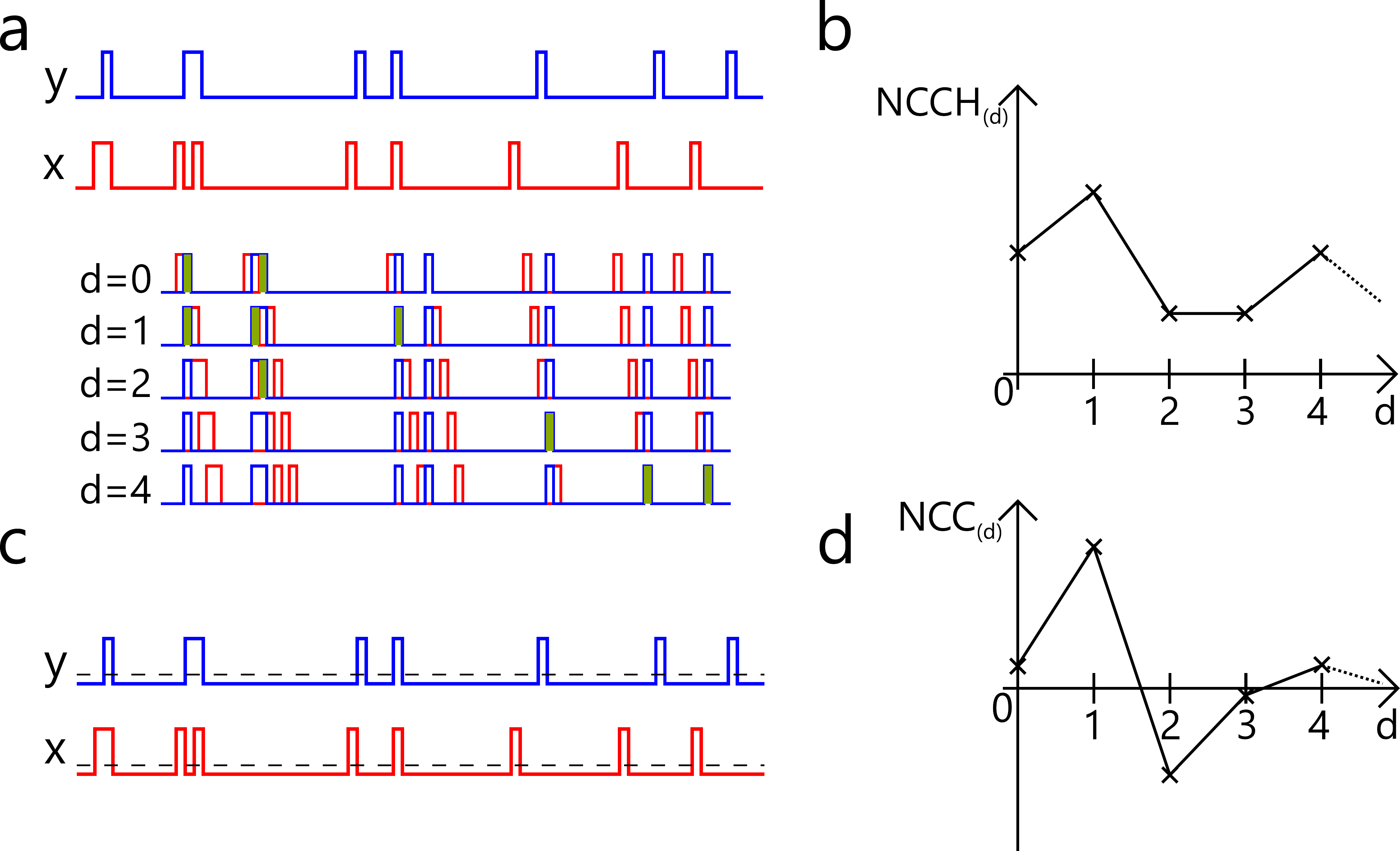}
	\caption[Cross Correlation]{\textbf{Cross Correlation:} 
		(a) Two spike trains are illustrated. The blue spike train $y$ is target while the red one $x$ is source spike train of the investigated connection $X \rightarrow Y$. The binary spike trains are shifted for delay times $d$ in range of 0 to 4. The green spots are summed up for their respective delay time. (b) After normalization the \ac{NCCH} is obtained. The peak at delay $d=1$ would be the maximal correlation. By subtracting the mean values of respective spike trains the mean of them is zero. (c) The changed spike trains for \ac{NCC} are illustrated. The dashed lines indicate their zero levels. Thus, in a sparse spike train most values are slightly negative. (d) The normalization of \ac{NCC} leads to a different result. The total mean of this function should be zero, where negative values indicates an inhibitory effect. Finding the maximum of absolute values would support the assumption of an excitatory connection of $X \rightarrow Y$ at $d=1$.
	}\label{fig:CC}
\end{figure}

Another method of normalization is the usage of \acp{SD} $\sigma_x \cdot \sigma_y$ and the recording length $N$ stated in bins. In addition, subtracting the mean values $\bar{x}$ and $\bar{y}$ of the respective spike trains before multiplying them is recommended for binary binned spike trains. The mean of all results should now be zero and inhibitory correlations will lead to negative peaks while excitatory connections produce positive peaks~\cite{Bedenbaugh.1997} (see Figure \ref{fig:CC}). The so defined type of \ac{CC} is called \ac{NCC} in the following,
\begin{equation}\label{equ:NCC}
NCC_{XY}(d) = \frac{1}{N} \sum_{i=-\infty}^{\infty} \frac{(y{(i)}-\bar{y}) \cdot (x{(i-d)}-\bar{x})}{\sigma_x \cdot \sigma_y } .
\end{equation}
\\
For sparse spike trains \ac{NCC} and \ac{NCCH} lead to similar results. However, the performance of \ac{NCC} increases with increasing spike train length~\cite{Ito.2011}. Thus, here only \ac{NCC} is used for evaluation.
Since the resulting matrix for the pairwise comparison of all combinations of spike trains is symmetric (see formula (\ref{equ:symNCC})), the number of independent calculations for \acp{NCC} is $\frac{K^2-K}{2}$, where $K$ is the number of spike trains.
\begin{equation}\label{equ:symNCC}
NCC_{XY}(d) = NCC_{YX}(-d)
\end{equation}
\subsubsection*{Coincidence Index}
\begin{figure}[!htbp]
	\centering
	\includegraphics[width=0.86\textwidth]{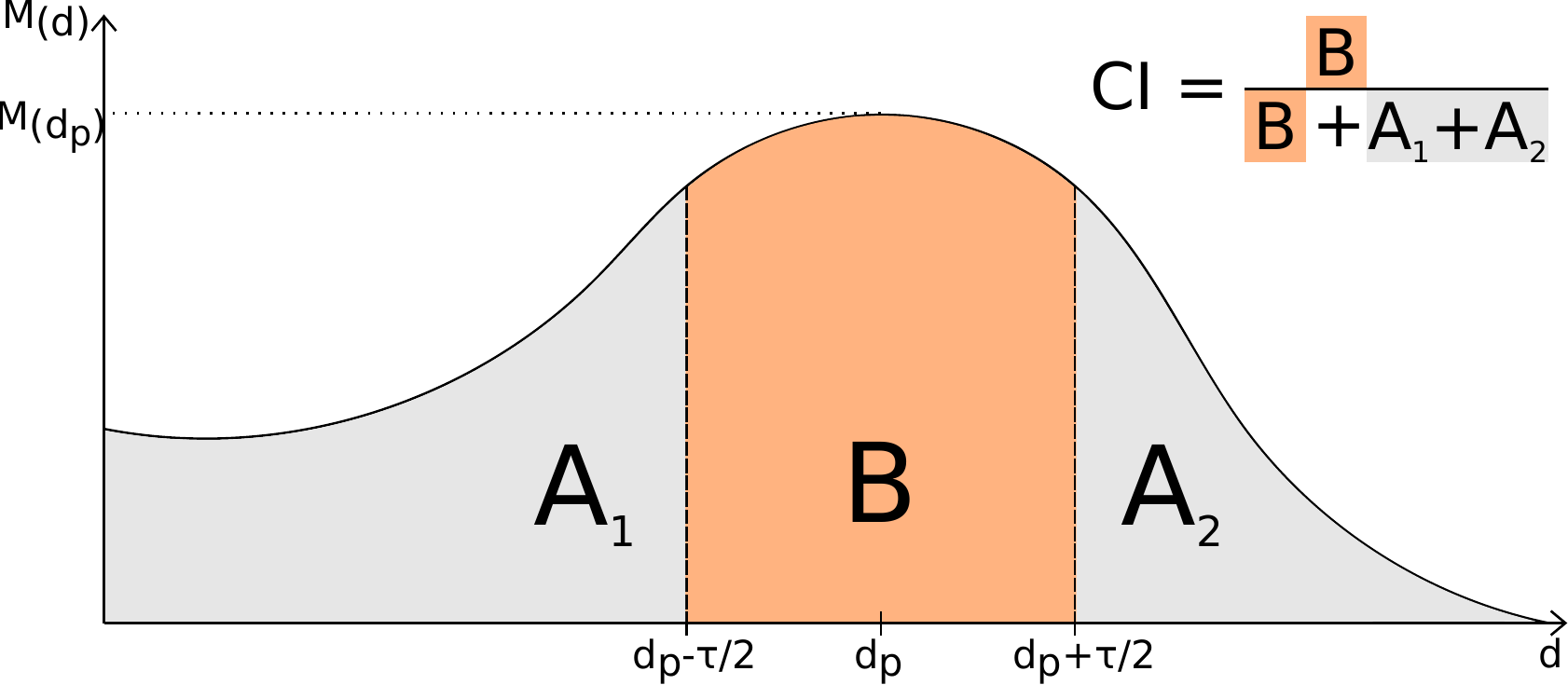}
	\caption[Principle of Coincidence Index]{\textbf{Principle of  Coincidence Index:} A delay-dependent function $M{(d)}$ (e.g. $NCCH_{XY}(d)$ or also $NCC_{XY}(d)$) has a peak value $M{(d_\mathrm p)}$. The integral B between $d_\mathrm p-\frac{\tau}{2}$ and $t_\mathrm p+\frac{\tau}{2}$ (orange area) will be divided by the total area under curve (orange plus grey areas). The result value is called \ac{CI}. An equation illustrates this function by using colors for respective integral values A$_1$, A$_2$ and B.
	}\label{fig:CI}
\end{figure}

A delay-dependent function $M{(d)}$ is normally analysed by looking for its peak value, like $NCC_{XY}(d)$ for example. In order to further improve the results of an analysis qualitatively, a widely used tool is introduced: The \ac{CI}~\cite{Ito.2011, Chiappalone.2006, Juergens.1997,  Jimbo.1999},
\begin{equation}\label{equ:CI}
CI=\frac{  \sum_{d=d_\mathrm p-\frac{\tau}{2}}^{d_\mathrm p+\frac{\tau}{2}}  M{(d)}     }{ \sum_{d=d_\mathrm{min}}^{d_\mathrm{max}} M{(d)}      } .
\end{equation}
The \ac{CI} algorithm integrates values in a range of $\tau$ around the maximum peak value and normalizes that integral (see Figure \ref{fig:CI} for an illustration of its function). In this context, for $M{(d)}$ $NCC_{XY}(d)$ was chosen, which improves the results. For reasons of overview, suffix -$CI$ will indicate the use of \ac{CI}, e.g. \ac{NCCCI} for the \ac{CI} of \ac{NCC}.

However, \ac{CI} normally handles only positive values, which is the reason to use the absolute values of \ac{NCC} for \ac{NCCCI}. The maximum correlation would now correspond to a value of one. An improvement of performance achieved with this modification was reported in an evaluation of connectivity estimation algorithms~\cite{Ito.2011}.

\section{Transfer Entropy}
Information theory provides many tools for potentially estimate connectivity. One of the simplest tools is \ac{MI}, which measures statistical dependency of two random processes, e.g. spike trains (further information can be found in Appendix \ref{A:MI}). By upgrading \ac{MI} with a third process under investigation the so-called \ac{TE} is introduced. Since this universal tool can be implemented and used in different ways, some relevant implementations are explained in this section beginning with its most common one.

\subsection{Delay One Transfer Entropy}
The source spike train is just like the target spike train one of the three observed processes for the \ac{TE}. The third process is chosen as the future bin of the target spike train. Therefore, spike trains are usually delayed by one bin, 
\begin{equation}\label{equ:D1TE}
TE_{X \rightarrow Y}=\sum_{x_{i} \in X}^{} \sum_{y_{i} \in Y}^{} \sum_{y_{i+1} \in Y}^{} p({y_{i+1}},{y_{i}},{x_{i}})  \cdot \log_2 \frac{ p({y_{i+1}} | {y_{i}},{x_{i}})}{ p({y_{i+1}} | {y_{i}})} .
\end{equation}
This form of \ac{TE} is called \ac{D1TE} and known as the original definition~\cite{Schreiber.2000}.
In Figure \ref{fig:TE} the original \ac{TE} and its considered bins are illustrated. \ac{TE} measures information flow by testing all possible patterns for any kind of dependency. Therefore, for each process there is a sum used in formula (\ref{equ:D1TE}) running through all combinations (e.g. for binary spike trains only spiking or non-spiking). If a combination occurs significantly more frequently than others, the information flow increases.
For instance, the binary pattern $x_i=1$, $y_i=0$, $y_{i+1}=1$ means all cases where neuron $X$ emitted at least one spike and neuron $Y$ did not, which then leaded to spiking of $Y$ in the next time bin. So discovered statistical dependency would support the assumption of an excitatory synapse $X \rightarrow Y$, because of its change from non-spiking to spiking.

The required probabilities in formula (\ref{equ:D1TE}) are normally calculated by counting the number of occurred patterns and divide it by the maximal possible number of existences. Thus, $p({y_{i+1}},{y_{i}},{x_{i}})$ is just the likelihood of occurrence for a given pattern $y_{i+1} y_{i}  x_{i}$. \ac{TE} makes also use of conditional probabilities, which are probabilities of occurrence for a pattern whenever a special condition is giving. For example, observing for a pattern $x_{i}=1$ and $y_{i}=1$, $p{(x_{i}|y_i)}$ is the likelihood of occurrence of $x_{i}=1$ whenever $y_{i}=1$ is given. This can be calculated by using the probability that $x_{i}=1$ and $y_{i}=1$ occur together, divided by the likelihood of an occurring $y_{i}=1$, 
\begin{equation}\label{equ:prop1|1}
p{(x_{i}|y_i)}=\frac{p{(x_{i},y_i)}}{p{(y_{i})}} .
\end{equation}
It is a similar calculation for the conditional probability of three processes. Taking this third process in account is the only variation, 
\begin{equation}\label{equ:prop1|2}
p{(x_{i}|y_i,z_i)}=\frac{p{(x_{i},y_i,z_i)}}{p{(y_{i},z_i)}} .
\end{equation}
Like \ac{MI} also \ac{TE} is able to detect linear and nonlinear correlations~\cite{Gourevitch.2007, Garofalo.2009}. The advantage of \ac{TE} taking own history into account is tremendous for neuronal data because of refractory periods after spiking and causality statements~\cite{Ito.2011}.
The formula (\ref{equ:D1TE}) of \ac{D1TE} takes for signals with base $n$ (e.g. binary spike trains have the base two) $n^3$ independent calculations. Thus, by increasing the base of spike train data from binary to a higher base $n$ computing time will increase exponential just like the number of patterns. Using logarithms with base two leads to bits as result unit in the \ac{CM}. Higher values indicate a stronger information flow from source to target neuron.

\ac{D1TE} was often able to show good results for \textit{in silico} evaluations~\cite{Garofalo.2009, Pastore.2016, Poli.2016}. Unfortunately, these models did not take variable delay times of axonal conductions into account, which is reason for an easy but unrealistic selection of optimal bin sizes (e.g. in their studies 1\,ms). In contrast, for an \textit{in silico} evaluation with variable delay times by~\cite{Ito.2011} the best performance of \ac{D1TE} was reached with a bin size of approximately 15\,ms. Thus, in reality, it is a problematic issue to choose, rather guess a good value because it depends on each pair of spike trains.

\subsection{Higher Order Transfer Entropy}
Normally \ac{TE} is used with an order of one bin for both spike trains~\cite{Lungarella.2006, Garofalo.2009, Pastore.2016, Poli.2016}. However, \ac{D1TE} can be extended to an \ac{HOTE} by increasing its temporal range~\cite{Stetter.2012}.~\cite{Ito.2011} studied different combinations from one bin variable up to five bins to improve its performance. To understand the modification it is recommended to study Figure \ref{fig:TE} combined with formula (\ref{equ:HOTE}).
\begin{equation}\label{equ:HOTE}
HOTE_{X \rightarrow Y}=\sum_{x_{i}^{(l)} \in X}^{} \sum_{{y_{i}^{(k)}} \in Y}^{}  \sum_{y_{i+1} \in Y}^{} p({y_{i+1}},y_{i}^{(k)},x_{i}^{(l)})  \cdot \log_2 \frac{    p(y_{i+1} | y_{i}^{(k)},x_{i}^{(l)})  }{    p(y_{i+1} | y_{i}^{(k)})    }
\end{equation}
Parameters $k$ and $l$ are the orders of history bins of target and source spike train taken into account. For $k=1$ and $l=1$ \ac{HOTE} would be equal to \ac{D1TE}. The number of patterns is $2^{1+k+l}$ and rises with the chosen order exponentially. Thus, \ac{HOTE} can be computationally intensive compared with \ac{D1TE} which can be uncomfortable in context with an increasing amount of recorded neurons.
\subsection{Delayed Higher Order Transfer Entropy}
Besides \ac{HOTE} another modification of \ac{TE} was introduced to neuroscience~\cite{Ito.2011}, which was already used in other fields of research~\cite{Overbey.2009}. By shifting the source spike train with a delay $d$ in the past, it is possible to consider effects in a variable time window. This shifting process is similar to \ac{CC}. Therefore, the formula of \ac{HOTE} (\ref{equ:HOTE}) is further extended to
\begin{equation}\label{equ:DHOTE}
DHOTE_{X \rightarrow Y (d)}=\sum_{x_{i}^{(l)} \in X}^{} \sum_{{y_{i}^{(k)}} \in Y}^{}  \sum_{y_{i+1} \in Y}^{} p({y_{i+1}},y_{i}^{(k)},x_{i+1-d}^{(l)})  \cdot \log_2 \frac{    p(y_{i+1} | y_{i}^{(k)},x_{i+1-d}^{(l)})  }{    p(y_{i+1} | y_{i}^{(k)})    } .
\end{equation}

\begin{figure}[!htbp]
	\centering
	\includegraphics[width=1\textwidth]{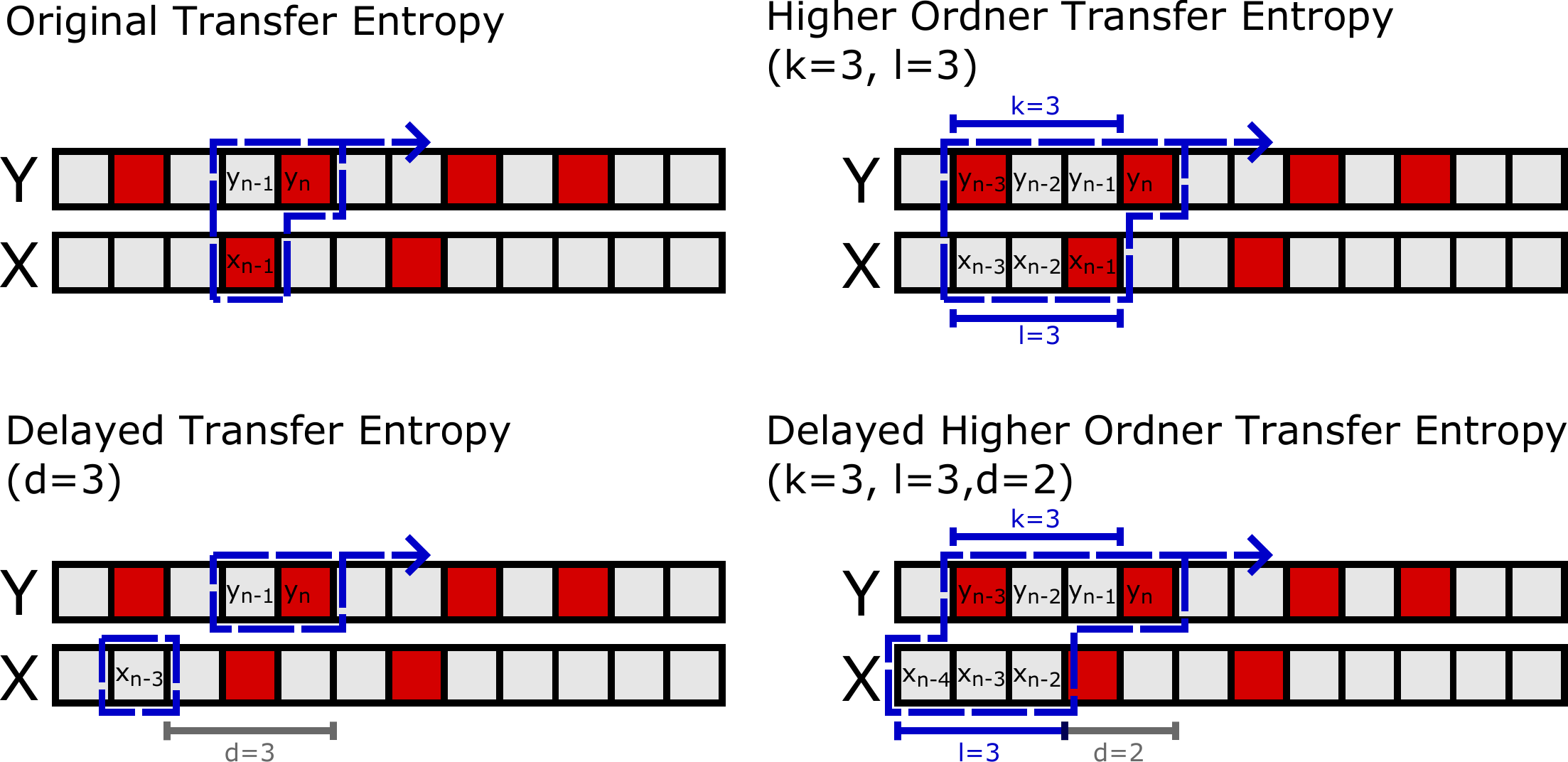}
	\caption[Principle of Delayed Transfer Entropy]{\textbf{Principle of Delayed Transfer Entropy:} Two binary binned spike trains Y and X are illustrated. Red bins mean spiking while gray bins mean no spiking. The patterns (dashed blue lines) for \ac{TE} are shifted over the whole spike trains. These patterns can have many forms like shown for four cases: Normal \ac{D1TE}, \ac{HOTE}, \ac{DTE} and \ac{DHOTE}. By calculating \ac{TE} for shifted spike trains of X with different delays $d$ a function of $d$ can be created. Its peak is the maximum of information transfer. Figure inspired by~\cite{Ito.2011}.
	}\label{fig:TE}
\end{figure}

In Figure \ref{fig:TE} \ac{DTE} for $d=3$ and \ac{DHOTE} for $d=2$ are exemplary comparable with their normal forms \ac{TE} and \ac{HOTE}. \ac{DHOTE} with $d=1$ is equal to the normal \ac{HOTE}. For varied delay times (e.g. 1 to 25\,ms in 1\,ms steps) \ac{DHOTE} can now be calculated. Investigating of so stored result values enable locating a maximum of flown information. This single peak value is then taken for the resulting \ac{CM} value of the examined connection $X \rightarrow Y$. 

Thus, in contrast to normal \ac{TE} or \ac{HOTE}, the information flow is observed for variable impact times. Since it is no longer necessary to guess a good bin size in order to take into account as many influences as possible, the selection of a small bin size in combination of a wide shifting range should always be able to process all relevant effects. Nevertheless, selecting the smallest possible bin size, which would be limited by the sampling frequency, leads to a longer computing time because of the increasing signal length for probability calculations.

However, besides the increased computing time of \ac{DHOTE}, there is another disadvantage of small bin sizes, which is even able to affect the quality of results negatively. This is due to the information flow function of delay times getting sensitive for outliers. Selecting always just the peak could be misleading. Therefore, the \ac{CI} is turned into account, which can be used for all delay-dependent functions like explained. Since the results of \ac{DTE} and \ac{DHOTE} are such functions, \ac{DTECI} and \ac{DHOTECI} are introduced.

\subsection{Combined Higher Order Transfer Entropy}
\begin{figure}[!htb]
	\centering
	\includegraphics[width=0.9\textwidth]{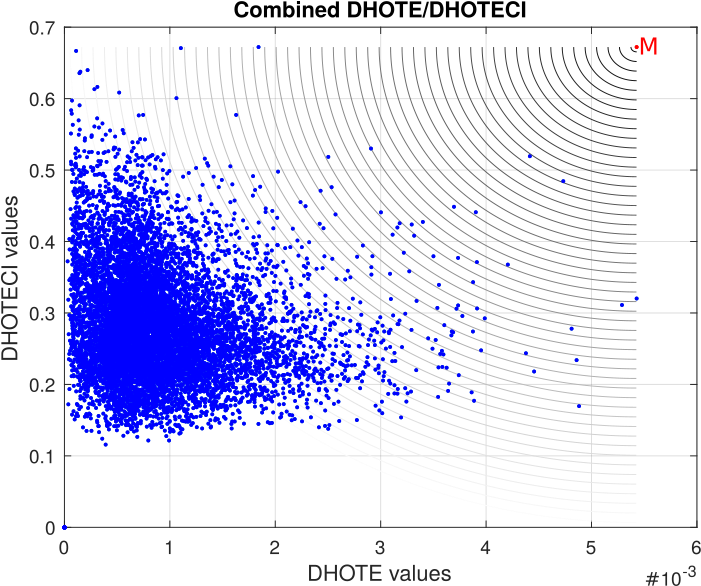} 
	\caption[Principle of CDHOTE]{\textbf{Principle of \ac{CDHOTE}}: 
		The \ac{DHOTECI} values are plotted against \ac{DHOTE} values. Each blue dot is a result pair. The coordinates of the red point $M$ are (max(\ac{DHOTE}); max(\ac{DHOTECI})), whereas euclidean distance from $M$ is indicated as gray quarter circles. The radius can be seem as threshold, which leads to encircled result pairs identified as connections. The nearest result pair to $M$ is the most likely connection.
		
	}\label{fig:chote}
\end{figure}

By combining the results of connectivity estimation algorithms, it is possible to get a different accuracy. Based on the idea of establishing the significance of \ac{DHOTE}~\cite{Shimono.2014}, an approach of \ac{CDHOTE} is implemented by plotting \ac{DHOTECI} values against \ac{DHOTE} values (see Figure~\ref{fig:chote}). 
In this plot the value pair $M$ (max(\ac{DHOTE}); max(\ac{DHOTECI})) is assumed as the place with highest possibility for a connection. 
The two dimensional euclidean distances between two points $P$ and $Q$ can be calculated by
\begin{equation}\label{equ:cdhote}
d(P,Q)=\sqrt{(P_1 - Q_1)^2 + (P_2 - Q_2)^2} .
\end{equation}
Calculating the euclidean distance from the point of interest $M$ to each value pair $V$ with coordinates (\ac{DHOTE}; \ac{DHOTECI}), a new \ac{CM} can be formed. 
Low values of formula (\ref{equ:D}) (distances to $M$) are more likely to indicate connections and the threshold is the Euclidean distance starting from 0.
\begin{equation}\label{equ:D}
CDHOTE_{X\rightarrow Y}=d(V_{X\rightarrow Y}, M)
\end{equation}
In this way the nearest result pair to $M$ is the most likely connection and points near to (0;0) are very unlikely.

\section{Total Spiking Probability Edges}
A novel effective connectivity estimation method, called \ac{TSPE}~\cite{DeBlasi.2019}, is based on the following assumptions:
\begin{itemize}
	\item If the spike rate of a neuron increases after it has received an input signal, the connection is considered excitatory. Due to refractory times of the receiving neuron, the spike rate drops again after the excitatory input signal. 
	The resulting cross-correlogram of emitting and receiving spike train shows a maximum (excitatory input) followed by low values (refractory time).
	\item If the spike rate of a neuron decreases after receiving an incoming action potential, the connection is considered inhibitory. The resulting cross-correlogram of emitting and receiving spike train shows a minimum (inhibitory input) surrounded by high values (activity before and after the inhibitory input). In this way inhibitory stimulation can only be identified if the receiving neuron is active before the stimulation. 

\end{itemize}

\begin{figure}[!htb]
	\centering
	\includegraphics[width=1\textwidth]{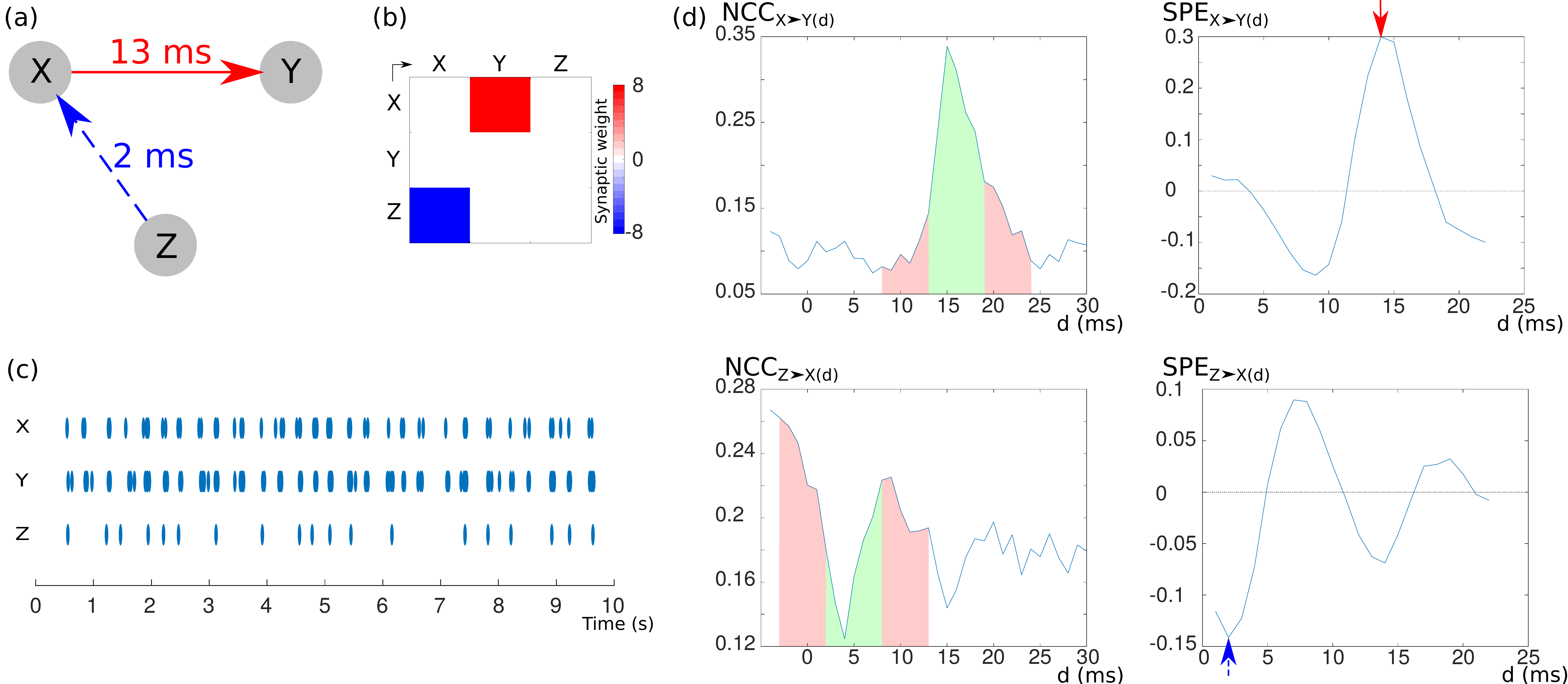}
	\caption[Principle of SPE]{\textbf{Principle of SPE:}
		An exemplary network of three neurons (a) with a \ac{SWM} (b). Neuron $X$ has an excitatory impact on $Y$ with a latency of 13\,ms after activation, while $X$ is influenced by an inhibitory input of $Z$ with a latency of 2\,ms. (c) Spike trains of the three neurons $X$, $Y$ and $Z$. 	
		(d) NCC of neuron pair $X \rightarrow Y$ and $Z \rightarrow X$ (left column). The convolution of the NCC with an edge filter $g{(i)}$ results in the respective SPE (right). Global maxima (excitatory) and minima (inhibitory) are indicated by red and blue arrows, respectively. The latency can be seen on the abscissa. 
		Areas in the left column show the corresponding areas for the calculation of the maximum and minimum (green: addition, red: subtraction).
	}\label{fig:spe}
\end{figure}

In order to capture these local maximum and minimum of the cross-correlogram, an edge filter is applied to the cross-correlogram. More precisely, the cross-correlation between spike train $X$ and spike train $Y$ (see Figure~\ref{fig:spe}.(c)) is calculated (see \ac{NCC} in the attachment) to obtain the cross-correlogram $NCC_{XY}(d)$, where $d$ is the temporal displacement (see Figure~\ref{fig:spe}.(d)). Next, the filter is applied by convolving $NCC_{XY}(d)$ with 1D edge filter $g{(i)}$ (Figure~\ref{fig:spe}), resulting in  \ac{SPE}
\begin{equation}\label{equ:conv}
SPE_{X\rightarrow Y(d)}=NCC_{XY}(d) * g{(i)} .
\end{equation}
\vspace{0.5cm}

\begin{figure}[!htb]
	\centering
	\includegraphics[width=0.78\textwidth]{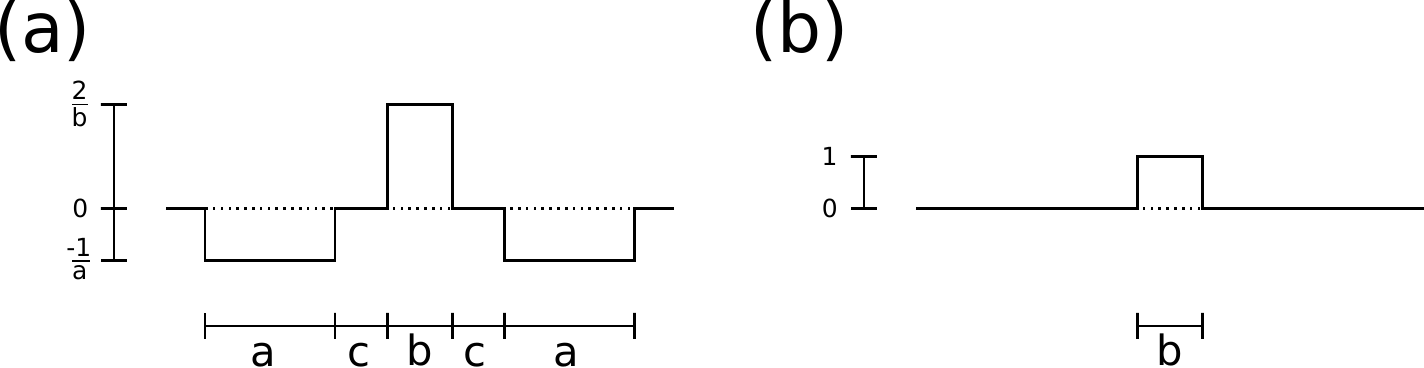}
	\caption[Design of edge and running total filters]{\textbf{Design of edge filter $g{(i)}$ and running total filter $h{(i)}$}: 
		(a) The designed edge filters have an arithmetic mean of zero and are applied to the cross-correlogram.
		(b) The running total filters are designed by the same parameter $b$ of the corresponding edge filter.
	}\label{fig:filters}
\end{figure}

The edge filters are defined as a function $g{(i)}$ with window size parameters $a$, $b$, $c$ (in sampling periods, see Figure~\ref{fig:filters} (a)). In our simulation, sampling frequency is set to 1\,kHz.
\begin{equation}\label{equ:g}
g{(i)}=\begin{cases}
-\frac{1}{a} &\mbox{if } 0 < i  \leq a \\ 
\frac{2}{b}&\mbox{if } a+c < i  \leq a+b+c \\ 
-\frac{1}{a} &\mbox{if } a+b+2c < i  \leq 2a+b+2c \\ 
0 &\mbox{else.} 
\end{cases}
\end{equation}
$a$ is the window size for the surrounding area of the point of interest which is used to calculate the local spiking probability average. $b$ is the window size of the observed area. Small values for $b$ increase the sensitivity for single outliers at the cross-correlogram. 
To avoid the including of overlapped spiking probabilities of interest with the local spiking probability average a soft crossover parameter $c$ can be used. For an \textit{in silico} evaluation with constant transmission times and a simple neuron model the usage of $c$ is not necessary. Note that without using $c$ the spiking probability edges of complex networks (\textit{in vitro} or \textit{in vivo}) can be smoothed, which is disadvantageous for an edge detection.

If the mean of function $g{(i)}$ is zero, the calculation results in an arithmetic mean of zero for $SPE{(d)}$, which prevents an offset for the resulting value range of $SPE{(d)}$.
If $NCC_{X \rightarrow Y}(d)$ shows a local maximum, $SPE_{X \rightarrow Y}(d)$ leads to a positive peak, while a local minimum for $NCC_{X \rightarrow Y}(d)$ results in a negative peak, see Fig~\ref{fig:spe}.(d).
Thus, negative peaks of $SPE_{X\rightarrow Y}(d)$ indicate an inhibitory effect of neuron $X$ to neuron $Y$ while positive peaks correspond to excitatory effects. By considering the highest absolute value of $SPE_{X\rightarrow Y}(d)$, the synaptic relation $X\rightarrow Y$ is obtained.

Since the network activity is significantly higher at periods of network bursts, this leads to a local offset of $SPE_{X\rightarrow Y}(d)$ and distort the calculation by overestimating the influences on receiving neurons.
Normalization can reduce this unwanted impact. For this purpose, each $SPE(d)$ is divided by the sum over all neuron pair results for the delay $d$,

\begin{equation}\label{eq}
SPE'_{X\rightarrow Y}(d)=\frac{SPE_{X\rightarrow Y}(d)}{\sum_{X=1}^{X=N}\sum_{Y=1}^{Y=N} SPE_{X\rightarrow Y}(d)} .
\end{equation}

Parameter values $a=5$, $b=4$, and $c=0$ (no smoothing) provided the best results to our simulated data. Thus, this combination captures the time constant of the used neural model. For more realistic applications different time scales should be considered because neurons are able to emit action potentials in several firing patterns.

To cover multiple spiking behaviours of neurons $SPE_{X\rightarrow Y}(d)'$ is extended by the integration of many combinations of filter parameters. $a=[3,4, 5, 6, 7, 8]$, $b=[2, 3, 4, 5, 6]$, $c=[0, 1]$ with vector length $N_a$, $N_b$, $N_c$ were chosen.
Low values of $a$ and $b$ increase the sensitivity to noise whereas high values do not affect results.

Further $N_a \cdot N_b \cdot N_c$ combinations were taken into account. This introduces different lengths of convolution results $SPE_{X\rightarrow Y}(d)'^{(n)}$ ($n$ is the index of the used edge filter). 
To obtain result vectors with same length 1D running total filters $h{(i)}^{(n)}$ (see Figure~\ref{fig:filters}.(b)) are applied to $SPE_{X\rightarrow Y}(d)'^{(n)}$.
For each edge filter $g{(i)}^{(n)}$ a corresponding running total filter is designed (\ref{equ:h}) by using the same parameter $b$ from (\ref{equ:g}).

\begin{equation}\label{equ:h}
h{(i)}=\begin{cases}
1&\mbox{if } 0 < i  \leq b \\ 
0 &\mbox{else.} 
\end{cases}
\end{equation}

\begin{figure}[H]
	\centering
	\includegraphics[width=1\textwidth]{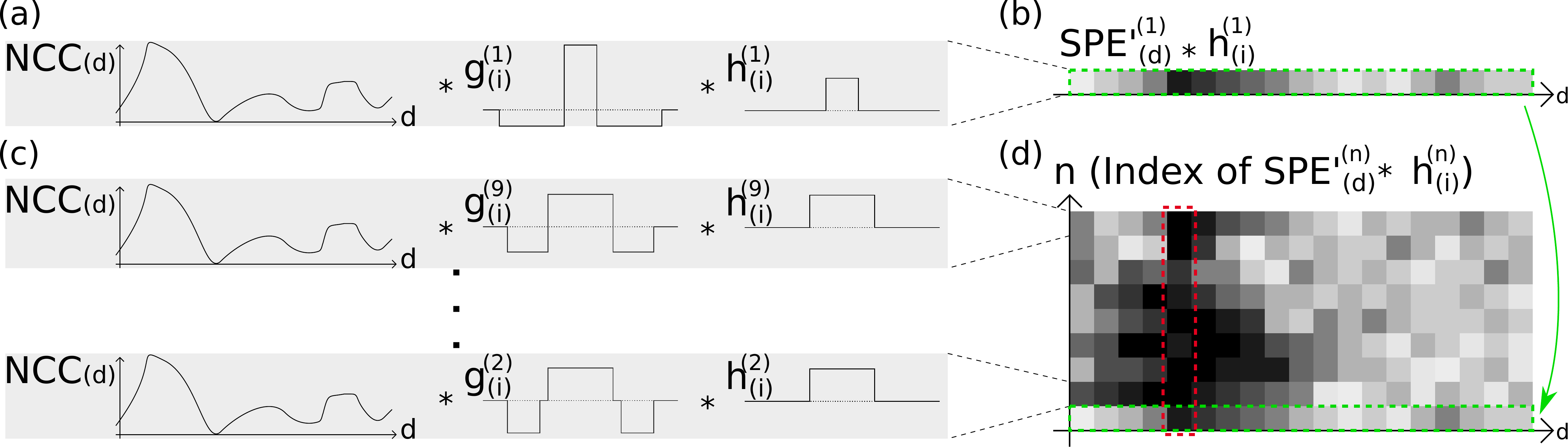}
	\caption[Principle of TSPE]{\textbf{Principle of TSPE}: 
		(a) After the $SPE{(d)}$ is calculated by a convolution of $NCC{(d)}$ with an edge filter $g{(i)}^{(1) }$ (see Figure~\ref{fig:spe}), a second convolution with a 1D running total filter $h{(i)}^{(1) }$ is performed.
		(b) The result of the second convolution $SPE{(d)}'^{(1) } * h{(i)}^{(1) }$ is plotted with a gray scale plot. Dark color indicates a high value at a certain delay time $d$. (c) In order to capture the variance of time constants of the neurons, $n$ edge filters $g{(i)}$ are calculated for different window parameters. By $n = 9$ filtering operations, a three dimensional representation is obtained (d). The abscissa is the delay time and the ordinate is the index of calculated $SPE{(d)}' * h{(i)}$, while the gray scale indicates the resulting value. The green marked convolution (b) can be found in function (d) as first row, which is also marked green. The resulting $TSPE{(d)}$ can be obtained by adding the values vertically. The absolute maximum is the most likely point of effect, which is marked red.
	}\label{fig:tspe}
\end{figure}

As all $SPE_{X\rightarrow Y}(d)'^{(n)}$ have the same length, a matrix is obtained by introducing a row for each calculated $SPE_{X\rightarrow Y}(d)'^{(n)}$ (see Figure~\ref{fig:tspe}.(d)). An vertically addition of this matrix enables the consideration of different time scales. 
\begin{equation}\label{equ:sumo}
TSPE_{X\rightarrow Y}(d)=\sum_{n=1}^{N_a \cdot N_b \cdot N_c} SPE_{X\rightarrow Y}(d)'^{(n)} * h{(i)}^{(n) }
\end{equation}
The resulting $TSPE_{X\rightarrow Y}(d)$ values are interpreted as described before for $SPE'_{X\rightarrow Y}(d)$. 
The sign of $TSPE_{X\rightarrow Y}(d)$ with $d$ at the 
absolute extreme value allows a discrimination between inhibitory from excitatory effects.    

\vspace{0.5cm}

\newpage
\section{Threshold calculation methods}
The resulting \acp{CM} of connectivity estimation algorithms contain values depending on the chosen method. These values have to be classified in order to distinguish between a 'real connection' and a 'statistical correlation'. There are different approaches to select a threshold, which are presented below. 

\subsection{Easy threshold calculation}
In some studies~\cite{Pastore.2017} the distribution of the resulting \ac{CM} is used to calculate the threshold. By obtaining the mean value and the \ac{SD} of all \ac{CM} values a threshold for the whole \ac{CM} is calculated. For example, a threshold is selected at $\overline{CM} + 4 \cdot \sigma_{CM}$~\cite{Pastore.2017}. Here, the computing time is negligible. 

\subsection{Threshold calculation with surrogate data}
\begin{figure}[htbp]
	\centering
	\includegraphics[width=0.8\textwidth]{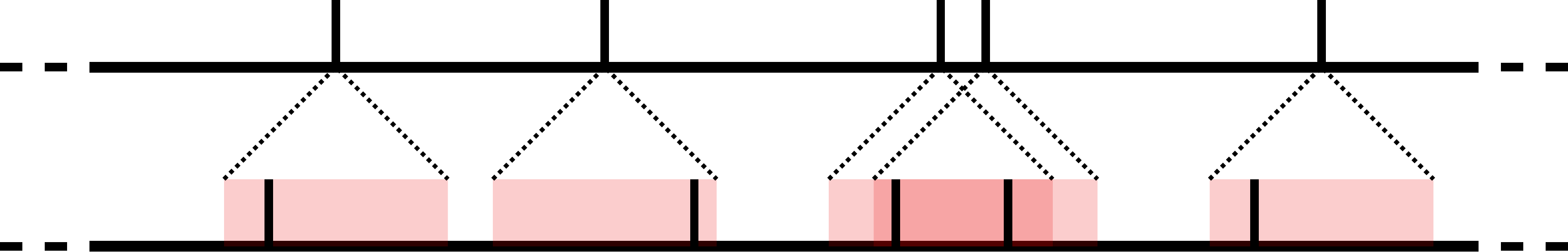}
	\caption[Spike dithering for generate surrogate data]{\textbf{Spike dithering for generate surrogate data:} Every detected spike of an original spike train will be shifted by a random time in a specific time window (shaded areas). The upper spike train is original and was used to generate the lower surrogate spike train. Close spikes (e.g., at bursting) can even be switched.}\label{fig:jitter}
\end{figure}

To generate surrogate data spike dithering (also known as jittering)~\cite{Date.1998} is used. The method of Date is a classic tool of neuroscience~\cite{Abeles.2001,Pazienti.2007,Pazienti.2008,Butts.2007} for testing the significance of results.
In a defined time window each spike of an original spike train will be individual and randomly displaced in order to generate spike trains with similar characteristics (see Figure~\ref{fig:jitter}). The distribution of shifting times is chosen uniform. In this way a set of slightly different spike trains is obtained.

Since the focus is on paired-wise examinations of connectivity, the threshold should be calculated for each pair of spike trains. Considering two spike trains $X$ and $Y$, which are both used for the generation of surrogate data several times (here $n$ is between 100 and 1000). A distribution of connectivity estimation results is obtained for this potential link between the neurons. The threshold is selected at values calculated with the surrogate value distribution.
This single threshold is only used for the classification of the relation between the two spike trains. For each relation of spike trains the generation of surrogate data and calculation of threshold is necessary. Thus, the computation time of this classification step is about $n$ times the time required by the estimation algorithm for connectivity.

\section{Implementations}
For \ac{DTE}, \ac{DHOTE}, \ac{DTECI} and \ac{DHOTECI} the toolbox described in~\cite{Ito.2011} is used, which is a \ac{MEX} application. \ac{CDHOTE} is based on the calculations of \ac{DTECI} and \ac{DHOTECI}. The implementation of \ac{NCC}, \ac{NCCCI} and \ac{TSPE} are realized with sparse matrices multiplications
(unbinned simulation data, approx. 0.2\% filled).
All spike trains are stored in a common sparse matrix with the length that equals the sampling number. This matrix is multiplied with the transposed and time shifted matrix. The normalization matrix is calculated by multiplying the \ac{SD} vector with the its transposed version. For a faster calculation of \ac{TSPE} the mean values of spike trains (0.005 to 0.04 at bin size of 1\,ms) were omitted, because they did not affect the accuracy.

\chapter{Evaluation of connectivity estimation}
Based on the findings of Chapter~\ref{cha:sim} an evaluation framework was used to evaluate the connectivity estimation methods of Chapter~\ref{cha:alg}. First, the necessary tools for an appropriate comparison of estimation accuracy are presented as well as the results of the evaluation. Furthermore, the calculation time of each method were measured and compared with each other. The issue of interpreting the result values of the tested methods is analysed with multiple methods. At the end of this chapter, the results are concluded and summarized.

\section{Evaluation tools}
\subsection{Simulation}
All simulations were performed in \textit{2017a MATLAB}, MathWorks, with a modified version of the published code described in~\cite{Izhikevich.2006}. 
Each network type was generated and simulated ten times with different seed values for the random number generator of our simulation. 
The \textit{in silico} networks were designed according the guide described in~\cite{DeBlasi.2018} for evaluation applications in neuroscience. 
The spike train subset of 100 neurons was recorded for 60 minutes, while studies were performed also for shorter time frames to analyse the impact of this parameter.
It was demonstrated that long recording times improve the estimation results~\cite{Ito.2011}. 

\subsection{ROC Curve}
The accuracy of a connectivity estimation algorithm is evaluated by comparing the results with the properties of the simulated network, described in the \ac{SWM}.
Since value ranges of \ac{CM} are strongly dependent on the measured signal, these values are not directly comparable.
In the best case, values of \ac{CM} should be proportional to real synaptic weights of \ac{SWM}. A threshold to distinguish between a 'real connection' and a 'statistical correlation' is used to calculate the \ac{TCM}. This binary pattern of connection or non-connection will be used for the comparison with the \ac{SWM}.
The matches and mismatches between \ac{TCM} and \ac{SWM} are stored in four groups. Matches of connections are \ac{TP}, mismatches are \ac{FP}, matches for non existing synapses are \ac{TN}, and mismatches are \ac{FP}.
A standard method to evaluate the performance of classifiers is the \ac{ROC} curve, which is a plot of \ac{TPR}
\begin{equation}\label{equ:TPR}
TPR=\frac{TP}{TP + FN}
\end{equation}
depending on the \ac{FPR}
\begin{equation}\label{equ:FPR}
FPR=\frac{FP}{FP + TN} .
\end{equation}
A perfect reconstruction of the \ac{SWM} is indicated by a \ac{TPR} of 1 and a \ac{FPR} of 0. In case of equality of both rates classification is a random guess.

Because of the sparse \ac{SWM} a low \ac{FPR} means a larger amount of wrongly estimated connections than correctly estimated connections even with large \acp{TPR}. 
To prevent this misleading impression, the evaluation focuses on the \ac{TPR} values at 1\% \ac{FPR}.

\subsection{Confusion matrix}
A widely used visualization tool for the classification performance is the confusion matrix, or error matrix. It is a specific table layout that allows visualization of the performance of an algorithm. Each column of the matrix represents the labels while each row represents the predicted class. This visualization allows to see easily which classes are classified with high or low accuracy and which classes are often confused. In this study, inhibitory, excitatory, and no connection were used as class labels. 
The columns of a 2D confusion matrix are real labels obtained by the \ac{SWM}. 
The rows contain the predicted classes of the classification algorithm. 
On the very right column the percentage of correctly classified connections of the output class is shown, at the bottom the percentage of correctly classified connections of the target classes is displayed. The general classification accuracy can be found in the lower right box.
The classification ability of \ac{TSPE} for distinguishing inhibitory from excitatory synapses is evaluated at a 1\% \ac{FPR} level.

\section{Accuracy of functional connectivity estimation}
\begin{figure}[H]
	\centering
	\includegraphics[width=0.95\textwidth]{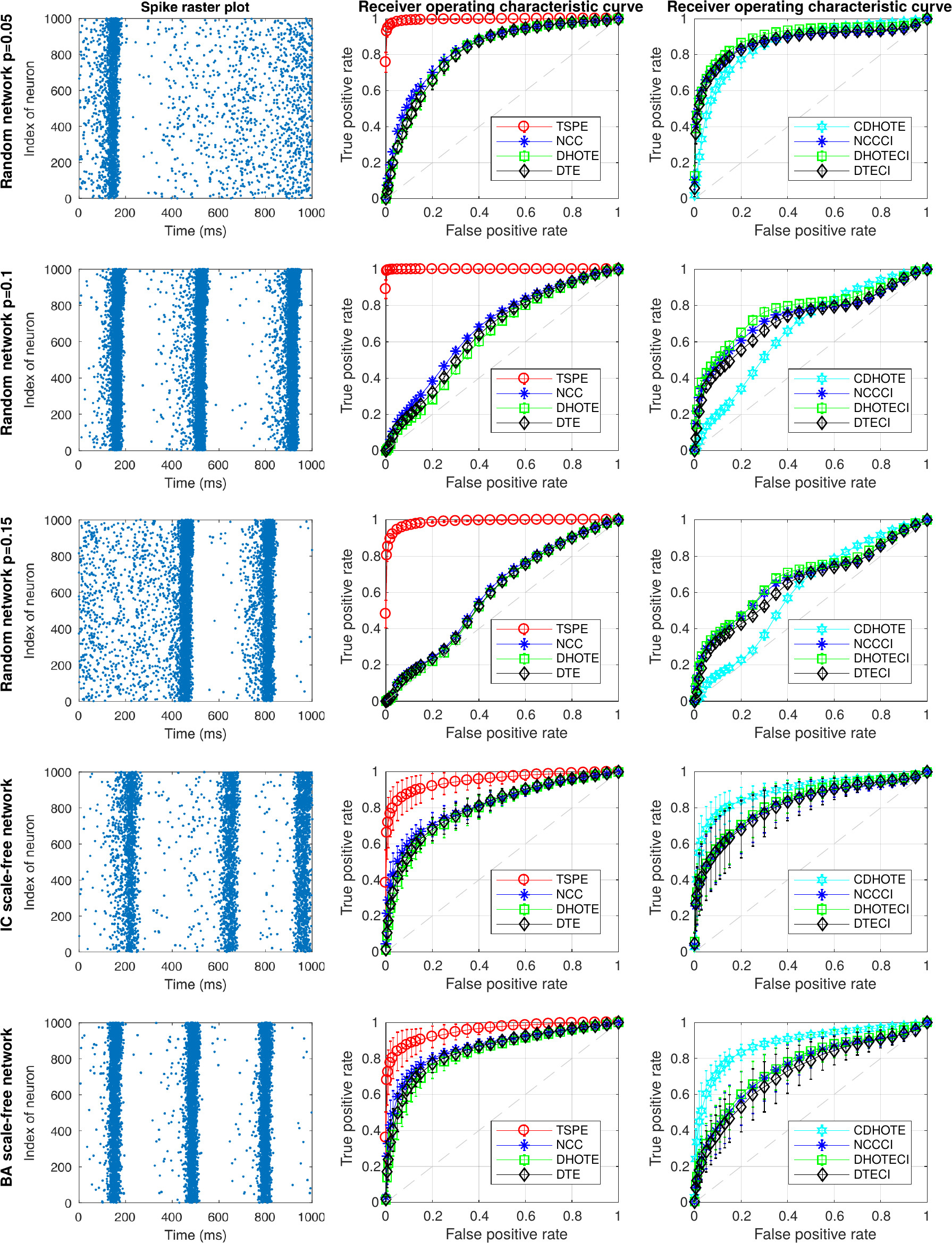} 
	\caption[Evaluation of connectivity estimation algorithms for different network topologies]{\textbf{Evaluation of connectivity estimation algorithms for different network topologies:} \textbf{Left column:}
		Spike trains of the simulated networks. For the evaluation a subset of only 100 spike trains with a simulation duration of 30 minutes were used. Network bursts appeared for all network topologies. 
		\textbf{Middle and right column:} \ac{ROC} curves of all tested algorithms for $n=10$ simulations per network topology. With increased complexity of the \textit{random networks} the accuracy of all algorithms decreased. TSPE outperformed all evaluated algorithms.}\label{fig:rocs}
\end{figure}

Generally, spike raster plots of all simulated topologies show spikes and network bursts with varying rate (see Figure~\ref{fig:rocs} left). For \textit{random networks} the spike density within network bursts increases with connection probability $p$. For \textit{scale-free networks}, spike density within network bursts is lower for the \ac{IC} version than for the \ac{BA} version.

The accuracy of the connectivity estimation methods \ac{TSPE}, \ac{NCC}, \ac{NCCCI}, \ac{DHOTE}, \ac{DHOTECI}, \ac{DTE}, \ac{DTECI} and \ac{CDHOTE} was calculated for signals generated with \textit{random} and \textit{scale-free networks} and compared.
The results of all tested algorithms are depicted in Figure~\ref{fig:rocs} and better than random guessing (grey dashed line). The accuracy strongly depends on network topologies.
For spike raster plots generated with \textit{random networks}, the performances of the tested algorithms deteriorates with higher connection probability $p$.
Methods based on a coincidence index (\ac{NCCCI}, \ac{DHOTECI}, \ac{DTECI}) perform  significantly better than their corresponding basic algorithms (\ac{NCC}, \ac{DHOTE}, \ac{DTE}). The results of the \ac{TSPE} algorithm shows a \ac{ROC} curve with a \ac{TPR} up to 99.5\% for $p = 0.05$ and $p = 0.1$. For $p = 0.15$ the performance accuracy decreased.
The \ac{CDHOTE} algorithm was inferior to \ac{NCCCI}, \ac{DHOTECI}, and \ac{DTECI}. In contrast, for \textit{scale-free network} topologies \ac{CDHOTE} was superior. 

The \ac{SD} of the results varied depending on topology. For \textit{random networks}, the \ac{SD} was lower than for \textit{scale-free networks}.
At a \ac{FPR} of 1\% \ac{TSPE} estimated the connectivity more precisely than any other tested algorithms. For both \textit{scale-free networks} the accuracy of \ac{TSPE} is inferior to the performance for random networks.

\begin{figure}[H]
	\centering
	\includegraphics[width=0.6\textwidth]{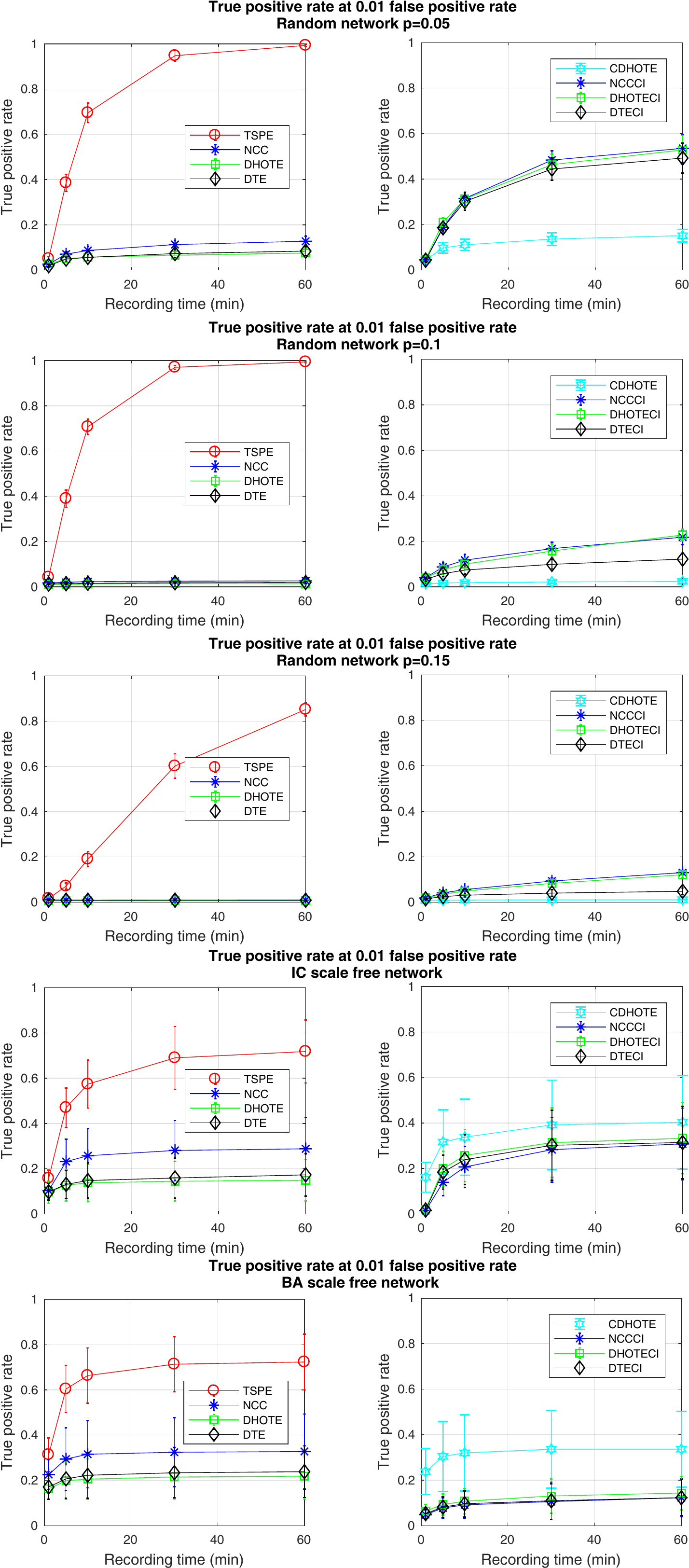}
	\caption[Effects of the recording time on the accuracy of the connectivity estimation]{\textbf{Effects of the recording time on the accuracy of the connectivity estimation}. 
		Accuracy estimation increases until a saturation is reached. Within the first ten minutes, the increase is strongest which is prominent for the \ac{TSPE} algorithm.
	}\label{fig:time_FPR01}
\end{figure}

To study the performance dependency on the recording duration of spike train data, the \ac{TPR} was measured at \ac{FPR} of 1\% for simulation durations of 1\,min, 5\,min, 10\,min,\,30\,min and 60\,min (see Figure~\ref{fig:time_FPR01}). The results show that accuracy increases with recording time. In case of \textit{random networks}, \ac{TSPE} is able to reach almost a \ac{TPR} of 100\% for a simulation duration of 60\,min ($p = 0.05$ and $p = 0.1$). Connectivity estimation saturates and therefore no accuracy increase is assumed for simulation durations longer than 60 minutes which is also true for \textit{scale-free networks}.
\newpage

\newpage
\section{Accuracy of effective connectivity estimation by TSPE}
\begin{figure}[!htb]
	\centering
	\includegraphics[width=\textwidth]{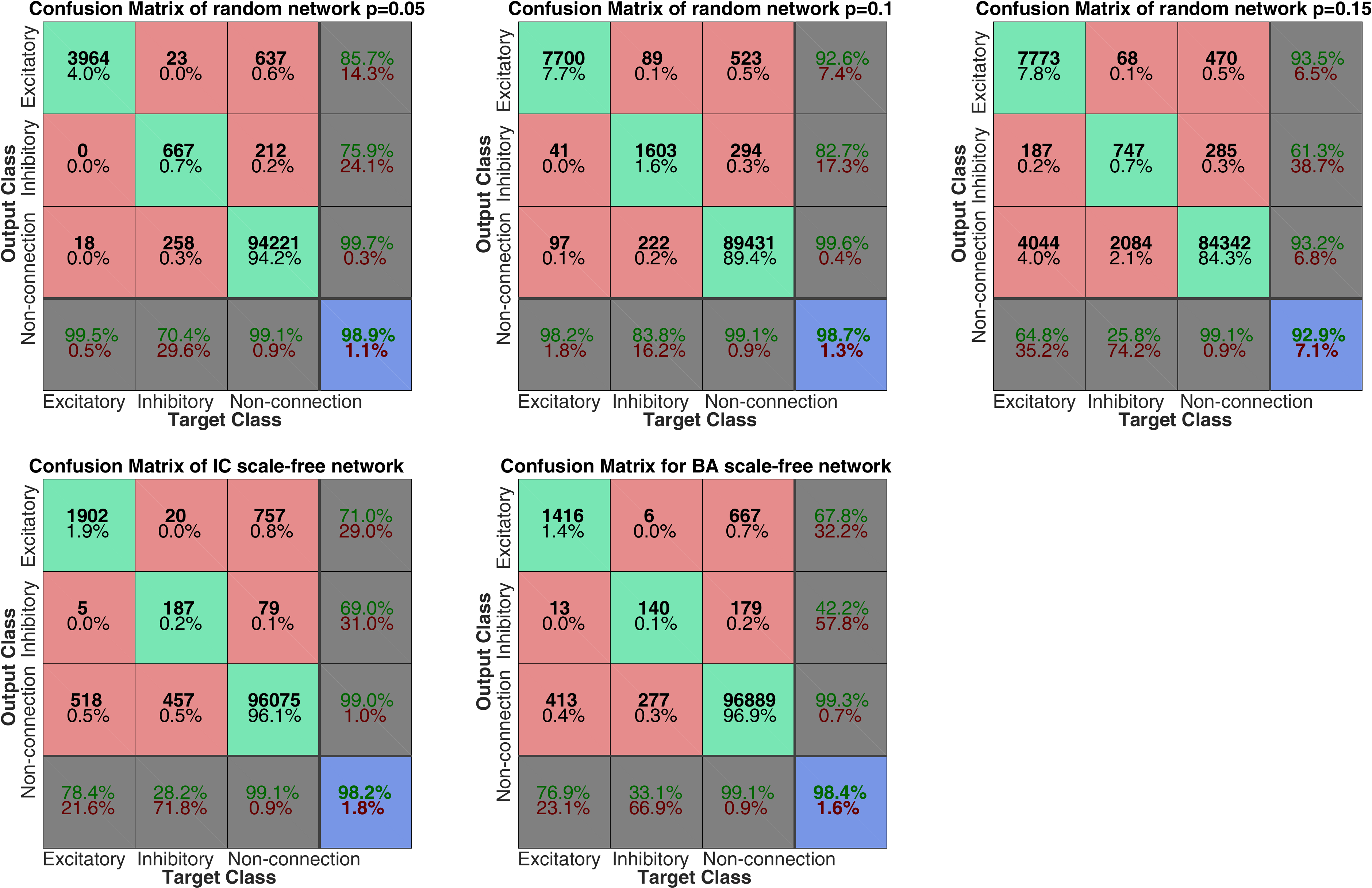}
	\caption[Confusion matrices for connection types classified by TSPE]{\textbf{Confusion matrices for connection types classified by \ac{TSPE}}: 
		Green (red) fields: correct (incorrect) classification (upper number: absolute, lower number: in percentage). Bottom row (grey): estimated connections to actual connections; Right column (grey): correctly estimated connections to all estimated connections.  Blue field: total accuracy of classification. Grey and blue fields: green number percentage of correctness; red number percentage of incorrectness.	
	}\label{fig:conplots}
\end{figure}

For effective connectivity not just connection strength but also information about causality and the synaptic effect is required. In contrast to the other tested algorithms, \ac{TSPE} offers information about excitation and inhibition.
In Figure~\ref{fig:conplots}, the confusion matrices are plotted for TSPE at a FPR threshold level of 1\%.

For \textit{random networks} with $p = 0.05$, the total classification accuracy was 98.9\%, which is 0.1\% below the maximal achievable classification accuracy. 
By increasing the complexity of \textit{random networks} to $p = 0.1$, the total accuracy decreased by 0.2\% (see blue boxes in Figure~\ref{fig:conplots}).
The classification accuracy of excitatory connections decreased and the detection rate of inhibitory effects increased.
Further increase of complexity to $p = 0.15$ resulted in a collapse of the classification performance and many effects were not detected instead of classified as a synaptic connection.

For both \textit{scale-free network} types, the accuracy was 98.2 and 98.4\% (78.4\% and 76.9\% for excitatory effects). However, about 70\% of all inhibitory effects were not classified correctly (see Figure~\ref{fig:conplots} dark grey, bottom-center boxes). 

To summarize, the estimation accuracy for excitatory synapses was between $64.8$ and $99.5$\%, for inhibitory effects it was between $28.2$ and $83.8$\%.

\vspace{0.5cm}

\newpage
\section{Calculation time}
\begin{figure}[H]
	\centering
	\includegraphics[width=\textwidth]{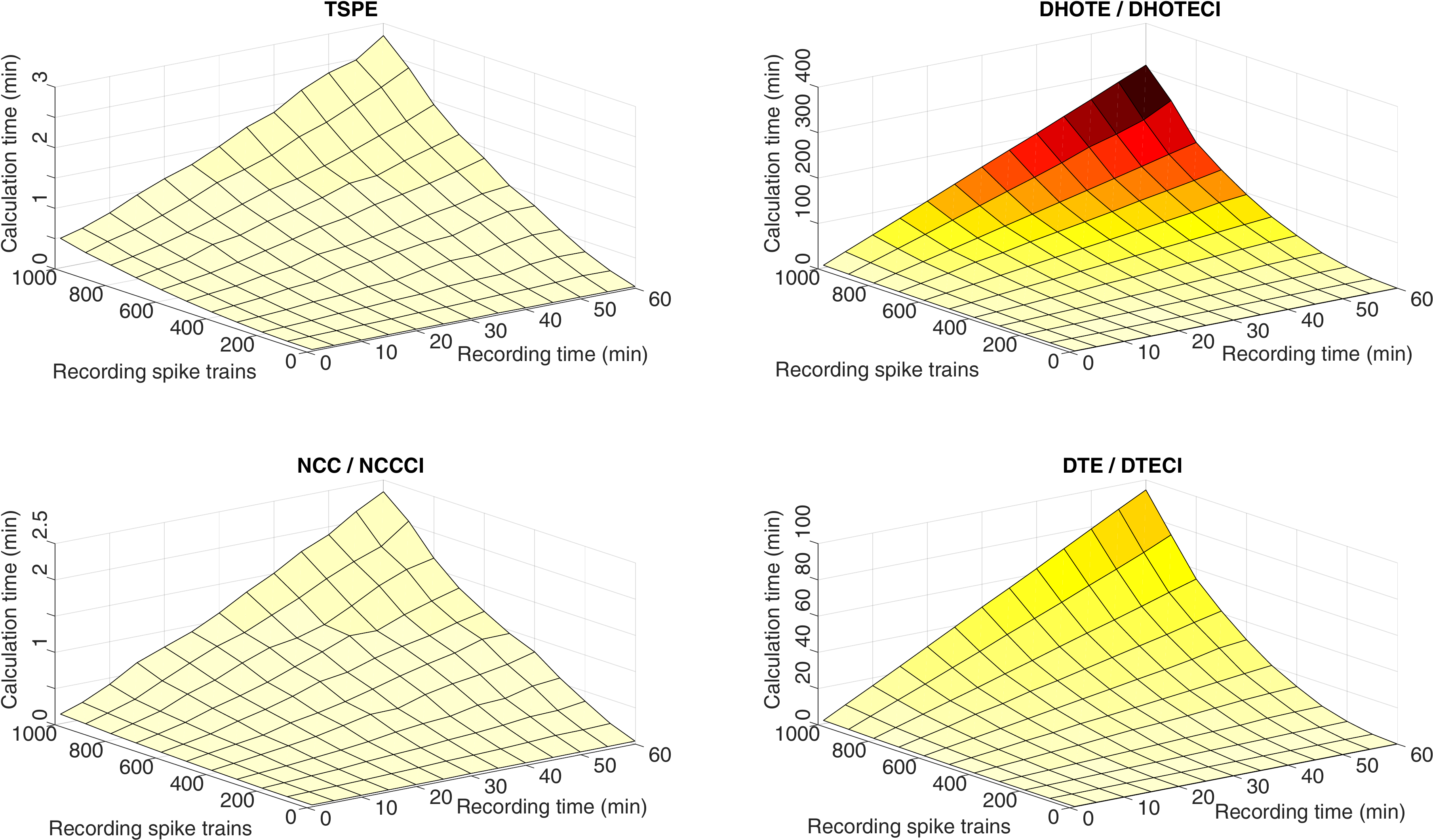}
	\caption[omparison of calculation times of tested algorithms]{\textbf{Comparison of calculation times of tested algorithms:} 
		The connectivity estimation algorithms were evaluated using data generated by \ac{IC} \textit{scale-free networks} for different lengths of recording and a variable number of spike trains. Red colored areas indicated long and light areas indicated small calculation times.
		Calculation time increased linearly with duration but exponentially with number of spike trains. 
		Calculation time was in the following ascending order for algorithm groups: 1) \ac{NCC}, \ac{NCCCI}, \ac{TSPE}  2) \ac{DTE}, \ac{DTECI} 3) \ac{DHOTE}, \ac{DHOTECI}.
	}\label{fig:calctime}
\end{figure}

For a comparison of calculation efficiency, the processing time was measured for the calculation of connectivity estimation for \ac{IC} \textit{scale-free} networks with different number of spike trains (between 2 and 1000) and different simulation durations (between 1 and 60 minutes, see Figure~\ref{fig:calctime}).
Calculations were conducted using \textit{MATLAB Distributed Computing Server} toolbox on a high-performance computer, which is equipped with 2 Intel Xeon 'Broadwell' E5-2680v4 processors, 8x32\,GB DDR4 2400\,MHz RAM, SSD and 4 SXM-2 P100 GPUs.

The results show that the calculation time increased linearly with recording time (visible in Fig~\ref{fig:calctime}).
Due to the pairwise comparison of spike trains, the computing time increased exponentially (power of two) with the number of spike trains. For large numbers of recorded spike trains, the calculation of \ac{TE} based algorithms, like \ac{DTE} or \ac{DHOTE}, was longer than the calculation of \ac{CC} based algorithms. \ac{NCC} and \ac{TSPE} were parallelized (matrix operation based algorithms). 
For example, the calculation time for a \ac{IC} \textit{scale-free} network with 1000 recorded spike trains and a simulation duration of 10 minutes was approx. 25 seconds for \ac{NCC} or \ac{NCCCI}, 45 seconds for \ac{TSPE}, 16 minutes for \ac{DTE} or \ac{DTECI}, and 51 minutes for \ac{DHOTE} or \ac{DHOTECI}.

\newpage
\section{Evaluation of threshold calculation}
Since resulting \acp{CM} of connectivity estimation algorithms have different value ranges, different thresholds have to be calculated in order to obtain a \ac{TCM} with the best combination of \ac{TPR} and \ac{FPR} (optimal case: \ac{TPR}~=~1 at \ac{FPR}~=~0). This is a crucial step in connectivity estimation because even the best connectivity estimation method have two random guess points: \ac{TPR}~=~0 at \ac{FPR}~=~0 (no estimated connection) and \ac{TPR}~=~1 at \ac{FPR}~=~1 (all estimation values are identified as connection). In the following the threshold calculation methods are evaluated.

\subsection{Easy threshold calculation}
The \acp{TPR} and \acp{FPR} are illustrated for \ac{TSPE} and \ac{DHOTE} with easy threshold calculation with $4 \cdot \sigma_{CM}$ in Table~\ref{tab:easythreshold4}. To keep clarity other connectivity estimation methods are not mentioned here. All \ac{FPR} are smaller than 1 percent. For \ac{TSPE} it is even zero in all examined cases. However, the result is not satisfying because of the low \acp{TPR}. At least for \ac{ER} \textit{random networks} with $p=0.05$ and the \textit{scale-free networks} \ac{TSPE} is able to archive mean \acp{TPR} between 0.357 and 0.448. This means the selected threshold is too large. For \ac{DHOTE} the easy threshold calculation with $\overline{CM} + 4 \cdot \sigma_{CM}$ works fine in order to set the \ac{FPR} at one percent (with low \acp{TPR} because its \ac{ROC}).

\begin{table}[H]
	\caption[Evaluation of easy threshold calculation (four times std)]{\textbf{Evaluation of easy threshold calculation (four times \ac{SD}):\\} For \ac{TSPE} and DHOTE \acp{FPR} is low with easy threshold calculation ($\overline{CM} + 4 \cdot \sigma_{CM}$). Especially for \textit{random networks} with connection probability larger than 0.05, \acp{TPR} are also low. \acp{FPR} of \ac{DHOTE} are around one percent.} \label{tab:easythreshold4}	\vspace{-0.5cm}
	\begin{center}	
		\begin{tabular}{ l | c | c | c | c}
			\hline
			Network type & \ac{TPR} of \ac{TSPE} & \ac{FPR} of \ac{TSPE} & \ac{TPR} of \ac{DHOTE} & \ac{FPR} of \ac{DHOTE}\\
			\hline
			ER \textit{random network} $p$=0.05 &  0.392 $\pm$ 0.028 & 0.000 $\pm$ 0.000 & 0.045 $\pm$ 0.011 & 0.007 $\pm$ 0.001 \\
			ER \textit{random network} $p$=0.1  &  0.104 $\pm$ 0.012 & 0.000 $\pm$ 0.000 & 0.001 $\pm$ 0.003 & 0.007 $\pm$ 0.001 \\
			ER \textit{random network} $p$=0.15 &  0.030 $\pm$ 0.005 & 0.000 $\pm$ 0.000 & 0.009 $\pm$ 0.002 & 0.011 $\pm$ 0.001 \\
			IC \textit{scale-free network}      &  0.357 $\pm$ 0.075 & 0.000 $\pm$ 0.000 & 0.110 $\pm$ 0.055 & 0.008 $\pm$ 0.002 \\
			BA \textit{scale-free network}      &  0.448 $\pm$ 0.109 & 0.000 $\pm$ 0.000 & 0.149 $\pm$ 0.056 & 0.006 $\pm$ 0.002 \\

			\hline
		\end{tabular}
	\end{center}
\end{table}

By decreasing the threshold to $\overline{CM} + 3 \cdot \sigma_{CM}$ the probability of increasing \acp{FPR} is larger. Since \acp{FPR} of \ac{TSPE} were small in Table~\ref{tab:easythreshold4}, larger \acp{FPR} can be allowed in order to archive better \acp{TPR}. In Table~\ref{tab:easythreshold3} the results for $\overline{CM} + 3 \cdot \sigma_{CM}$ are illustrated. While \acp{FPR} of \ac{DHOTE} are here already too large, \acp{FPR} of \ac{TSPE} are still low. In this way the estimation accuracy of \ac{TSPE} was able to be further increase by up to 25.3 percent.

\begin{table}[H]
	\caption[Evaluation of easy threshold calculation (three times std)]{\textbf{Evaluation of easy threshold calculation (three times \ac{SD}):\\} While \acp{FPR} of \ac{DHOTE} strongly increased, \acp{FPR} of \ac{TSPE}  are still low with easy threshold calculation ($\overline{CM} + 3 \cdot \sigma_{CM}$). } \label{tab:easythreshold3}	\vspace{-0.5cm}
	\begin{center}	
		\begin{tabular}{ l | c | c | c | c}
			\hline
			Network type & \ac{TPR} of \ac{TSPE} & \ac{FPR} of \ac{TSPE} & \ac{TPR} of \ac{DHOTE} & \ac{FPR} of \ac{DHOTE}\\ 
			\hline
			ER \textit{random network} $p$=0.05 & 0.645 $\pm$ 0.017 & 0.000 $\pm$ 0.000 & 0.115 $\pm$ 0.002 & 0.017 $\pm$ 0.001 \\
			ER \textit{random network} $p$=0.1  & 0.349 $\pm$ 0.020 & 0.000 $\pm$ 0.000 & 0.032 $\pm$ 0.005 & 0.019 $\pm$ 0.002 \\
			ER \textit{random network} $p$=0.15 & 0.130 $\pm$ 0.014 & 0.000 $\pm$ 0.000 & 0.021 $\pm$ 0.001 & 0.025 $\pm$ 0.002 \\
			IC \textit{scale-free network}      & 0.469 $\pm$ 0.079 & 0.001 $\pm$ 0.002 & 0.173 $\pm$ 0.079 & 0.013 $\pm$ 0.002 \\
			BA \textit{scale-free network}      & 0.581 $\pm$ 0.095 & 0.002 $\pm$ 0.001 & 0.240 $\pm$ 0.078 & 0.013 $\pm$ 0.003 \\

			\hline
		\end{tabular}
	\end{center}
\end{table}

For  $\overline{CM} + 2 \cdot \sigma_{CM}$, all \acp{TPR} increase even further. Like before \ac{DHOTE} gains too large \acp{FPR}. Results of Table~\ref{tab:easythreshold2} substantiate that each method the optimal multiplication factor of \ac{SD} can be different. Even lower thresholds increase the \acp{FPR} of \ac{TSPE} too much. 

In summary, the optimal easy calculation threshold for \ac{TSPE} is $\overline{CM} + 2 \cdot \sigma_{CM}$ and for \ac{DHOTE} $\overline{CM} + 4 \cdot \sigma_{CM}$.
By using one threshold for the whole \ac{CM} the resulting estimation accuracy will always be a point of the \ac{ROC} function.

\begin{table}[H]
	\caption[Evaluation of easy threshold calculation (two times std)]{\textbf{Evaluation of easy threshold calculation (two times \ac{SD}):\\} While \acp{FPR} of \ac{DHOTE} strongly increased, \acp{FPR} of \ac{TSPE} are still low with easy threshold calculation. } \label{tab:easythreshold2}	\vspace{-0.5cm}
	\begin{center}	
		\begin{tabular}{ l | c | c | c | c}
			\hline
			Network type & \ac{TPR} of \ac{TSPE} & \ac{FPR} of \ac{TSPE} & \ac{TPR} of \ac{DHOTE} & \ac{FPR} of \ac{DHOTE}\\ 
			\hline
			ER \textit{random network} $p$=0.05 & 0.828 $\pm$ 0.020 & 0.000 $\pm$ 0.000 & 0.224 $\pm$ 0.022 & 0.037 $\pm$ 0.002 \\
			ER \textit{random network} $p$=0.1  & 0.700 $\pm$ 0.019 & 0.000 $\pm$ 0.000 & 0.090 $\pm$ 0.008 & 0.039 $\pm$ 0.003 \\
			ER \textit{random network} $p$=0.15 & 0.372 $\pm$ 0.021 & 0.001 $\pm$ 0.000 & 0.057 $\pm$ 0.008 & 0.039 $\pm$ 0.001 \\
			IC \textit{scale-free network}      & 0.604 $\pm$ 0.074 & 0.005 $\pm$ 0.007 & 0.272 $\pm$ 0.095 & 0.027 $\pm$ 0.004 \\
			BA \textit{scale-free network}      & 0.718 $\pm$ 0.088 & 0.013 $\pm$ 0.008 & 0.376 $\pm$ 0.102 & 0.032 $\pm$ 0.006 \\

			\hline
		\end{tabular}
	\end{center}
\end{table}

\subsection{Threshold calculation with surrogate data}
Referencing to the long computing time of \ac{DHOTE} and the high number of iterations (between 100 and 1000), this part of evaluation is only done for \ac{TSPE}.
The calculation was tested for three different thresholds:
\begin{itemize}
	\item The mean value +/- four times \ac{SD} (Table~\ref{tab:surrothreshold})
	\item The minimum/maximum value (Table~\ref{tab:surrothreshold1})
	\item The minimum/maximum value +/- \ac{SD} (Table~\ref{tab:surrothreshold2})
\end{itemize}
These mean, minimum and maximum values refer to the generated surrogate data of the examined pairwise correlation.
The accuracy of minimum and maximum based threshold calculation varies even with high numbers of iterations due to the randomly increasing thresholds. Even if the evaluation results of these methods are promising, the reproducibility of the results in the experimental environment may be negatively affected.
This is the reason for recommending the first method 'mean value +/- four times \ac{SD}'. Here, the reproducibility of results is good.

\begin{table}[!htb]
	\caption[Evaluation of surrogate threshold calculation (mean value +/- 4 SD)]{\textbf{Evaluation of surrogate threshold calculation (mean value +/- 4 SD) with window size 2\,ms:}
	For 100 iterations, \acp{FPR} are fine for all cases except \ac{BA} \textit{scale-free networks}. Here, the \ac{FPR} of 1.9\% was able to be improved only slightly to 1.7\% by increasing the number or iterations.
} \label{tab:surrothreshold}	\vspace{-0.0cm}
	\begin{center}	
		\begin{tabular}{ l | c | c | c}
			\hline
			Network type & TPR of TSPE & FPR of TSPE & Number of iterations \\
			\hline
			ER \textit{random network} p=0.05 & 0.329 $\pm$ 0.072 & 0.000 $\pm$ 0.000 & 100\\
			ER \textit{random network} p=0.1  & 0.643 $\pm$ 0.066 & 0.000 $\pm$ 0.000 & 100\\
			ER \textit{random network} p=0.15 & 0.045 $\pm$ 0.018 & 0.000 $\pm$ 0.000 & 100\\
			IC \textit{scale-free network}    & 0.623 $\pm$ 0.101 & 0.001 $\pm$ 0.003 & 100\\
			BA \textit{scale-free network}    & 0.918 $\pm$ 0.030 & 0.019 $\pm$ 0.020 & 100\\
										      & 0.929 $\pm$ 0.029 & 0.017 $\pm$ 0.018 & 500\\			
			\hline
		\end{tabular}
	\end{center}
\end{table}

\begin{table}[!htb]
	\caption[Evaluation of surrogate threshold calculation (minimum/maximum)]{\textbf{Evaluation of surrogate threshold calculation (minimum/maximum) with window size 2\,ms:}
	In contrast to all previous threshold selection methods, the \acp{TPR} are improved with the minimum/maximum version. The \acp{FPR} of all \textit{scale-free networks} are too large -- even with 1000 iterations. 
} \label{tab:surrothreshold1}	\vspace{-0.0cm}
	\begin{center}	
		\begin{tabular}{ l | c | c | c}
			\hline
			Network type & TPR of TSPE & FPR of TSPE & Number of iterations \\
			\hline
			ER \textit{random network} p=0.05 & 0.770 $\pm$ 0.049 & 0.003 $\pm$ 0.000 & 100 \\
			ER \textit{random network} p=0.1  & 0.959 $\pm$ 0.027 & 0.010 $\pm$ 0.003 & 100 \\
			ER \textit{random network} p=0.15 & 0.354 $\pm$ 0.078 & 0.004 $\pm$ 0.000 & 100 \\
			IC \textit{scale-free network}    & 0.818 $\pm$ 0.066 & 0.026 $\pm$ 0.037 & 100 \\
			  								  & 0.786 $\pm$ 0.076 & 0.016 $\pm$ 0.027 & 200 \\
			BA \textit{scale-free network}    & 0.954 $\pm$ 0.019 & 0.052 $\pm$ 0.044 & 1000\\
			\hline
		\end{tabular}
	\end{center}
\end{table}

\begin{table}[!htb]
	\caption[Evaluation of surrogate threshold calculation (minimum/maximum -/+ SD)]{\textbf{Evaluation of surrogate threshold calculation (minimum/maximum -/+ SD) with window size 2\,ms:}
	Except for \textit{random network} p=0.05, the \acp{TPR} are larger than for the easy threshold calculation methods. With 1000 iterations low \acp{FPR} are even ensured for \ac{BA} \textit{scale-free networks}. 
} \label{tab:surrothreshold2}	\vspace{-0.0cm}
	\begin{center}	
		\begin{tabular}{ l | c | c | c} 
			\hline
			Network type & TPR of TSPE & FPR of TSPE & Number of iterations \\
			\hline
			ER \textit{random network} p=0.05 & 0.463 $\pm$ 0.059 & 0.000 $\pm$ 0.000 & 100 \\
			ER \textit{random network} p=0.1  & 0.818 $\pm$ 0.090 & 0.000 $\pm$ 0.000 & 100 \\
			ER \textit{random network} p=0.15 & 0.130 $\pm$ 0.035 & 0.000 $\pm$ 0.000 & 100 \\
			IC \textit{scale-free network}    & 0.677 $\pm$ 0.107 & 0.001 $\pm$ 0.001 & 100 \\
			BA \textit{scale-free network}    & 0.948 $\pm$ 0.034 & 0.045 $\pm$ 0.043 & 100 \\
											  & 0.909 $\pm$ 0.041 & 0.005 $\pm$ 0.005 & 1000 \\
			\hline
		\end{tabular}
	\end{center}
\end{table}

\newpage
\section{Conclusion of evaluation}
A large framework of \textit{in silico} networks with different topologies was used to benchmark the performance of the connectivity estimation methods \ac{TSPE}, \ac{NCC}, \ac{NCCCI}, \ac{DHOTE}, \ac{DHOTECI}, \ac{DTE}, \ac{DTECI} and \ac{CDHOTE}. Influences of recording time on estimation accuracy and calculation time were analysed. Furthermore, the classification ability in terms of inhibitory and excitatory effects of \ac{TSPE} was evaluated.

The novel method \ac{TSPE} is able to outperform the accuracy of other connectivity estimation algorithms when applied on simulated neuronal network data with different topologies. Especially for spike trains with a long recording duration like 30 minutes \ac{TSPE} was outperforming.
With \ac{TSPE}, it was possible to discriminate between estimated excitatory and inhibitory connections, which is characteristic for effective connectivity. The total classification accuracy varied between 92.9 and 98.9\%, depending on complexity of network topology. 

Although these results are very promising, there are some critical aspects.
\ac{TSPE} is not able to detect effects of self-connections. Thus, the diagonal values of the \ac{CM} have to be neglected or set to zero.
Like for all other evaluated algorithms, the current implementation does not take multiple effects, e.g. driven by parallel connections, for a causal relation into account. Further, inhibition is more difficult to identify and to classify than excitation with estimation algorithms ~\cite{Ito.2011, Masud.2017} which is one aim of further improvements of \ac{TSPE}.

\ac{TSPE} is easy to implement and fast for large spike train datasets. 
Since new technologies of electrophysiological recording are able to record from thousands of electrodes, e.g. \ac{HDMEA} chips with 4096 electrodes~\cite{Berdondini.2009}, it is crucial to minimize the computation time for large numbers of recorded spike trains. In our studies, \ac{TSPE}, \ac{NCC} and \ac{NCCCI} were computed for 1000 spike trains (30 minutes duration) in less than 2 minutes, while Transfer Entropy based methods needed more than 45 minutes.
Nevertheless, also the Transfer Entropy based methods take less than 3 minutes for 30 minutes of 60 channel recordings (standard \ac{MEA} chip), which is considered acceptable for this application. 
The \textit{in silico} model for evaluation was sampled at 1\,kHz. It was also simulated for 10\,kHz (data not shown) but no difference in accuracy for the evaluated algorithms with the unbinned or binary binned (1\,ms bin size) spike trains were found. The binning for preprocessing is recommended because of the linearly increasing computing time for evaluated connectivity estimation algorithms with smaller bin sizes. 
The gradient of computing time is smaller for cross correlation based connectivity estimation algorithms than for the \ac{TE} based algorithms. Thus, long term experiments will benefit by applying \ac{TSPE} or \ac{NCC}. 

The activity dependent plasticity of connectivity should be considered for long term recordings of \textit{in vitro} or \textit{in vivo} neuronal networks. Since a long duration of recording improves the performance of connectivity estimation algorithms, a compromise between recording time and plasticity of connectivity has to be found. 
Our research will be continued with the development of innovative threshold selection methods for connectivity estimation and the application of \ac{SPE} and \ac{TSPE} for spike trains of \textit{in vitro} experiments. The ability of \ac{TSPE} to distinguish between excitatory and inhibitory effects could improve the meaningfulness of these experiments.

The determination of an accurate threshold was evaluated as well. This is a crucial step of connectivity estimation because it is not possible to select a threshold at 1\% \ac{FPR} for real spike trains without knowing the connectivity of biological \textit{in vitro} network, which is generally true for all reviewed algorithms. For \ac{TSPE} the easy threshold calculation with 3 times \ac{SD} obtain good results with small \acp{FPR}. Using some surrogate based methods can improve the \acp{TPR} at still small \acp{FPR}. However, these methods require a multiplication of the computation time for the connectivity estimation.

To summarize, the evaluation results show that the accuracy of connectivity estimation of large scale neuronal networks has been enhanced by the novel algorithm \ac{TSPE}. This advantage combined with the ability to distinguish between excitatory and inhibitory effects will help to improve the accuracy of future experiments. The used simulation framework for large scale neural networks with different topologies is available as well as the \textit{MATLAB} based \ac{TSPE} toolbox, which has the potential of parallelization, on the attached DVD. The \textit{MATLAB} code of \ac{TSPE} is also attached to this thesis (see Appendix~\ref{A:TSPE}).

\chapter{Graph theory}
The resulting network of connectivity estimation methods (see Chapter~\ref{cha:alg}) is a structure of correlated neurons, a so-called graph of neurons. Since graphs can be very complex depending on the analysed network, specific analysis methods were developed. These methods are not only used for the analysis of neuronal networks, but also for many graphs such as social networks or power grids. The study of graphs is called graph theory.

To compare experimental network properties, the graph obtained is mathematical analysed. In this way quantitative metrics are gained~\cite{Poli.2015}. The first introduced graph theory parameter is the \ac{MPL}, which is defined as the average distance between nodes in the whole network. A low \ac{MPL} means a high density of connections in the network. In that case quick communication between neurons is more likely. For the unweighted graphs each given connection has a distance of one. The distance between two nodes $i$ and $j$ (here neurons) is given by the function $d{(i,j)}$, which counts how many nodes must be passed through to reach the target node. The \ac{MPL} is calculated by the knowledge about all distances,
\begin{equation}\label{equ:MPL}
MPL =  \frac{2}{N \cdot (N-1)} \sum_{i \neq j}  d{(i,j)}.
\end{equation}
In graph theory, a clique is a subset of the graph in which all nodes are connected to each other, also called a full graph.
The local clustering coefficient for undirected graphs is a parameter quantifying the closeness of neighbour nodes to a clique. This coefficient is calculated by
\begin{equation}\label{equ:NC}
{C^{ws}} =  \frac{2 E }{k \cdot (k -1)}.
\end{equation}
Where $E$ is the number of edges between neighbours and $k$ is the number of connections.
To quantify the network communication of the neuronal network, the local clustering coefficient and the \ac{MPL} are calculated for a \textit{random network} with the same number of nodes and connections like the observed network: $MPL_{rand}$ and $C^{ws}_{rand}$. The generated \textit{random network} is then compared with the observed network in terms of graph theory.
The so-called small-world-ness~\cite{Humphries.2008} is expressed by
\begin{equation}\label{equ:SWN}
{S^{ws}} =  \frac{\gamma^{ws}_g}{\lambda_g}=\frac{C^{ws}_g / C^{ws}_{rand}  }{ MPL_g / MPL_{rand}}.
\end{equation}
While a $S^{ws}$ below 1 means that the observed network has \textit{random network} character, a network is identified as a \textit{small-world network} if $S^{ws}$ is larger than one. In this way a quantitative metric can be used to characterize a network or a network change.

Since the focus of this work is on connectivity estimation, no further parameters are explained in detail.
However, subsequent research work could include additional parameters like efficiency, hubs, centrality or robustness~\cite{Bullmore.2009}.

\chapter{Application of connectivity estimation}
Since the evaluation results of \ac{TSPE} are promising, the connectivity estimation method is applied to \textit{in vitro} experiments. Here, measured signals of neuronal cultures were obtained with \ac{MEA} chips. Due to setup difficulties no \ac{HDMEA} experiments could be performed. The work described in this chapter is based on a collaboration between \ac{UCI} and \ac{UAS} Aschaffenburg. This collaboration was founded by the \ac{BaCaTeC} project \textit{Estimation of Effective Connectivity in Neuronal Networks}.

\section{Setup of the experiment}
\begin{figure}[h]
	\centering
	\includegraphics[width=0.45\textwidth]{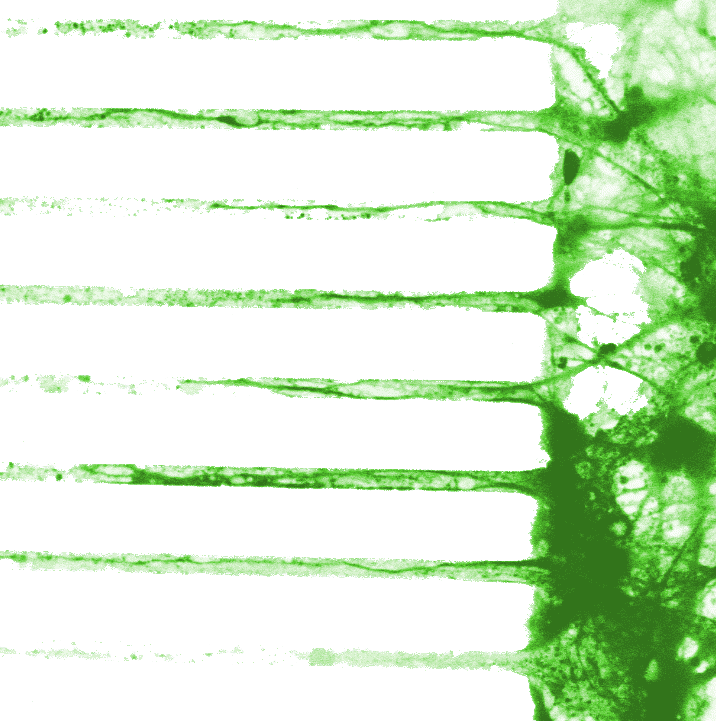}
	\caption[Neurites in a micro-tunnel device]{\textbf{Neurites in a micro-tunnel device:} The micro-tunnel device prevents that neuronal cells of the chamber (right-hand side) enter the area on the left-hand side. Only axonal connections and dendrites are able to pass the tunnels. Origin picture taken by Udit Narula in 2016 at \ac{UCI}.}\label{fig:tunnel}
\end{figure}

\begin{figure}[H]
	\centering
	\includegraphics[width=0.72\textwidth]{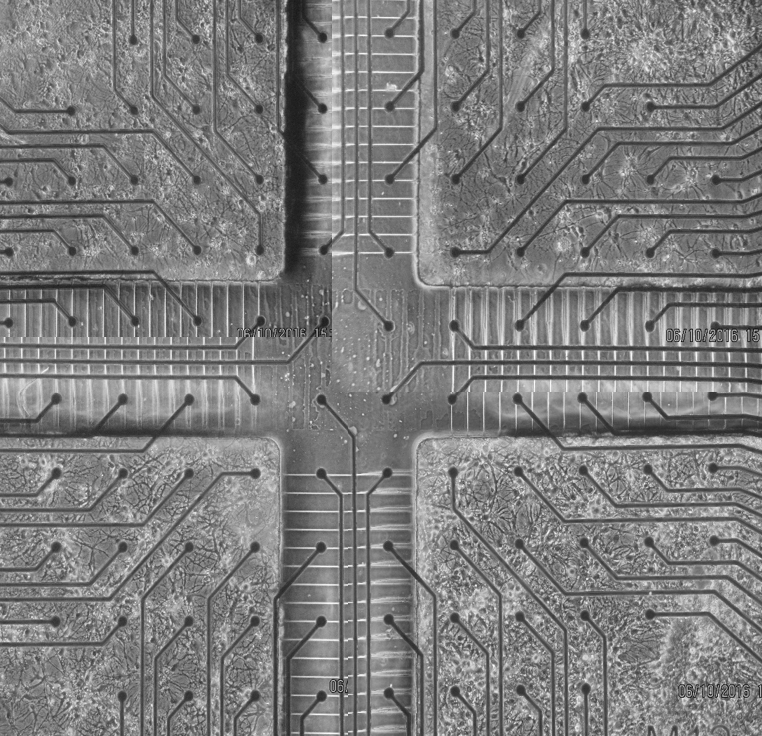}
	\caption[Experiment with four cultures]{\textbf{Experiment with four cultures:} The micro-tunnel device connects four subcultures \ac{EC}, \ac{DG}, \ac{CA3} and \ac{CA1} with each other. Black dots are electrodes and lines are conducting paths. For some micro-tunnels a pair of electrodes (all in all 120 electrodes) measured signals of axons in the tunnels (bright lines). Four electrodes in the middle of the \ac{MEA} are covered and isolated by the tunnel device. Neuronal cultures can be recognized in the chambers. Picture taken by Daniele Poli at \ac{UCI}.}\label{fig:setup}
\end{figure}

To analyse neuronal networks of the hippocampus and associated brain regions involved in the learning and memory of an 120-channel \ac{MEA} chip of \ac{MCS} (Reutlingen, Germany, \url{http://www.multichannelsystems.com}) is used. \ac{EC}, \ac{DG}, \ac{CA3} and \ac{CA1} sub-regions of the brain were cultured in a four-chamber system interconnected with micro-tunnels~\cite{Pan.2015}. The micro-tunnels enabled passages of neurites but not somata and communication between sub-regions~\cite{Pan.2015,DeMarse.2016} (see Figure~\ref{fig:tunnel}). 
To obtain these chambers a \ac{PDMS} device was used (see Figure~\ref{fig:setup}).

The hippocampal cells were cultured at densities equal to 1000 cells/mm\textsuperscript{2} for \ac{DG}, 330 cells/mm\textsuperscript{2} for \ac{CA3}, 410 cells/mm\textsuperscript{2} for \ac{CA1} and 330 cells/mm\textsuperscript{2} for \ac{EC}~\cite{Poli.2018}. In this way a realistic ratios of neuronal densities \textit{in vivo} (\ac{EC}-\ac{DG} 1:3, \ac{DG}-\ac{CA3} 3:1, \ac{CA3}-\ac{CA1} 1:1.25 and \ac{CA1}-\ac{EC} 1.25:1) is ensured~\cite{Braitenberg.1981}.
For further information about the used devices and their fabrication see~\cite{Pan.2011} and for information about the dissection process of the rat brain see~\cite{Brewer.2013}. The experiments were carried out at the \ac{MIND} of \ac{UCI} in 2015 and 2016. All datasets were recorded between 21 and 37 \ac{DIV}. For some experiments the order of cultures were changed. \ac{CW} means the order \ac{EC}, \ac{DG}, \ac{CA3}, \ac{CA1} (starting from the left upper corner clockwise) and \ac{CCW} means switched chambers \ac{CA1} and \ac{DG}. For the experiments each chamber was stimulated at three electrodes with different stimulation protocol:
\begin{itemize}
\item Single stimulation
\item \ac{PP} stimulation
\item $\theta$ burst stimulation~\cite{Bouteiller.2010}
\item 5\,Hz $\theta$ burst stimulation
\item \ac{HF} stimulation
\end{itemize}
For three experiments, \acp{LR} with a duration between 50 and 60 minutes were saved.
Due to the complex datasets, it was previously not possible to estimate the connectivity with significant results. In Table~\ref{tab:experiments} indicates which stimulation was applied to which experiment.

\begin{table}[h]
	\caption[Description of experiment datasets]{\textbf{Description of experiment datasets:}
	The first six digits of the experiment names is the used chip. The next six digits indicate the start date of the cultivation process and the last six digits indicate the date of the experiment with measurements.
	} \label{tab:experiments}	\vspace{-0.0cm}
	\begin{center}	
		\begin{tabular}{ c | c | c | c | c | c | c | c | c | c | c} 
			\hline
			Nr. & Experiment & \ac{DIV} & Order & Spon. & Single & \ac{PP} & $\theta$ & $5 \theta$ & \ac{HF} & \ac{HF} (\ac{LR}) \\
			\hline
			1 & 19908 160518 160610 & 22 & \ac{CCW} & \checkmark & \checkmark & \checkmark & \checkmark & \checkmark & \checkmark &\\
			2 & 19914 160127 160217 & 21 & \ac{CW}  & \checkmark & \checkmark & \checkmark & \checkmark & 		  & \checkmark &\\
			3 & 24574 160727 160818 & 22 & \ac{CW}  & \checkmark & \checkmark & \checkmark & \checkmark & \checkmark & \checkmark &\\
			4 & 19914 160127 160303 & 37 & \ac{CW}  & \checkmark &         &         &         &         &  \checkmark& \checkmark\\
			5 & 24088 160127 160302 & 36 & \ac{CW}  & \checkmark &         &         &         &         &  \checkmark& \checkmark\\
			6 & 19908 150729 150823 & 25 & \ac{CCW} & \checkmark & \checkmark & \checkmark & \checkmark &  &         & \\
			7 & 19914 150805 150828 & 25 & \ac{CCW} & \checkmark & \checkmark & \checkmark & \checkmark &  &         &\\
			8 & 24574 160127 160303 & 37 & \ac{CW}  & \checkmark &         &         &         &         & \checkmark & \checkmark\\
			\hline
		\end{tabular}
	\end{center}
\end{table}

\vspace{-0.5cm}

\section{Spike detection and sorting}
\begin{figure}[H]
	\centering
	\includegraphics[width=0.98\textwidth]{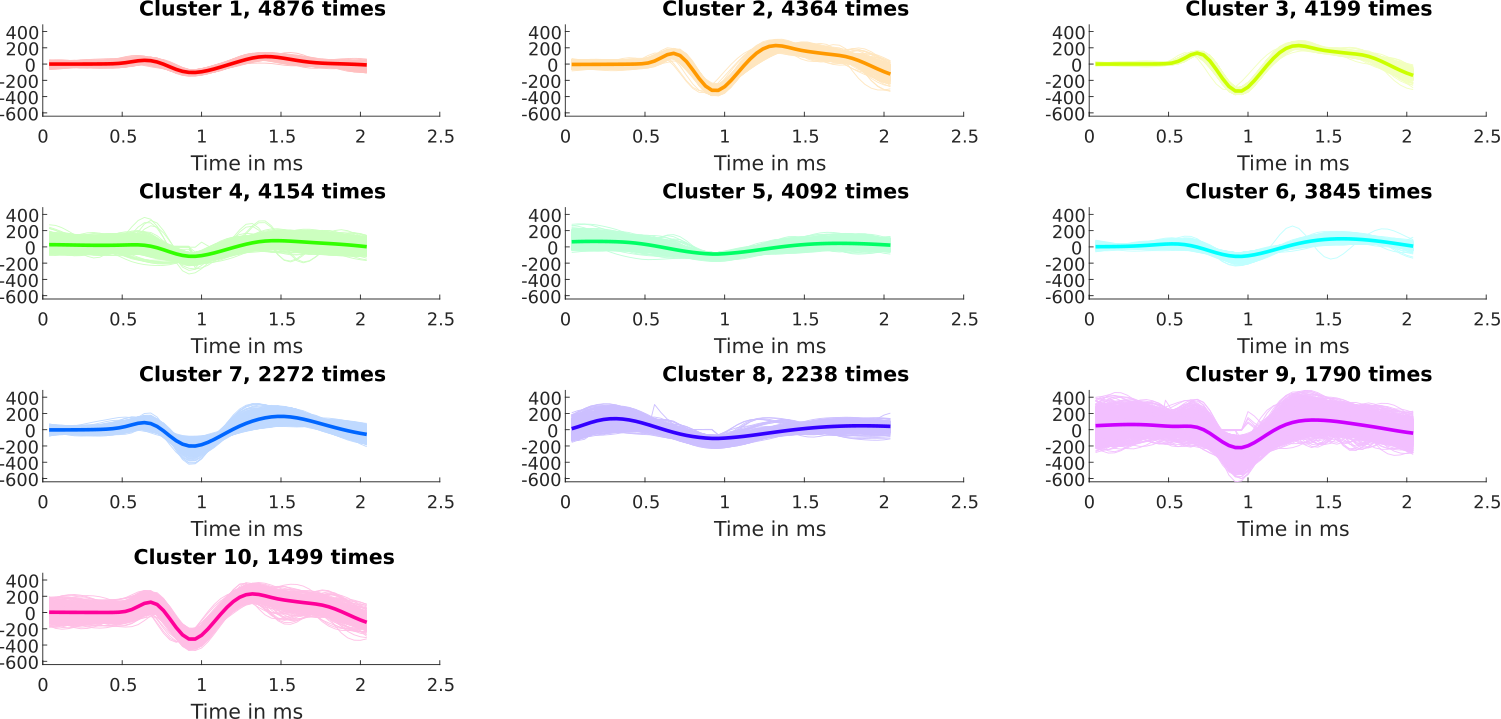} 
	\caption[Example of spike sorting]{\textbf{Example of spike sorting:} The measured signal in a micro-tunnel was filtered, spikes were detected and sorted. Ten different source neurons of these action potentials were identified. 4876 spikes were associated with cluster 1, which is the largest source of action potentials. The ordinates have normalized value ranges. }\label{fig:spikesorting}
\end{figure}

After filtering the measured raw data with a 500\,Hz Butterworth high pass filter, spikes were detected at a threshold 5.5 times the estimated \ac{SD} of the background noise for chamber electrodes. Since the \ac{SNR} for signals measured in the micro-tunnels is larger, here only 4.0 times the estimated \ac{SD} were necessary. To distinguish between different neurons at a single electrode the spike sorting toolbox \textit{FFMSpikeSorter}~\cite{Carlson.2014} was used with a 2\,ms window size for the waveforms of detected spikes. Each cluster had to have at least 20 action potentials in order to be considered as identified neuron for postprocessing. In Figure~\ref{fig:spikesorting} the results of spike sorting for one electrode in a tunnel is illustrated. 

To compare the network dynamics the spontaneous datasets before and after stimulations of an experiment were merged. After applying spike sorting the results were separated again. 

\begin{figure}[htbp]
	\centering
	\includegraphics[width=1\textwidth]{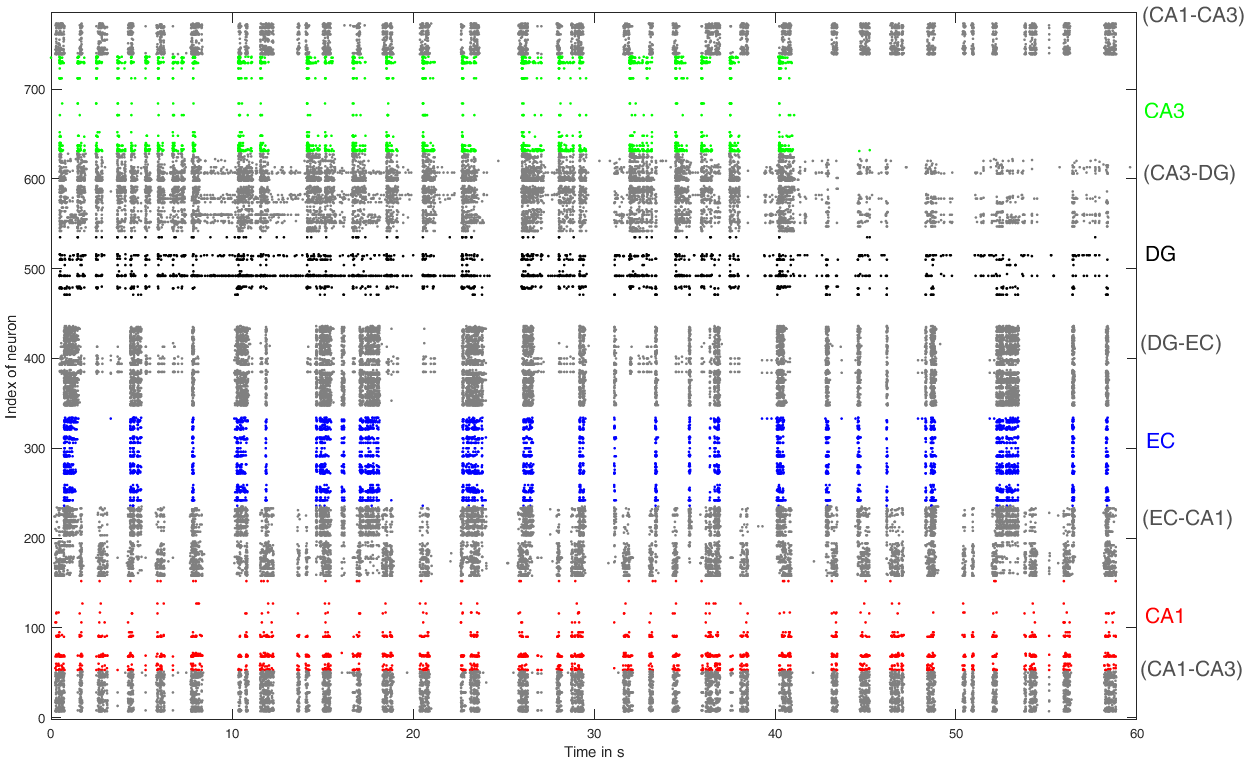}
	\caption[Example of measured spike trains (1 min)]{\textbf{Example of measured spike trains (1 min):} Over 700 neurons were identified by the spike sorting algorithm. Subcultures and tunnels are illustrated in different colors (see right hand side).}\label{fig:spiketrains}
\end{figure}

The obtained raster-plot in Figure~\ref{fig:spiketrains} illustrates different network behaviour in the chambers (each chamber marked with different colour). For example, chamber \ac{CA3} has a large abstinence of spikes in the last 18 seconds of the sample while all other chambers and the tunnels are still very active. The used spike detection and sorting identified over 700 neurons with the 120 electrode \ac{MEA}. Especially in the tunnels many clusters were found because of the large \ac{SNR}.

Since some clusters seem to have similar waveforms (cluster 2 and 3 in Figure~\ref{fig:spikesorting}), the spike sporting algorithm could be too aggressive. As the focus of this work is on the estimation of connectivity, this topic has not been further investigated. For further research different spike sorting algorithms should be evaluated in order to gain the best accuracy of connectivity estimation.

\section{Connectivity estimation with TSPE}
\begin{figure}[htbp]
	\centering
	\includegraphics[width=1\textwidth]{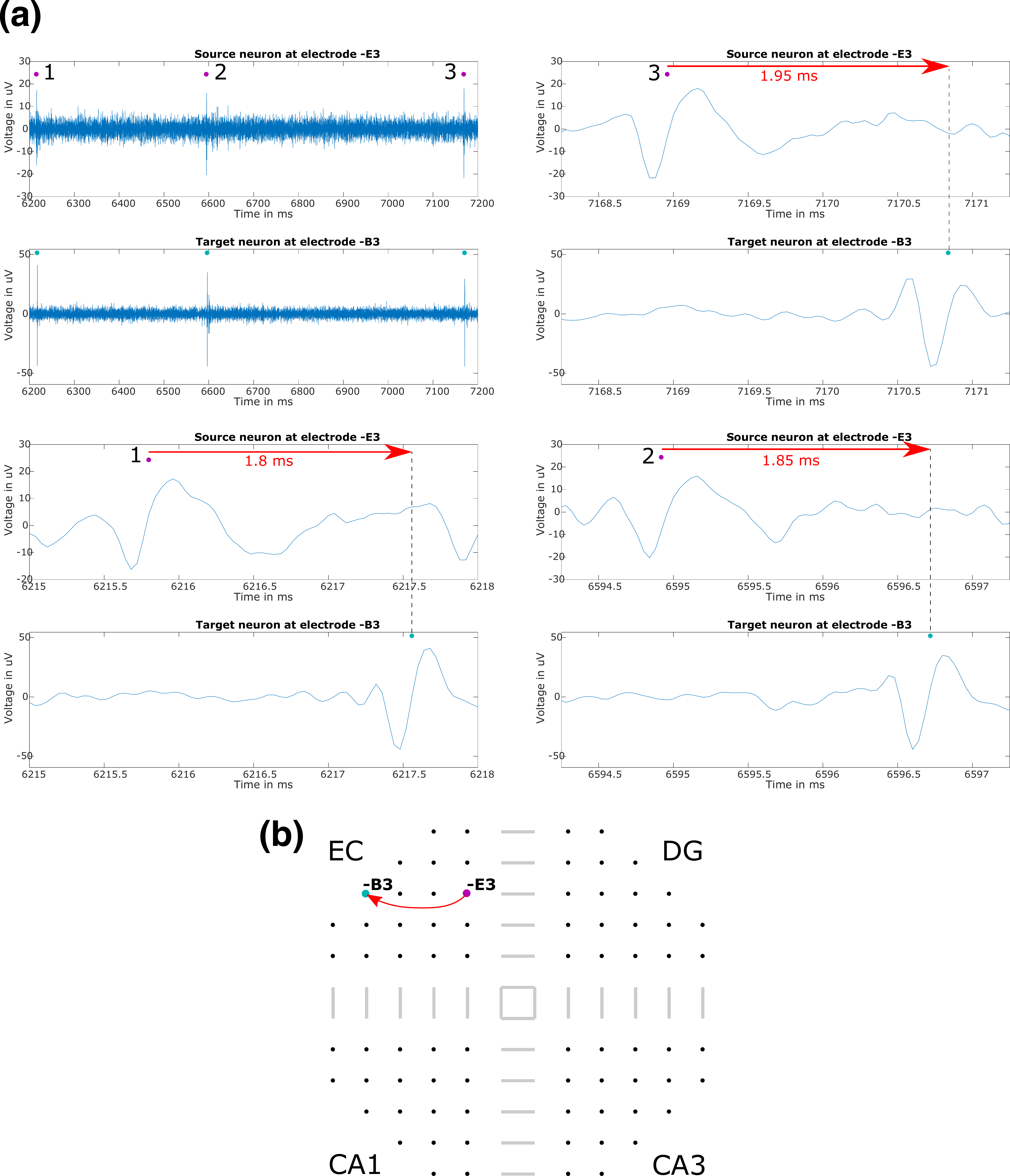}
	\caption[Example of an estimated connection]{\textbf{Example of an estimated connection:} The measured signals (a) of electrodes -E3 and -B3  is zoomed for the three action potential pairs. The action potentials of -E3 are marked with a purple dot and a green dot for -B3. After each action potential of -E3 it takes approx. 1.9\,ms that an action potential at -B3 can be measured. This indicates an excitatory effect, which is illustrated at the \ac{MEA} (b) with an red arrow.}\label{fig:sampleconnectivityesti}
\end{figure}

The novel connectivity estimation method \ac{TSPE} was applied to 5 minute recordings before and after the different stimulation types. For the operation parameters of \ac{TSPE} the delay window was selected to be 30\,ms, $a=[3,4,5,6,7,8]$, $b=[2,3,4,5,6]$ and $c=[0]$ (all in ms). 
Optional scaling disabled because of different network bursts in each chamber. Estimated interchamber connections were only allowed between neighboured chambers. Furthermore, the four electrodes in the middle of the \ac{MEA} were not considered for the estimation. Thresholds were calculated with 100 surrogate data generated with a 2\,ms jittering window (threshold for inhibitory effects is the mean value of all negative values minus their \ac{SD} and the mean value of all positive values plus their \ac{SD} for excitatory effects). Each result value that passes these threshold calculation successfully is identified as connection. In Figure~\ref{fig:sampleconnectivityesti} an example is illustrated. A measured neuron at electrode -E3 excites another measured neuron at electrode -B3 with a transmission time of approx. 1.9\,ms.

\section{Graph theory analysis}
\begin{figure}[htbp]
	\centering
	\includegraphics[width=0.88\textwidth]{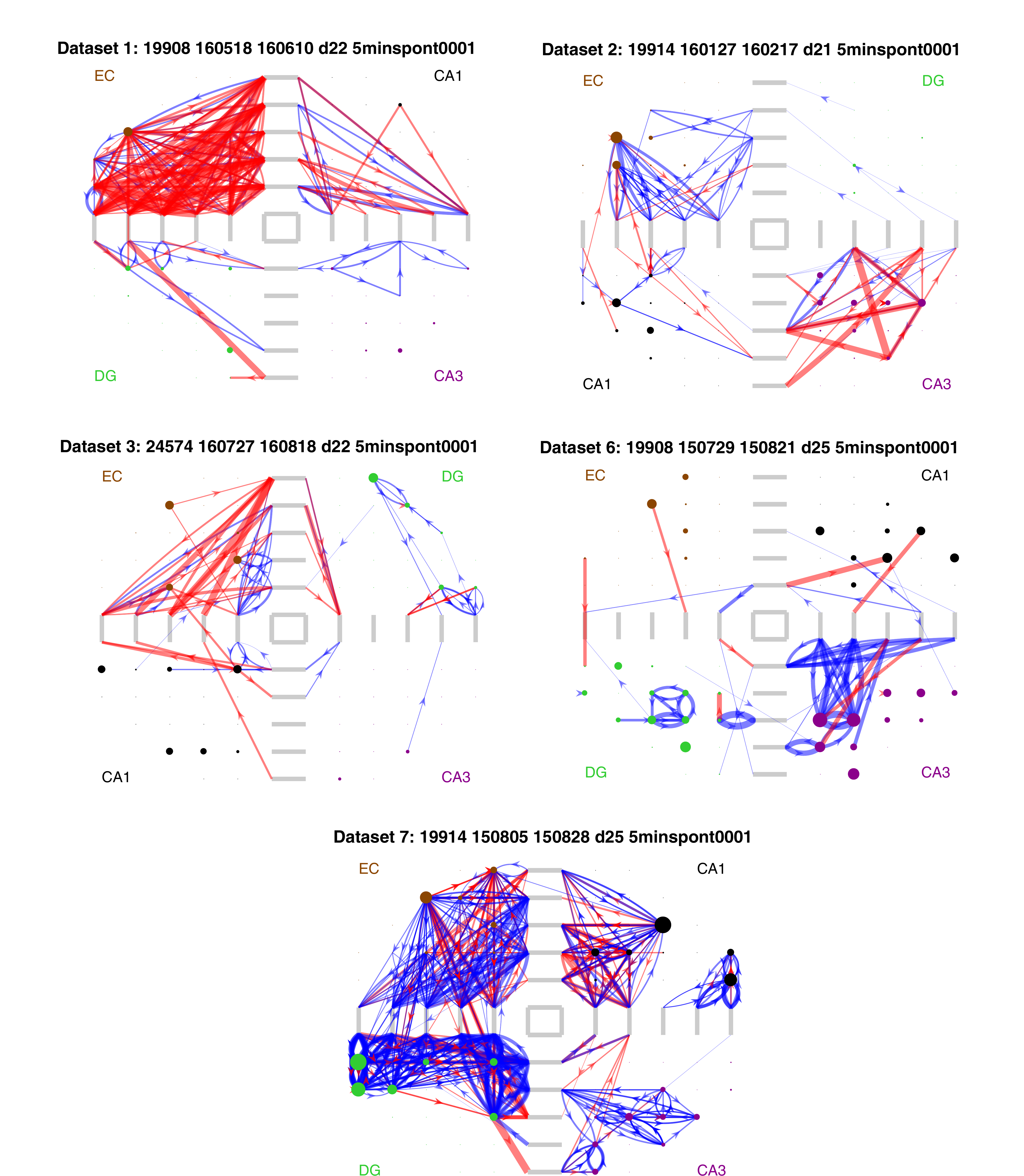}
	\caption[Results of estimated graphs]{\textbf{Results of estimated graphs:} The \ac{MEA} is illustrated with different sizes of electrodes depending on the activity (large means more active). The width of a line indicates the strength of estimated correlation and the causality specified by arrowheads. Inhibitory effects are marked blue and excitatory effects red.  }\label{fig:connectivityestiplots}
\end{figure}

For each result, the estimated connectivity was analysed in terms of graph theory. The \ac{MPL} was calculated as well as the small-world-ness. A toolbox was used for the calculation of latter (\url{https://github.com/mdhumphries/SmallWorldNess}).
Figure~\ref{fig:connectivityestiplots} illustrates differences in the estimated connectivity between the experiments. Inhibitory effects are marked blue and excitatory effects red. Unfortunately, the differences in activity between the experiments made it impossible to use statistical groupings. Except for dataset 7 (bottom in Figure~\ref{fig:connectivityestiplots}) all experiments have at least one very densely connected chamber.

\newpage
\section{Network dynamics}
As the analysis was not successful for the whole graph, individual identified connections were grouped and examined individually for each chamber. The observation of status changes of a network in time is called studying the network dynamics. 

In order to gain more detailed knowledge about network dynamics, synaptic effects are divided into four groups: strong/weak inhibitory and strong/weak excitatory. Strong effects were defined through the 75\% quantile of all appearing effects.
The network activity changes due to the applied stimulus. Each group of effects is considered separately with percentage changes. There are five change possibilities:
\begin{itemize}
\item Gained effect -- during the two recordings to be compared, this effect appeared new
\item Stronger effect -- during the two recordings to be compared, this effect became more powerful than before
\item Same effect -- during the two recordings to be compared, this effect did not change
\item Weaker effect -- during the two recordings to be compared, this effect became less powerful than before
\item Lost effect -- during the two recordings to be compared, this effect disappeared
\end{itemize}
To identify significant differences the \ac{K-S test} was used ($\alpha=0.05$). The null hypothesis is always that stimuli have no influence on the behaviour of effective connectivity in a neuronal network.

\subsection{Comparison of different stimuli}
\begin{figure}[htbp]
	\centering
	\includegraphics[width=0.78\textwidth]{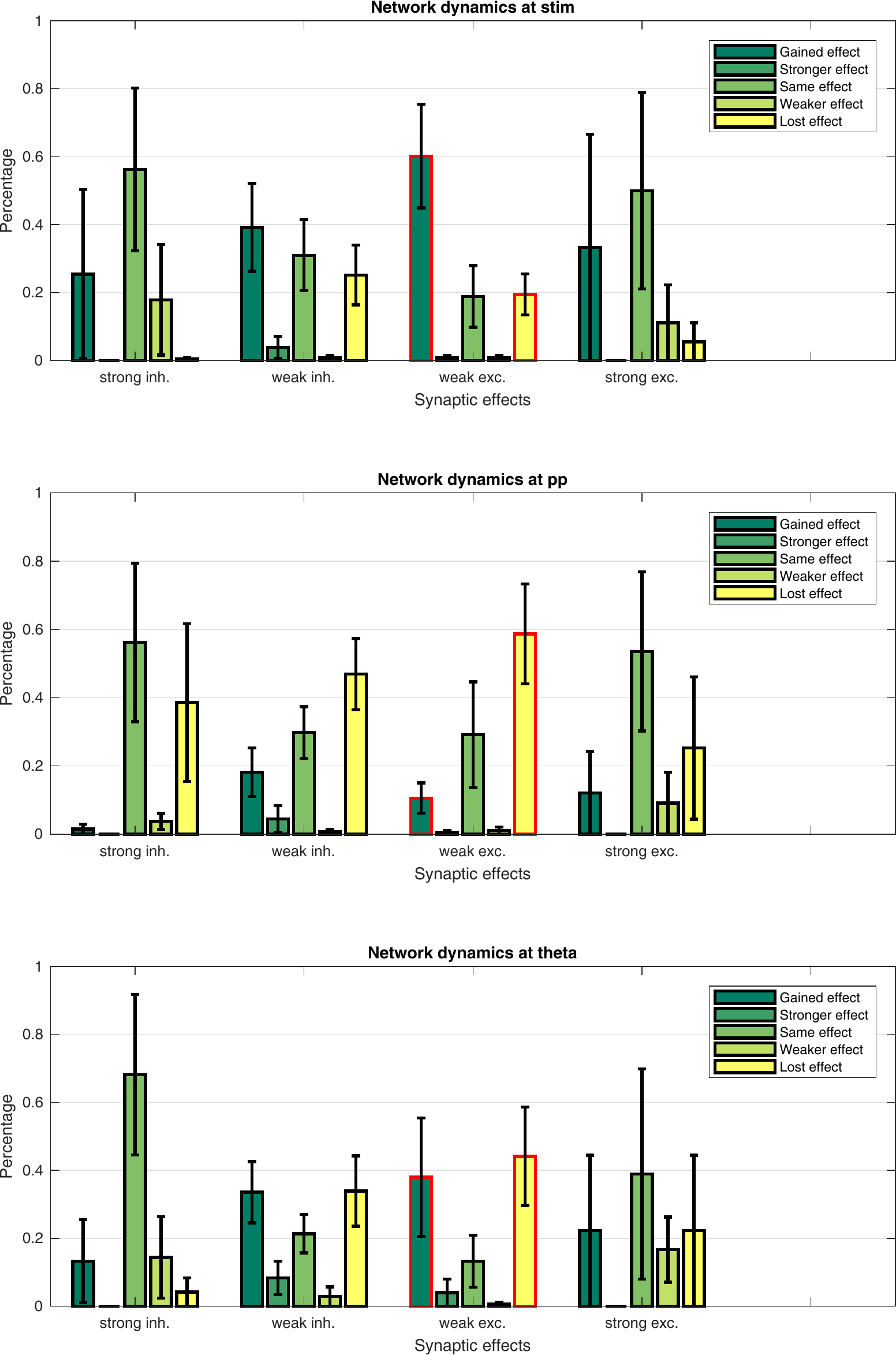}
	\caption[Network dynamics for different stimuli at DG]{\textbf{Network dynamics for different stimuli at \ac{DG}:} The representation with standard error bars is based on five experiments. Synaptic effects are divided into four groups: strong/weak inhibitory and strong/weak excitatory. The network activity changes due to the applied stimulus. Each group of effects is considered separately with percentage changes. There are five change possibilities for each appeared effect. Significant differences are marked in red.}\label{fig:DG_short}
\end{figure}

This analysis focuses on influences of various stimuli (single stim, \ac{PP} and $\theta$ burst). For the analysis, $n=5$ (dataset 1, 2, 3, 6 and 7). Because of this low number, it is hard to disprove the null hypothesis that there is no correlation of stimuli on connectivity. However, in \ac{DG} it was possible to successfully prove an influence on the behaviour (see Figure~\ref{fig:DG_short}, cases marked red). The stimulation with \ac{PP} reduced the generation of weak excitatory effects and simultaneously enhanced the lost of weak excitatory effects in contrast to single and $\theta$ burst stimulation.

\subsection{Long recording based analysis} \label{sec:longrec}
To increase the statistical power one considers longer recordings. The one hour recordings (dataset 4, 5 and 8) were divided into five minutes pieces to obtain a larger number of connectivity states. No overlap was used. In this way the network dynamics without influences can be studied and compared with the dynamics after a simulation protocol. With the three \ac{LR} experiments it is possible to gain $n_\mathrm {before}=31$, $n_\mathrm {stim}=3$ and $n_\mathrm {after}=31$. In the following the observations are listed and sorted by chamber.

\subsubsection*{Dynamics in DG}
In \ac{DG}, strong synaptic effects rarely appeared. However, after the \ac{HF} stimulation the strong and weak excitatory effects were less stable than before. The effective connectivity of excitatory synapses was then more dynamic.
Shortly after stimulation, less weak excitatory synaptic effects appeared new then before stimulation or later.

\begin{figure}[H]
	\centering
	\includegraphics[width=0.78\textwidth]{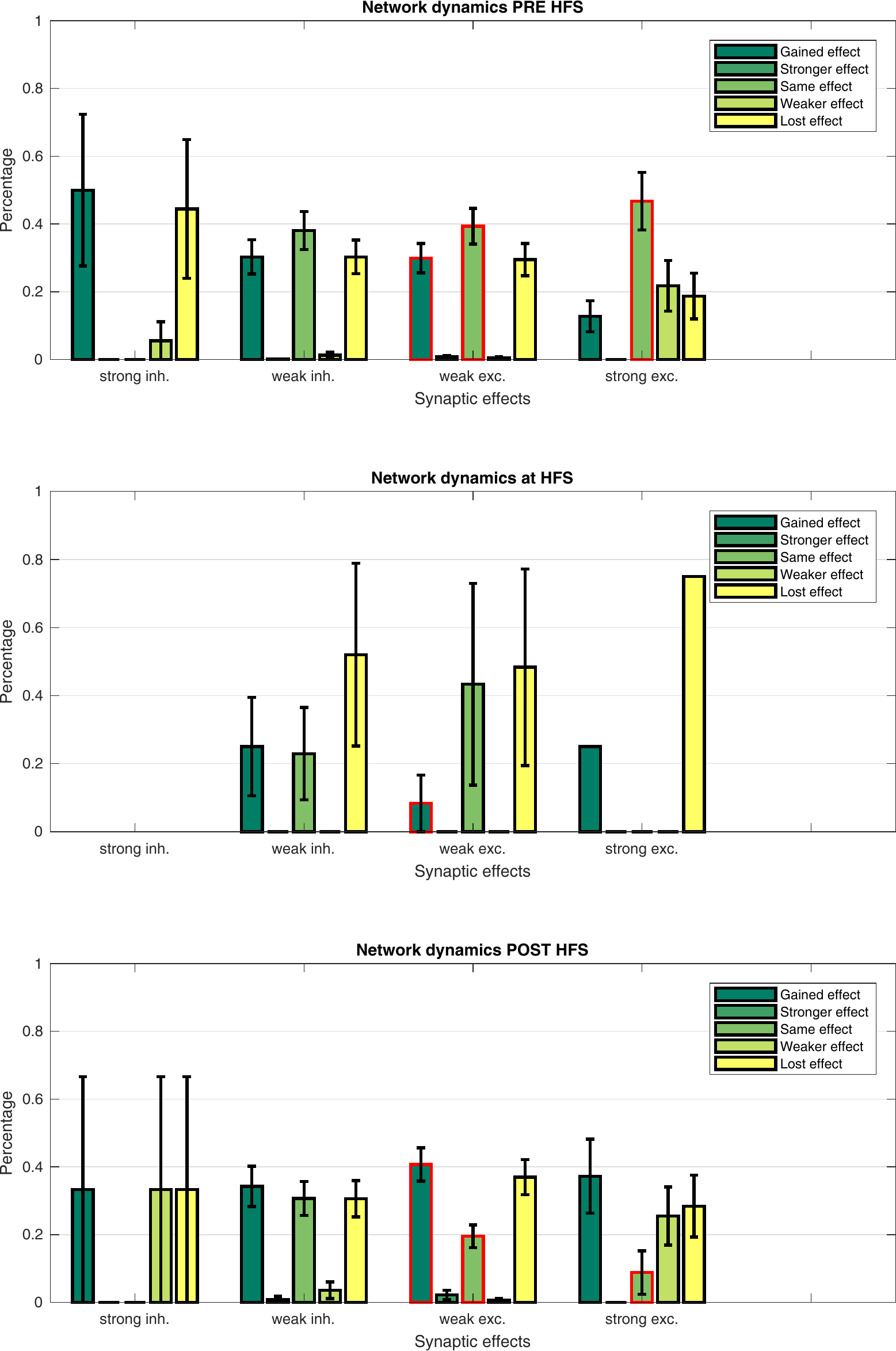}
	\caption[Network dynamics for HF stimulation at DG with long recordings]{\textbf{Network dynamics for \ac{HF} stimulation at \ac{DG} with long recordings:} The representation with standard error bars is based on three experiments. Significant differences are marked in red.}\label{fig:DG_LT}
\end{figure}

\newpage
\subsubsection*{Dynamics in CA1}
In \ac{CA1}, more weak inhibitory effects disappeared with \ac{HF} stimulation than before and after stimulation. It is also interesting that the dynamics before and after the stimulation are very similar. If the neural network is not externally influenced, the behaviour of dynamics is predictable.

\begin{figure}[H]
	\centering
	\includegraphics[width=0.78\textwidth]{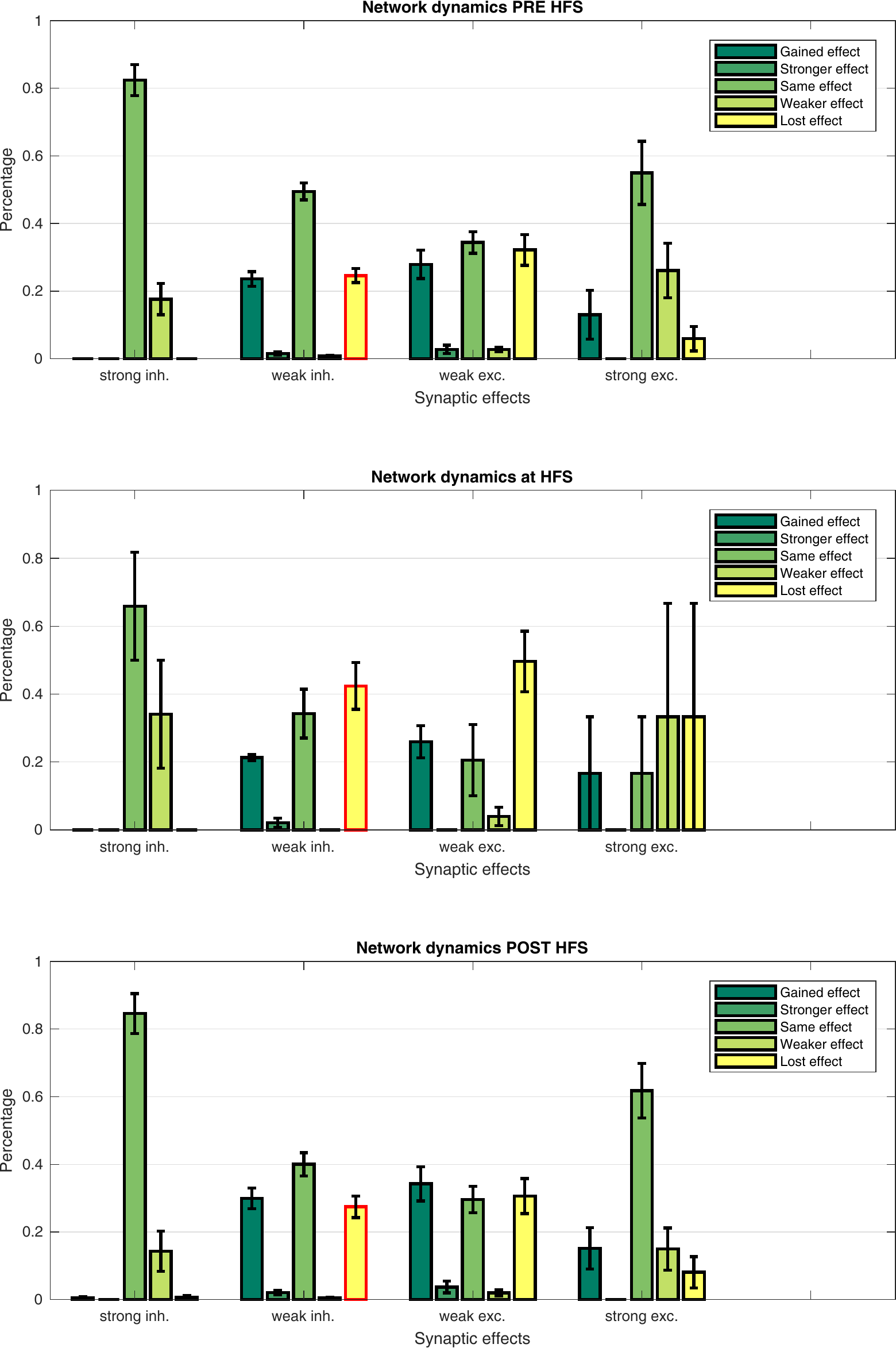}
	\caption[Network dynamics for HF stimulation at CA1 with long recordings]{\textbf{Network dynamics for \ac{HF} stimulation at \ac{CA1} with long recordings:} The representation with standard error bars is based on three experiments. Significant differences are marked in red.}\label{fig:CA1_LT}
\end{figure}

\newpage
\subsubsection*{Dynamics in CA3}
In \ac{CA3}, strong synaptic effects rarely appeared. For weak synaptic effects, \ac{CA3} generally is more flexible than \ac{EC}, \ac{DG} and \ac{CA1}. The effective connectivity became even more unstable thru the \ac{HF} stimulation.

\begin{figure}[H]
	\centering
	\includegraphics[width=0.78\textwidth]{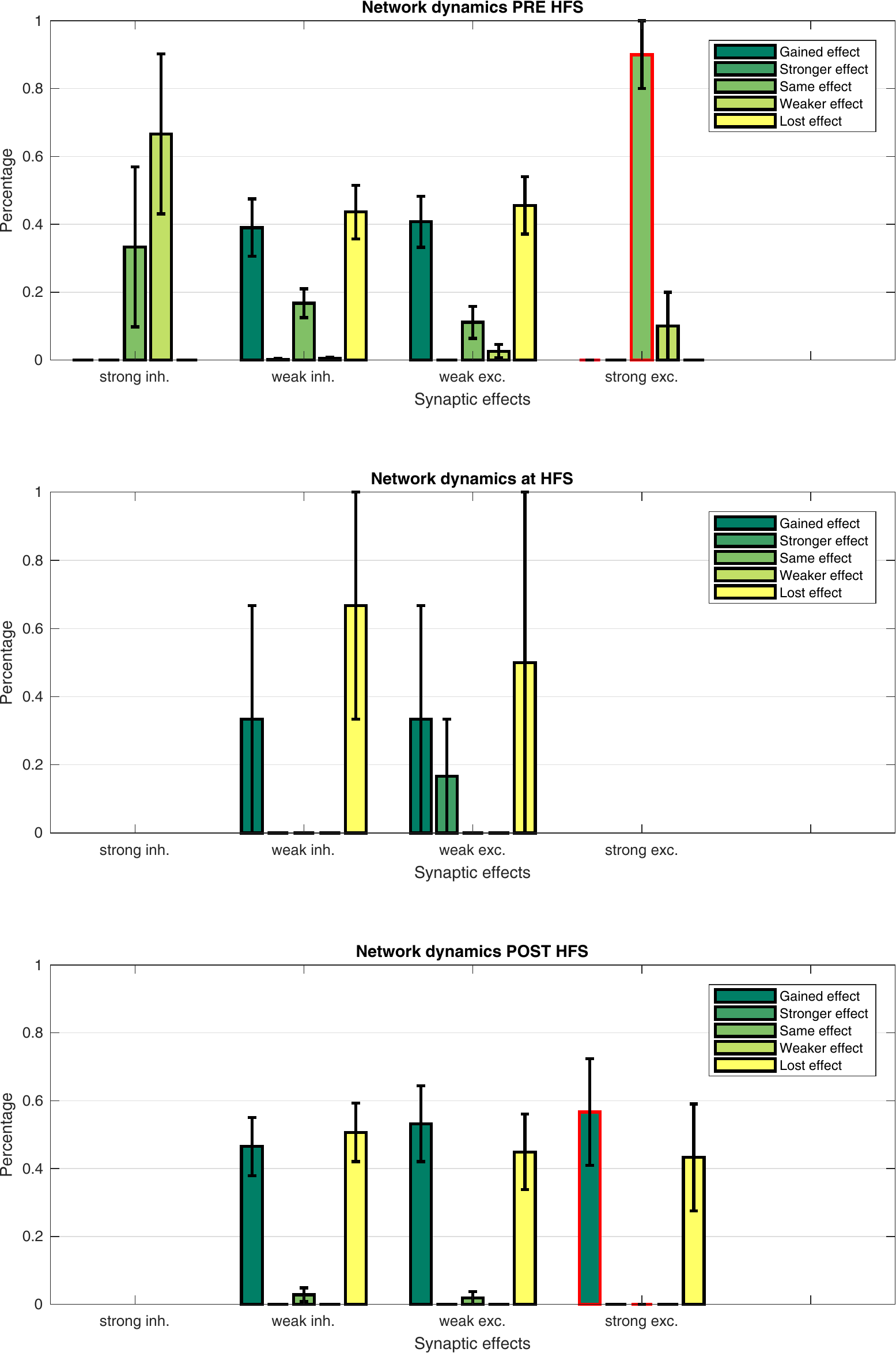}
	\caption[Network dynamics for HF stimulation at CA3 with long recordings]{\textbf{Network dynamics for \ac{HF} stimulation at \ac{CA3} with long recordings:} The representation with standard error bars is based on three experiments. Significant differences are marked in red.}\label{fig:CA3_LT}
\end{figure}

\newpage
\subsubsection*{Dynamics in EC} 
In \ac{EC}, strong inhibitory effects were rather lost during \ac{HF} stimulation.
However, after the stimulation weak inhibitory effects are more stable than before.

\begin{figure}[H]
	\centering
	\includegraphics[width=0.78\textwidth]{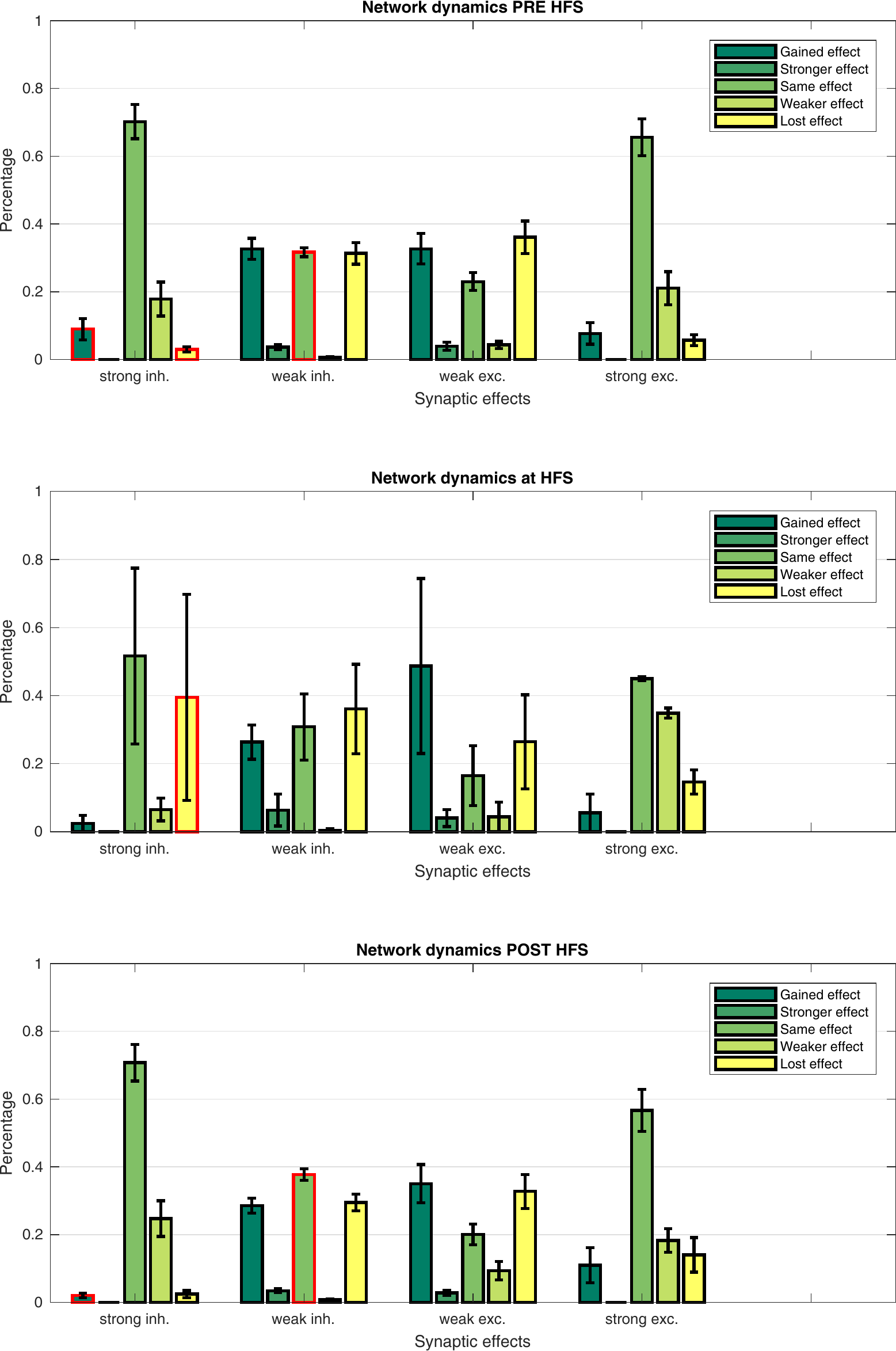}
	\caption[Network dynamics for HF stimulation at EC with long recordings]{\textbf{Network dynamics for \ac{HF} stimulation at \ac{EC} with long recordings:} The representation with standard error bars is based on three experiments. Significant differences are marked in red.}\label{fig:EC_LT}
\end{figure}

\subsection{Interchamber comparison}
To this point, the \acp{K-S test} have been applied to the comparison of states in each chamber. Another exciting aspect is the comparison between the chambers, the so-called interchamber comparison. Do the chambers generally behave differently? Does the stimulation influence the chambers in different ways?

To answer these questions, the \ac{K-S test} is applied to the same states of different chambers ($\alpha=0.05$). The analysis of recording datasets were used (for description see Section~\ref{sec:longrec}). In this way there were multiple significant statements obtained, which are listed below.

\subsubsection*{CA3}
Weak inhibitory and excitatory effects were less stable in \ac{CA3} than in all other chambers. This observation was not changed for weak inhibitory effects by the \ac{HF} stimulation. After 5 minutes also the weak excitatory effects were less stable again. 
In contrast, the strong excitatory effects were more stable in \ac{CA3} than in all other chambers before stimulation. 
This changed after \ac{HF} stimulation, since strong excitatory effects were less stable in \ac{CA3} than \ac{EC} and \ac{CA1}. 
The instability of weak inhibitory effects is justified by the amount of gained effects in \ac{CA3} (larger than in \ac{EC} and \ac{CA1}) before stimulation and the amount of lost effects in \ac{CA3} (larger than in \ac{EC} and \ac{CA1}) after stimulation. 
Furthermore, after the stimulation there were more often weak inhibitory and strong excitatory effects gained in \ac{CA3} than \ac{CA1} and \ac{EC}. Also weak excitatory effects were more often gained in \ac{CA3} (here in contrast to \ac{CA1} and \ac{DG}). 

\subsubsection*{CA3 and DG}
Weak inhibitory effects became less often stronger in \ac{CA3} and \ac{DG} than in \ac{EC} and \ac{CA1} before the stimulation.

\subsubsection*{DG}
Strong effects were less stable in \ac{DG} than in \ac{EC} and \ac{CA1} after stimulation.

\subsubsection*{EC}
Strong inhibitory effects were gained more often in \ac{EC} than in \ac{CA3} and \ac{CA1} before the stimulation. The stimulation affected the weak inhibitory effects, which became more often stronger in \ac{EC} than in all other chambers after stimulation. Furthermore, these effects were gained less often in \ac{EC} than in DG and \ac{CA3}.

\subsubsection*{EC and CA1}
Weak excitatory effects became more often weaker in \ac{EC} and \ac{CA1} than in \ac{DG} and \ac{CA3} before stimulation. This observation was not changed for \ac{EC} by the \ac{HF} stimulation. 

\subsubsection*{CA1}
Weak inhibitory effects were more stable in \ac{CA1} than in all other chambers before stimulation. The stimulation affected excitatory effects (weak and strong), which were more stable in \ac{CA1} than in \ac{DG} and \ac{CA3} after stimulation.

\newpage
\section{Discussion}
Within the scope of the research project of the biomems lab (\ac{UAS} Aschaffenburg) and the \ac{MIND} (\ac{UCI}) \ac{TSPE} was successfully used to estimate the connectivity of neuronal networks in complex experiments. These experiments were designed with a four-chamber system interconnected with micro-tunnels~\cite{Pan.2015} for different sub-regions of the brain.

The workflow creating the connectivity graphs from raw data measured by 120 channel \ac{MEA} chips from neuronal cells was explained and implemented. Eight experiments were analysed in terms of the influence on connectivity depending on stimulation with different protocols. Statistical methods were used to find and prove such influences.

Due to the complex experimental setup, it was not possible to use the estimated connectivity for statistically significant statements about graph theory. However, the changing of estimated connectivity with \ac{TSPE} proved different behaviour for the chambers. In \ac{DG} weak excitatory effects became unstable by the $\theta$ and \ac{HF} stimulation, while the single stimulation gained new effects. Furthermore, the interchamber comparisons leaded to many statements. These could improve the understanding of the differences in brain regions. The continuation of this project would include the biological interpretation of these results.

\chapter{Conclusion and outlook}
In this thesis the three types of neuronal network connectivity were presented as well as methods for estimating the functional connectivity according to the state of the art. Only algorithms with a higher potential of good performances are considered for a scenario measuring signals of a small subset of a large scale neuronal network, which is more realistic for most \textit{in vitro} and \textit{in vivo} applications. 
Furthermore, a novel estimation algorithm for effective connectivity was proposed called \textit{Total Spiking Probability Edges} (TSPE). The new algorithm is based on cross-correlation and detects correlation by edge filtering on different time scales of the cross-correlogram. 
Since the number of recorded neurons can be tremendous with the usage of new technologies like the \ac{HDMEA} chip, used algorithms have to be highly computationally effective.

A large framework of \textit{in silico} networks was designed to benchmark the performance of all selected approaches. As the topology of an \textit{in silico} network can affect the results and accuracy of algorithms~\cite{Kadirvelu.2017}, such a multi-topology evaluation is essential to evaluate the performance of connectivity estimation. Usage of at least one \textit{scale-free network} implementation is necessary for good, sufficient biological plausibility due to more realistic \ac{MFR}.
The findings of the modelling of this framework were presented at the \textit{POSTER 2018} on May 10, 2018 at Faculty of Electrical Engineering, \ac{CTU} Prague~\cite{DeBlasi.2018}. 
For future evaluations, a standardised method improves an effective research of neurocomputational algorithms. Widely used and uniform benchmarking makes it easier to compare newly developed methods with previous methods. Moreover, the further development and improvement of intergroup research is possible in a simpler way.

The novel method \ac{TSPE} is able to outperform the accuracy of all tested state of the art connectivity estimation algorithms (\ac{NCC}, \ac{NCCCI}, \ac{DHOTE}, \ac{DHOTECI}, \ac{DTE}, \ac{DTECI} and \ac{CDHOTE}) when applied to simulated neuronal network data with different topology complexity. 
In addition to improved accuracy, \ac{TSPE} is able to distinguish between  distinguish inhibitory and excitatory synaptic effects.
This ability is one of the current challenges of connectivity estimation methods~\cite{Pastore.2018,Liu.2017}. In this way \ac{TSPE} will help to understand neural communication. 
\ac{TSPE} and the evaluation results were published at the \textit{Journal of Neuroscience Methods}~\cite{DeBlasi.2019}.

Different approaches to select the threshold for the resulting \ac{CM} have been evaluated, which is the final step of the connectivity estimation. Depending on the topology and complexity of the network, different methods perform better. This makes it difficult to reach a general recommendation. The results of easy threshold calculation (absolute mean value + 2 \ac{SD}) and the surrogate threshold calculation (mean value +/- 4 \ac{SD}) were promising with reproducible results.

For the \ac{BaCaTeC} project \textit{Estimation of Effective Connectivity in Neuronal Networks}, \ac{TSPE} was applied to complex experiments to understand the communication between brain regions. This application example was also used to explain the processing of the raw signals step by step up to the connectivity graphs.
Graph theoretically based analyses were not successful due to quality differences between the cultures. However, by the connectivity estimation results of \ac{TSPE} different network dynamics were observed. This contributes to a better understanding of how learning and memory are staged operations.


Experimental setups, which include \ac{HDMEA} chips, enable measurements of thousands of neurons in a network~\cite{Berdondini.2009} and thus improved graph-theoretical calculations. Repeating the experimental work with \ac{HDMEA} chips could lead to further understanding with calculated \ac{MPL} and small-world-ness for each chamber. Since the number of spike train comparisons and the associated computational effort will increase exponential, the use of a high performance computer is recommended.

All tested connectivity estimation methods are based on analysis of spike train data. The collection of additional data (f.e. optical data) in combination with the spike train data can be used to improve the methods. Since effective and functional connectivity is a subset of structural connectivity, optical methods can be used to discard connections that are not visible~\cite{Ullo.2014}. 
Furthermore, there are approaches to combine \ac{ROC} based classifiers~\cite{Mwaffo.2017} that can further improve the connectivity estimation accuracy.

Experimental applications such as neurotoxicity measurements will benefit from connectivity estimation as a number of new parameters can be calculated to observe changes, e.g. \ac{MPL}, connection density or cost, hubs, centrality or robustness~\cite{Bullmore.2009}.

\begin{appendices}
	\addtocontents{toc}{\protect\setcounter{tocdepth}{1}}
	\makeatletter
	\addtocontents{toc}{%
		\begingroup
		\let\protect\l@chapter\protect\l@subsubsection 
		\let\protect\l@section\protect\l@subsection
	}
	\makeatother

\chapter{Hudgkin-Huxley model}\label{A:HH}
One of the most realistic mathematical descriptions of neurons is the \ac{HH} model, which was introduced by Hodgkin and Huxley in 1952~\cite{Hodgkin.1952} and honored with the Nobel Prize in 1963. They measured at a giant axon of a squid different influences of ion channels and pumps on the resulting membrane current $I$. In detail they took basically the ions sodium $Na^{+}$ and potassium $K^{+}$ into account. While these ions contribute to the membrane current directly in form of $I_{K}$ and $I_{Na}$ there are also other influences which are summarized to a leakage current $I_l$ (mainly consists of $Cl^{-}$ ions).  
\begin{equation}\label{equ:HH_curr}
I=I_l + I_{K} + I_{Na}
\end{equation} 
This current passes through the channels but there is also a capacitive current $I_C$. The unknown membrane capacity $C$ will then be charged or uncharged. This causes a membrane voltage change.
\begin{equation}\label{equ:HH_C}
C \cdot \dot{u} = - I + I_{C}
\end{equation} 
Mathematically so called gating variables $n$, $m$ and $h$ define the activation and inactivation of ion channels. Together with experimental defined conductances of ion channels $\hat{g}_{K}$ and $\hat{g}_{Na}$ influences are more precisely realized. 
\begin{equation}\label{equ:HH_IK}
I_{K} = \hat{g}_{K} \cdot n^{4} \cdot (u-u_{K})
\end{equation} 
\begin{equation}\label{equ:HH_INa}
I_{Na} = \hat{g}_{Na} \cdot m^{3} \cdot h \cdot (u-u_{Na})
\end{equation} 
Each of the gating variables obeys its own dynamics which are described by a difference equation.
\begin{equation}\label{equ:HH1}
\dot{n}  = \alpha_{n}(u) \cdot (1-n) - \beta_{n}(u) \cdot n
\end{equation}
\begin{equation}\label{equ:HH2}
\dot{m} = \alpha_{m}(u) \cdot (1-m) - \beta_{m}(u) \cdot m
\end{equation}
\begin{equation}\label{equ:HH3}
\dot{h}  = \alpha_{h}(u) \cdot (1-h) - \beta_{h}(u) \cdot h
\end{equation}
The here used voltage-dependent transition rates $\alpha_{n}(u)$, $\alpha_{m}(u)$, $\alpha_{h}(u)$, $\beta_{n}(u)$,
$\beta_{m}(u)$ and $\beta_{h}(u)$ are exponential functions of the membrane voltage $u$. Their definition was also part of the experimental work of Hudgkin and Huxley. This model is able to imitate and explain the function of a single neuron in many ways. Simulate a whole network of \ac{HH} models leads to a realistic behaviour but for each neuron there would be the need of multidimensional difference equations.
Especially taking the runtime of simulations into account the \ac{HH} model seems to be inflated with its four difference equations. For this simple reason many researchers tried to decrease dimensions in order to simulate larger networks. For example, a famous two-dimensional simplification of the \ac{HH} model is called FitzHugh-Nagumo model~\cite{Izhikevich.2006b}.
Since for this work only the spiking times are of interest, easier neuron models should work out and be less computationally intensive.
\newpage
\chapter{Integrate-and-Fire model}\label{A:IF}
The \ac{IF} model is one of the most easiest neuron models to understand the main function of a single neuron. Every receiving spike of a neuron integrates the membrane voltage with specific weights until a threshold is reached, than the receiving neuron will also fire, which means it will emit a spike. Someone could imagine a neuron as simple repeater of neuronal signals. Basically there are two operating principles of \ac{IF} models: \ac{COBA} and \ac{CUBA} impacts. One of the most famous \ac{IF} models was designed to study the propagation of signals~\cite{Vogels.2005}. Let us concentrate on their \ac{COBA} version. Excitatory and inhibitory neurons obeys the same dynamics with time  $\tau = 20$\,ms.
\begin{equation}\label{equ:CUBA_v}
\tau \cdot \dot{v}  = (V_{rest} - v)+g_{ex} \cdot (E_{ex}-V_{rest}) + g_{inh} \cdot (E_{inh}-V_{rest})
\end{equation}
It is also possible to make the equation \ref{equ:CUBA_v} easier by summarizing constant values and using a different factor for the effective synaptic conductances $g_{ex}$ and $g_{inh}$.
\begin{equation}\label{equ:CUBA_v_easy}
\tau \cdot \dot{v}  = (V_{rest} - v)+g_{ex} + g_{inh}
\end{equation}
There are also used dynamics for both conductances with different time constants, where $\tau_{ex}$ = 5\,ms and $\tau_{inh}$ = 10\,ms.
\begin{equation}\label{equ:CUBA_ge}
\tau_{ex} \cdot\dot{g}_{ex} = -g_{ex}
\end{equation}
\begin{equation}\label{equ:CUBA_gi}
\tau_{inh} \cdot\dot{g}_{inh}  = - g_{inh}
\end{equation}
Reversal potentials are chosen as $E_{ex}$ = 0\,mV and $E_{inh}$ = -80\,mV. After each spike the membrane potential of the firing neuron will be overwrite with these values. The original model does not use synaptic distances in form of delays. Thus, \ac{DM} was introduced, which delivers a specific delay in range of 1 to 20\,ms for each existing connection. Nevertheless, since network bursts can be measured at real data, non-existing of tonic spiking is a problem for realistic testing algorithms with this model.

\chapter{Comparison of neuronal models}\label{A:Izh_plot}
Izhikevich reviewed eleven widely used neuron models for 20 neuro-computational features and biological plausibility. He also evaluated the implementation cost in form of \acp{FLOP}. In Figure \ref{fig:Aihzplot} the original summary of his studies is shown. Simulating \ac{HH} models in a realistic number for networks is too computationally intensive. For thousands of simulated spiking neurons the \ac{IF} model is most efficient. In contrast, the Izhikevich model is also efficient but even able to exhibit most firing patterns~\cite{Izhikevich.2004}.

\begin{figure}[htbp]
	\begin{center}
		\includegraphics[width=0.8\textwidth]{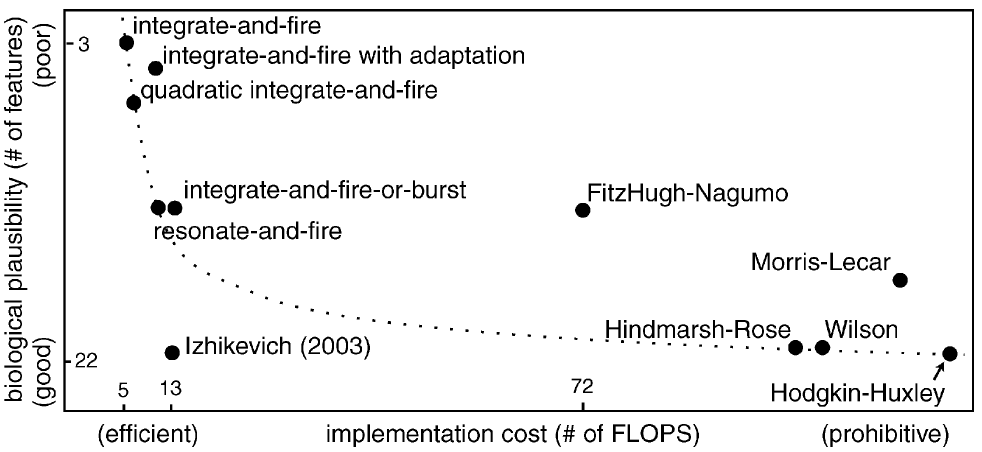}
	\end{center}
	\caption[Comparison of neuronal models]{\textbf{Comparison of neuronal models:} Eleven neuronal models are evaluated taking into account the implementation cost in form of \acp{FLOP} and biological plausibility measured by the number of possible features (e.g. tonic spiking, tonic bursting or integrator). Figure and information by Izhikevich~\cite{Izhikevich.2004}. }\label{fig:Aihzplot}
\end{figure}

\chapter{Simulation software}\label{A:SimSoft}
Many research groups work with simulated biological neuronal networks or even try to improve them further. Over years lot of simulators were published and evaluated. For example, Vitay compared the most widly used simulators, see Figure \ref{fig:simsoft}~\cite{Vitay.2015}. 
Since \textit{Brain 2} is implemented to be easy to use and learn, it is recommended for starter to build networks. Another reason is the fact \textit{Auryn}, \textit{ANNarchy} and \textit{NEST} are not designed for Microsoft Windows, which is the most often used operation system on desktop computers in 2017. For \textit{NEST} there exist workarounds with virtual machines or \textit{Cygwin}. In that case the computer is not able to use its full resources.
For professional long term simulation projects of complex neuronal networks with more than 5000 neurons and \ac{STDP} for example C++ based \textit{Auryn} on a Linux system is highly recommended. Since the used simulations are based on the Izhikevich model, there is also an option of using \textit{MATLAB} of MathWorks. Easier modifying and debugging are possible and an optional visualization every second is noteworthy, while a longer runtime is a disadvantage. At this state of the project we use \textit{MATLAB} because there is just the need of a few simulation runs. A future bachelor thesis by a colleague of mine will deal with further implementation ideas.

\begin{figure}[htbp]
	\begin{center}
		\includegraphics[width=0.6\textwidth]{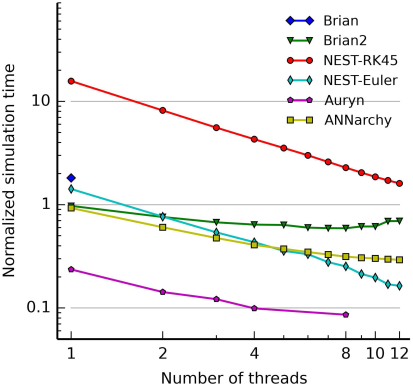} 
	\end{center}
	\caption[Comparison of simulation software]{\textbf{Comparison of simulation software:} \textit{Brian} (version 1.4.1), \textit{Brian 2} (version 2.0b3), \textit{NEST} (with \textit{Python} bindings, version 2.4.2), \textit{Auryn} (version 0.4.1) and \textit{ANNarchy} (version 4.4.0) are compared in context of simulation times depending on threads of a shared-memory system. Simulation times are normalized: One runtime second per simulated second means one in normalized simulation time. Figure and information by Vitay~\cite{Vitay.2015}. }\label{fig:simsoft}
\end{figure}

\chapter{Mutual Information}\label{A:MI}
\ac{MI} is a classic tool of probability theory, which measures the dependence between two random processes, which are in this case spike trains~\cite{Xu.1997}. The standard \ac{MI} of formula (\ref{equ:MI}) is able to detect also nonlinear correlations but does not provide any information of causality because of its symmetry, see formula (\ref{equ:MI_xchange}). Thus, it is more often used to estimate the synchrony. Using two as base of the logarithm leads to bits as result unit. 
\begin{equation}\label{equ:MI}
MI_{XY}=\sum_{x \in X}^{}\sum_{y \in Y }^{} P{(x,y)} \cdot \log_2 \frac{P{(x,y)}}{P{(x)} \cdot P{(y)}}
\end{equation}
\begin{equation}\label{equ:MI_xchange}
MI_{XY}=MI_{YX}
\end{equation}
However, by introducing a time shift for one spike train (\ref{equ:dMI}) and calculating \ac{MI} for many delays, statements about causality are possible~\cite{Xu.1997, Chavez.2003}. Each peak value of $MI_{XY}(d)$ for all neuron pairs is stored in the resulting \ac{CM}. 
\begin{equation}\label{equ:dMI}
MI_{XY}(d)=\sum_{x \in X}^{}\sum_{y \in Y }^{} P{(x_i,y_{i-d})} \cdot \log_2 \frac{P{(x_i,y_{i-d})}}{P{(x)} \cdot P{(y_{i-d})}}
\end{equation}
Binary or multistage binning is possible, but leads to an exponential gain of calculation operations. Garofalo evaluated \ac{MI} with delay shifting in 2009 and came to the conclusion of a bad performance in comparison to \ac{CC} and \ac{TE}~\cite{Garofalo.2009}. 

\chapter{TSPE MATLAB Code}\label{A:TSPE}
\begin{lstlisting}
function [CMres, DMres] = TSPE(sdf, d, neg_wins, co_wins, pos_wins, FLAG_NORM)
% Parameters:
%   sdf         - Time series in Spike Data Format (SDF)
%   d           - Maximal delay time (default 25)
%   neg_wins    - Windows for before and after area of interest (default [3, 4, 5, 6, 7, 8])
%   co_wins     - Cross-over window size (default 0)
%   pos_wins    - Sizes of area of interest (default [2, 3, 4, 5, 6])
%   jitt        - Jitter-Window-Size (default 4)
%   FLAG_NORM   - 0 - no usage of normalization (default)
%               - 1 - usage of normalization
%
% Returns:
%   CMres       - NxN matrix where N(i, j) is the total spiking probability edges (TSPE) i->j
%   DMres       - NxN matrix where N(i, j) is the transmission time with highest TSPE Value i->j
%
% Wrote by Stefano De Blasi, UAS Aschaffenburg in 2018

switch nargin
  case 1
    d = [];
    neg_wins = [];
    co_wins = [];
    pos_wins = [];
    FLAG_NORM = [];
  case 2
    neg_wins = [];
    co_wins = [];
    pos_wins = [];
    FLAG_NORM = [];
  case 3
    co_wins = [];
    pos_wins = [];
    FLAG_NORM = [];
  case 4
    pos_wins = [];
    FLAG_NORM = [];
  case 5
    FLAG_NORM = [];
  case 6 
      % all parameters are already set
  otherwise
    error('Input error.')
end
if isempty(pos_wins)
  pos_wins=[2, 3, 4, 5, 6];
end
if isempty(co_wins)
  co_wins=0;
end
if isempty(neg_wins)
  neg_wins=[3, 4, 5, 6, 7, 8];
end
if isempty(d)
  d=25;
end
if isempty(FLAG_NORM)
  FLAG_NORM=0;
end

%% Generation of sparse matrices
    a=sdf{end};
    NrC = a(1);
    vec1=[];
    vec2=[];
    for i=1:NrC
        vec1=[vec1 sdf{i}];
        vec2=[vec2 i*ones(1,length(sdf{i}))];
    end
    mat=sparse(vec1(vec1>0 & vec1 <= a(2)),vec2(vec1>0 & vec1 <= a(2)),1,a(2),a(1));
    NrS=a(2);   
    
%% Calculation of std deviation and mean values   
    l=ones(1,NrS);
    u_mean=l*mat/NrS;
    u_0=mat-u_mean;
    r=std(u_0);
   
%% Fast Cross-Correlation  
    ran=1-max(neg_wins)-max(co_wins):max(neg_wins)+d;
    CM=(zeros(length(ran),NrC,NrC));
    ind=max(neg_wins)+max(co_wins);                                                
    if(ind <= 0)
        ind=1;
    end
    for i=0:d+max(neg_wins)
        CM(ind,:,:)=(mat(1+i:end,:)'*mat(1:end-i,:))./(r'*r)/NrS;
        
        % Correct form: 
        % CM(ind,:,:)=(u_0(1+i:end,:)'*u_0(1:end-i,:))./(r'*r)/NrS;
        % takes longer, no performance impact

        ind=ind+1;
    end
    
% Usage of symmetric construction of cross correlation for faster
% calculation:
    if(max(neg_wins)+max(co_wins) > 0)
        bufCM=zeros(NrC);
        ind=0;
        for j=max(neg_wins)+max(co_wins)-1:-1:1
            bufCM(:)=CM(max(neg_wins)+max(co_wins)+j,:,:);
            ind=ind+1;
            CM(ind,:,:)=bufCM';
        end
    end
    
% Additional scaling for reduction of network burst impacts:
    if FLAG_NORM
      s=zeros(length(ran),1);
      for i=1:length(ran)
          zwi=CM(i, ~ diag(ones(NrC,1)));
          s(i)=sum(sum(  zwi(~ isnan(CM(i, ~ diag(ones(NrC,1)))))));
      end
      CM=CM./s;
    end  
    
%% Generation of edge filters
    WB=max(neg_wins)+max(co_wins);
    sumWin=zeros(d+WB,NrC,NrC);
    in=0;
    for win_before=neg_wins
      for win_p1=co_wins 
        for win_in=pos_wins
            in=in+1;
            win_p2=win_p1;
            win_after=win_before;
            windows{in}=[-1*ones(win_before,1) /win_before; zeros(win_p1,1);2/win_in* ones(win_in,1);zeros(win_p2,1);-1*ones(win_after,1)/win_after];
            beginnings{in}=1+WB-win_before-win_p1;
            win_inner{in}=win_in;
        end
      end
    end
    m=d+max(neg_wins)+max(co_wins)+max(pos_wins);    
   
%% Usage of edge filters: 
    for j=1:in         
            CM3=convn(convn(CM(beginnings{j}:end, : , :),windows{j},'valid'),[ones(win_inner{j},1)],'full');
            m=min(m,length(CM3(:,1,1)));
            sumWin(1:size(CM3,1),:,:)=CM3+sumWin(1:size(CM3,1),:,:);
    end
  
%% Only look at valid window
    sumWin=sumWin(1:m,:,:);
    
%% Adjustment and looking for maximum at each delay time 
    sumWin=permute(sumWin, [1 3 2]);
    [~,index] = max(abs(sumWin));
    CMres=zeros(NrC);
    DMres=index;
    for i=1:size(sumWin,2)
        CMres(:,i)=sumWin(sub2ind(size(sumWin), index(1,1:size(sumWin,2),i), 1:size(sumWin,2), i * ones(1,size(sumWin,2)))    );
    end

end
\end{lstlisting}

\chapter{Content of DVD}
The attached DVD contains several data related to the work during the candidature for a research degree at \ac{UAS} Aschaffenburg:
\begin{itemize}
\item Documents	
	\begin{itemize}
		\item Thesis (PDF)
		\item Submitted journal paper (PDF)
		\item Presented conference poster (PDF)
		\item Submitted conference paper (PDF)
		\item Literature (PDF)
	\end{itemize}
\item Written functions and code	
	\begin{itemize}
		\item \textit{MATLAB} code
		\item \textit{Python} code
		\item \textit{C++} code
	\end{itemize}
\item Data
	\begin{itemize}
		\item Evaluation data
		\item Experimental data
	\end{itemize}
\end{itemize}

\addtocontents{toc}{\endgroup}
\end{appendices}

\backmatter

\pagenumbering{Roman}
\setcounter{page}{\value{verz}}
\pagestyle{frontback}



\clearpage

\printbibliography

@article{DeBlasi.2018,
	title={Simulation of Large Scale Neural Networks for Evaluation Applications},
	author={De Blasi, Stefano},
	journal={arXiv preprint arXiv:1805.08626},
	year={2018}
}

@article{DeBlasi.2019,
	title={Total spiking probability edges: A cross-correlation based method for effective connectivity estimation of cortical spiking neurons},
	author={De Blasi, Stefano and Ciba, Manuel and Bahmer, Andreas and Thielemann, Christiane},
	journal={Journal of neuroscience methods},
	volume={312},
	pages={169--181},
	year={2019},
	publisher={Elsevier}
}

@inproceedings{Bouteiller.2010,
	title={Paired-pulse stimulation at glutamatergic synapses-pre-and postsynaptic components},
	author={Bouteiller, Jean-Marie C and Allam, Sushmita L and Greget, Renaud and Ambert, Nicolas and Hu, Eric Y and Bischoff, Serge and Baudry, Michel and Berger, Theodore W},
	booktitle={Engineering in Medicine and Biology Society (EMBC), 2010 Annual International Conference of the IEEE},
	pages={787--790},
	year={2010},
	organization={IEEE}
}

@article{Pastore.2018,
	title={Identification of excitatory-inhibitory links and network topology in large-scale neuronal assemblies from multi-electrode recordings},
	author={Pastore, Vito Paolo and Massobrio, Paolo and Godjoski, Aleksandar and Martinoia, Sergio},
	journal={PLoS computational biology},
	volume={14},
	number={8},
	pages={e1006381},
	year={2018},
	publisher={Public Library of Science}
}

@article{Pastore.2017,
	title={SpiCoDyn: A Toolbox for the Analysis of Neuronal Network Dynamics and Connectivity from Multi-Site Spike Signal Recordings},
	author={Pastore, Vito Paolo and Godjoski, Aleksandar and Martinoia, Sergio and Massobrio, Paolo},
	journal={Neuroinformatics},
	volume={16},
	number={1},
	pages={15--30},
	year={2018},
	publisher={Springer}
}

@article{Ullo.2014,
	author = {Ullo, Simona and Nieus, Thierry R. and Sona, Diego and Maccione, Alessandro and Berdondini, Luca and Murino, Vittorio},
	year = {2014},
	title = {Functional connectivity estimation over large networks at cellular resolution based on electrophysiological recordings and structural prior},
	pages = {137},
	volume = {8},
	issn = {1662-5129},
	journal = {Frontiers in neuroanatomy},
	doi = {10.3389/fnana.2014.00137}
}

@inproceedings{Pazienti.2007,
	title={Bounds of the ability to destroy precise coincidences by spike dithering},
	author={Pazienti, Antonio and Diesmann, Markus and Gr{\"u}n, Sonja},
	booktitle={International Symposium on Brain, Vision, and Artificial Intelligence},
	pages={428--437},
	year={2007},
	organization={Springer}
}

@article{Carlson.2014,
  title={Multichannel electrophysiological spike sorting via joint dictionary learning and mixture modeling},
  author={Carlson, David E and Vogelstein, Joshua T and Wu, Qisong and Lian, Wenzhao and Zhou, Mingyuan and Stoetzner, Colin R and Kipke, Daryl and Weber, Douglas and Dunson, David B and Carin, Lawrence},
  journal={IEEE Transactions on Biomedical Engineering},
  volume={61},
  number={1},
  pages={41--54},
  year={2014},
  publisher={IEEE}
}

@article{Mwaffo.2017,
	title={Analysis of pairwise interactions in a maximum likelihood sense to identify leaders in a group},
	author={Mwaffo, Violet and Butail, Sachit and Porfiri, Maurizio},
	journal={Frontiers in Robotics and AI},
	volume={4},
	pages={35},
	year={2017},
	publisher={Frontiers}
}

@article{Liu.2017,
	title={Use of a Neural Circuit Probe to Validate in silico Predictions of Inhibitory Connections},
	author={Liu, Honglei and Bridges, Daniel and Randall, Connor and Solla, Sara A and Wu, Bian and Hansma, Paul and Yan, Xifeng and Kosik, Kenneth S and Bouchard, Kristofer},
	journal={bioRxiv},
	year={2017},
	publisher={Cold Spring Harbor Laboratory}
}

@article{Butts.2007,
	title={Temporal precision in the neural code and the timescales of natural vision},
	author={Butts, Daniel A and Weng, Chong and Jin, Jianzhong and Yeh, Chun-I and Lesica, Nicholas A and Alonso, Jose-Manuel and Stanley, Garrett B},
	journal={Nature},
	volume={449},
	number={7158},
	pages={92},
	year={2007},
	publisher={Nature Publishing Group}
}

@article{Erdos.1959,
	title={Graph theory and probability},
	author={Erdos, Paul},
	journal={canad. J. Math},
	volume={11},
	number={11},
	pages={34--38},
	year={1959}
}

@article{Kadirvelu.2017,
	title={Inferring structural connectivity using Ising couplings in models of neuronal networks},
	author={Kadirvelu, Balasundaram and Hayashi, Yoshikatsu and Nasuto, Slawomir J},
	journal={Scientific reports},
	volume={7},
	number={1},
	pages={8156},
	year={2017},
	publisher={Nature Publishing Group}
}

@article{Shimono.2014,
	title={Functional clusters, hubs, and communities in the cortical microconnectome},
	author={Shimono, Masanori and Beggs, John M},
	journal={Cerebral Cortex},
	volume={25},
	number={10},
	pages={3743--3757},
	year={2014},
	publisher={Oxford University Press}
}

@techreport{Hagberg.2008,
	title={Exploring network structure, dynamics, and function using NetworkX},
	author={Hagberg, Aric and Swart, Pieter and S Chult, Daniel},
	year={2008},
	institution={Los Alamos National Lab.(LANL), Los Alamos, NM (United States)}
}

@article{Abeles.2001,
	title={Detecting precise firing sequences in experimental data},
	author={Abeles, Moshe and Gat, I},
	journal={Journal of neuroscience methods},
	volume={107},
	number={1-2},
	pages={141--154},
	year={2001},
	publisher={Elsevier}
}

@article{Pazienti.2008,
	title={Effectiveness of systematic spike dithering depends on the precision of cortical synchronization},
	author={Pazienti, Antonio and Maldonado, Pedro E and Diesmann, Markus and Gr{\"u}n, Sonja},
	journal={Brain research},
	volume={1225},
	pages={39--46},
	year={2008},
	publisher={Elsevier}
}

@article{Abbott.1997,
 author = {Abbott, L. F.},
 year = {1997},
 title = {Synaptic Depression and Cortical Gain Control},
 pages = {221--224},
 volume = {275},
 number = {5297},
 issn = {00368075},
 journal = {Science},
 doi = {10.1126/science.275.5297.221}
}

@article{Poli.2018,
	title={Synchronicity among hippocampal co-cultures in a four-chamber in vitro system},
	author={Poli, D and Vakilna, YS and DeMarse, TB and Wheeler, BC and Brewer, GJ},
	year = {2018},
	journal = {GNB2018}
}

@incollection{Braitenberg.1981,
	title={Anatomical basis for divergence, convergence, and integration in the cerebral cortex},
	author={Braitenberg, Valentino},
	booktitle={Sensory Functions},
	pages={411--419},
	year={1981},
	publisher={Elsevier}
}

@article{Brewer.2013,
	title={Toward a self-wired active reconstruction of the hippocampal trisynaptic loop: DG-CA3},
	author={Brewer, Gregory J and Boehler, Michael D and Leondopulos, Stathis and Pan, Liangbin and DeMarse, Thomas and Wheeler, Bruce C},
	journal={Frontiers in neural circuits},
	volume={7},
	pages={165},
	year={2013},
	publisher={Frontiers}
}

@article{Pan.2011,
	title={Propagation of action potential activity in a predefined microtunnel neural network},
	author={Pan, Liangbin and Alagapan, Sankaraleengam and Franca, Eric and Brewer, Gregory J and Wheeler, Bruce C},
	journal={Journal of neural engineering},
	volume={8},
	number={4},
	pages={046031},
	year={2011},
	publisher={IOP Publishing}
}

@article{DeMarse.2016,
	title={Feed-forward propagation of temporal and rate information between cortical populations during coherent activation in engineered in vitro networks},
	author={DeMarse, Thomas B and Pan, Liangbin and Alagapan, Sankaraleengam and Brewer, Gregory J and Wheeler, Bruce C},
	journal={Frontiers in neural circuits},
	volume={10},
	pages={32},
	year={2016},
	publisher={Frontiers}
}

@article{Bestel.2012,
	title={A novel automated spike sorting algorithm with adaptable feature extraction},
	author={Bestel, Robert and Daus, Andreas W and Thielemann, Christiane},
	journal={Journal of neuroscience methods},
	volume={211},
	number={1},
	pages={168--178},
	year={2012},
	doi = {https://doi.org/10.1016/j.jneumeth.2012.08.015},
	url = {http://www.sciencedirect.com/science/article/pii/S0165027012003263},
	publisher={Elsevier}
}

@article{Barabasi.2003,
 author = {Barab{\'a}si, Albert-L{\'a}szl{\'o} and Bonabeau, Eric},
 year = {2003},
 title = {Scale-free networks},
 pages = {60--69},
 volume = {288},
 number = {5},
 issn = {0036-8733},
 journal = {Scientific American}
}

@book{Barabasi.2016,
 author = {Barab{\'a}si, Albert-L{\'a}szl{\'o} and P{\'o}sfai, M{\'a}rton},
 year = {2016},
 title = {Network science},
 price = {Hardback},
 isbn = {1-107-07626-9}
}

@article{Bartho.2004,
 author = {Barth{\'o}, Peter and Hirase, Hajime and Monconduit, Lena{\"i}c and Zugaro, Michael and Harris, Kenneth D. and Buzs{\'a}ki, Gy{\"o}rgy},
 year = {2004},
 title = {Characterization of neocortical principal cells and interneurons by network interactions and extracellular features},
 pages = {600--608},
 volume = {92},
 number = {1},
 issn = {0022-3077},
 journal = {Journal of neurophysiology},
 doi = {10.1152/jn.01170.2003}
}

@article{Bedenbaugh.1997,
 author = {Bedenbaugh, P. and Gerstein, G. L.},
 year = {1997},
 title = {Multiunit normalized cross correlation differs from the average single-unit normalized correlation},
 pages = {1265--1275},
 volume = {9},
 number = {6},
 issn = {0899-7667},
 journal = {Neural computation}
}

@article{Berdondini.2009,
 author = {Berdondini, Luca and Imfeld, Kilian and Maccione, Alessandro and Tedesco, Mariateresa and Neukom, Simon and Koudelka-Hep, Milena and Martinoia, Sergio},
 year = {2009},
 title = {Active pixel sensor array for high spatio-temporal resolution electrophysiological recordings from single cell to large scale neuronal networks},
 pages = {2644--2651},
 volume = {9},
 number = {18},
 issn = {1473-0197},
 journal = {Lab on a chip},
 doi = {10.1039/b907394a}
}

@article{Berdondini.2009b,
 author = {Berdondini, L. and Massobrio, P. and Chiappalone, M. and Tedesco, M. and Imfeld, K. and Maccione, A. and Gandolfo, M. and Koudelka-Hep, M. and Martinoia, S.},
 year = {2009},
 title = {Extracellular recordings from locally dense microelectrode arrays coupled to dissociated cortical cultures},
 pages = {386--396},
 volume = {177},
 number = {2},
 issn = {0165-0270},
 journal = {Journal of neuroscience methods},
 doi = {10.1016/j.jneumeth.2008.10.032}
}

@book{Bhatti.2017,
 author = {Bhatti, Asim and Lee, Kendall H. and Garmestani, Hamid},
 year = {2017},
 title = {Emerging Trends in Neuro Engineering and Neural Computation},
 url = {http://dx.doi.org/10.1007/978-981-10-3957-7},
 publisher = {{Springer Singapore}},
 isbn = {978-981-10-3957-7},
 series = {Series in BioEngineering}
}

@article{Bonifazi.2009,
 author = {Bonifazi, P. and Goldin, M. and Picardo, M. A. and Jorquera, I. and Cattani, A. and Bianconi, G. and Represa, A. and Ben-Ari, Y. and Cossart, R.},
 year = {2009},
 title = {GABAergic hub neurons orchestrate synchrony in developing hippocampal networks},
 pages = {1419--1424},
 volume = {326},
 number = {5958},
 issn = {1095-9203},
 journal = {Science (New York, N.Y.)},
 doi = {10.1126/science.1175509}
}

@article{Brosch.1999,
 author = {Brosch, M. and Schreiner, C. E.},
 year = {1999},
 title = {Correlations between neural discharges are related to receptive field properties in cat primary auditory cortex},
 pages = {3517--3530},
 volume = {11},
 number = {10},
 issn = {0953-816X},
 journal = {The European journal of neuroscience}
}

@article{Buchs.1996,
 author = {Buchs, P. A. and Muller, D.},
 year = {1996},
 title = {Induction of long-term potentiation is associated with major ultrastructural changes of activated synapses},
 pages = {8040--8045},
 volume = {93},
 number = {15},
 issn = {0027-8424},
 journal = {Proceedings of the National Academy of Sciences},
 doi = {10.1073/pnas.93.15.8040}
}

@article{Bullmore.2009,
 author = {Bullmore, Ed and Sporns, Olaf},
 year = {2009},
 title = {Complex brain networks: graph theoretical analysis of structural and functional systems},
 pages = {186--198},
 volume = {10},
 number = {3},
 issn = {1471-0048},
 journal = {Nature reviews. Neuroscience},
 doi = {10.1038/nrn2575}
}

@article{Catanzaro.2005,
 author = {Catanzaro, Michele and Bogu{\~n}{\'a}, Mari{\'a}n and Pastor-Satorras, Romualdo},
 year = {2005},
 title = {Generation of uncorrelated random scale-free networks},
 pages = {027103},
 volume = {71},
 number = {2 Pt 2},
 issn = {1539-3755},
 journal = {Physical review. E, Statistical, nonlinear, and soft matter physics},
 doi = {10.1103/PhysRevE.71.027103}
}

@article{Chan.2008,
 author = {Chan, Hsiao-Lung and Lin, Ming-An and Wu, Tony and Lee, Shih-Tseng and Tsai, Yu-Tai and Chao, Pei-Kuang},
 year = {2008},
 title = {Detection of neuronal spikes using an adaptive threshold based on the max-min spread sorting method},
 pages = {112--121},
 volume = {172},
 number = {1},
 issn = {0165-0270},
 journal = {Journal of neuroscience methods},
 doi = {10.1016/j.jneumeth.2008.04.014}
}

@article{Chavez.2003,
 author = {Ch{\'a}vez, Mario and Martinerie, Jacques and {van Quyen}, Michel},
 year = {2003},
 title = {Statistical assessment of nonlinear causality: Application to epileptic EEG signals},
 pages = {113--128},
 volume = {124},
 number = {2},
 issn = {0165-0270},
 journal = {Journal of neuroscience methods},
 doi = {10.1016/S0165-0270(02)00367-9}
}

@article{Chiappalone.2006,
 author = {Chiappalone, Michela and Bove, Marco and Vato, Alessandro and Tedesco, Mariateresa and Martinoia, Sergio},
 year = {2006},
 title = {Dissociated cortical networks show spontaneously correlated activity patterns during in vitro development},
 pages = {41--53},
 volume = {1093},
 number = {1},
 issn = {0006-8993},
 journal = {Brain research},
 doi = {10.1016/j.brainres.2006.03.049}
}

@article{Cohen.2003,
 author = {Cohen, Reuven and Havlin, Shlomo},
 year = {2003},
 title = {Scale-free networks are ultrasmall},
 volume = {90},
 number = {5},
 issn = {1079-7114},
 journal = {Physical review letters},
 doi = {10.1103/PhysRevLett.90.058701}
}

@article{Date.1998,
 author = {Date, Akira and Bienenstock, Elie and Geman, Stuart},
 year = {1998},
 journal={Society for Neuroscience Abstracts (Brown University, Division of Applied Mathematics)},
 title = {On the temporal resolution of neural activity}
}

@book{Dayan.2005,
 author = {Dayan, Peter and Abbott, L. F.},
 year = {2005},
 title = {Theoretical neuroscience: Computational and mathematical modeling of neural systems},
 price = {{\pounds}24.95},
 address = {Cambridge Mass. and London},
 publisher = {MIT},
 isbn = {0-262-54185-8},
 series = {Computational neuroscience}
}

@article{Deisseroth.2011,
 author = {Deisseroth, Karl},
 year = {2011},
 title = {Optogenetics},
 pages = {26--29},
 volume = {8},
 number = {1},
 issn = {1548-7105},
 journal = {Nature methods},
 doi = {10.1038/nmeth.f.324}
}

@article{Diekman.2014,
 author = {Diekman, Casey and Dasgupta, Kohinoor and Nair, Vijay and Unnikrishnan, K. P.},
 year = {2014},
 title = {Discovering functional neuronal connectivity from serial patterns in spike train data},
 pages = {1263--1297},
 volume = {26},
 number = {7},
 issn = {0899-7667},
 journal = {Neural computation},
 doi = {10.1162/NECO_a_00598}
}

@incollection{Ding.2006,
 author = {Ding, Mingzhou and Chen, Yonghong and Bressler, Steven L.},
 title = {Granger Causality: Basic Theory and Application to Neuroscience},
 pages = {437--460},
 publisher = {Wiley-VCH},
 isbn = {9783527609970},
 editor = {Schelter, Bj{\"o}rn and Timmer, Jens and Winterhalder, Matthias},
 booktitle = {Handbook of time series analysis},
 year = {2006},
 address = {Weinheim},
 doi = {10.1002/9783527609970.ch17}
}

@article{Eytan.2004,
 author = {Eytan, Danny and Minerbi, Amir and Ziv, Noam and Marom, Shimon},
 year = {2004},
 title = {Dopamine-induced dispersion of correlations between action potentials in networks of cortical neurons},
 pages = {1817--1824},
 volume = {92},
 number = {3},
 issn = {0022-3077},
 journal = {Journal of neurophysiology},
 doi = {10.1152/jn.00202.2004}
}

@article{Flachs.2016,
 author = {Flachs, Dennis and Ciba, Manuel},
 year = {2016},
 title = {Cell-Based Sensor Chip for Neurotoxicity Measurements in Drinking Water},
 url = {https://ojs.cvut.cz/ojs/index.php/CTJ/article/download/4195/4049},
 pages = {46--50},
 volume = {46},
 number = {2},
 issn = {2336-5552},
 journal = {L{\'e}ka{\v{r}} a technika - Clinician and Technology}
}

@article{Friston.1994,
 author = {Friston, Karl J.},
 year = {1994},
 title = {Functional and effective connectivity in neuroimaging: A synthesis},
 pages = {56--78},
 volume = {2},
 number = {1-2},
 issn = {10659471},
 journal = {Human Brain Mapping},
 doi = {10.1002/hbm.460020107}
}

@article{Friston.2011,
 author = {Friston, Karl J.},
 year = {2011},
 title = {Functional and effective connectivity: A review},
 pages = {13--36},
 volume = {1},
 number = {1},
 issn = {2158-0022},
 journal = {Brain connectivity},
 doi = {10.1089/brain.2011.0008}
}

@article{Friston.2011c,
 author = {Friston, Karl J. and Li, Baojuan and Daunizeau, Jean and Stephan, Klaas E.},
 year = {2011},
 title = {Network discovery with DCM},
 pages = {1202--1221},
 volume = {56},
 number = {3},
 issn = {1095-9572},
 journal = {NeuroImage},
 doi = {10.1016/j.neuroimage.2010.12.039}
}

@article{Garofalo.2009,
 author = {Garofalo, Matteo and Nieus, Thierry and Massobrio, Paolo and Martinoia, Sergio},
 year = {2009},
 title = {Evaluation of the performance of information theory-based methods and cross-correlation to estimate the functional connectivity in cortical networks},
 pages = {e6482},
 volume = {4},
 number = {8},
 issn = {1932-6203},
 journal = {PloS one},
 doi = {10.1371/journal.pone.0006482}
}

@book{Gerstner.2014,
 author = {Gerstner, Wulfram},
 year = {2014},
 title = {Neuronal dynamics: From single neurons to networks and models of cognition},
 address = {Cambridge},
 edition = {1. publ},
 publisher = {{Cambrige Univ. Press}},
 isbn = {978-1-107-06083-8}
}

@article{Gourevitch.2007,
 author = {Gour{\'e}vitch, Boris and Eggermont, Jos J.},
 year = {2007},
 title = {Evaluating information transfer between auditory cortical neurons},
 pages = {2533--2543},
 volume = {97},
 number = {3},
 issn = {0022-3077},
 journal = {Journal of neurophysiology},
 doi = {10.1152/jn.01106.2006}
}

@book{Hebb.1949,
 author = {Hebb, D. O.},
 year = {1949},
 title = {The organization of behavior a neuropsychological approach.},
 url = {http://worldcatlibraries.org/wcpa/oclc/821634185},
 address = {New York  NY},
 publisher = {{John Wiley {\&} Sons}}
}

@article{Hodgkin.1952,
 author = {Hodgkin, A. L. and Huxley, A. F.},
 year = {1952},
 title = {A quantitative description of membrane current and its application to conduction and excitation in nerve},
 pages = {500--544},
 volume = {117},
 number = {4},
 issn = {00223751},
 journal = {The Journal of Physiology},
 doi = {10.1113/jphysiol.1952.sp004764}
}

@article{Isomura.2014,
 author = {Isomura, Takuya and Takeuchi, Akimasa and Shimba, Kenta and Kotani, Kiyoshi and Jimbo, Yasuhiko},
 year = {2014},
 title = {Connection-Strength Estimation of Neuronal Networks by Fitting for Izhikevich Model},
 pages = {42--50},
 volume = {187},
 number = {4},
 issn = {04247760},
 journal = {Electrical Engineering in Japan},
 doi = {10.1002/eej.22517}
}

@article{Isomura.2015,
 author = {Isomura, Takuya and Ogawa, Yutaro and Kotani, Kiyoshi and Jimbo, Yasuhiko},
 year = {2015},
 title = {Accurate connection strength estimation based on variational bayes for detecting synaptic plasticity},
 pages = {819--844},
 volume = {27},
 number = {4},
 issn = {0899-7667},
 journal = {Neural computation},
 doi = {10.1162/NECO_a_00721}
}

@article{Ito.2011,
 author = {Ito, Shinya and Hansen, Michael E. and Heiland, Randy and Lumsdaine, Andrew and Litke, Alan M. and Beggs, John M.},
 year = {2011},
 title = {Extending transfer entropy improves identification of effective connectivity in a spiking cortical network model},
 pages = {e27431},
 volume = {6},
 number = {11},
 issn = {1932-6203},
 journal = {PloS one},
 doi = {10.1371/journal.pone.0027431}
}

@article{Izhikevich.2003,
 author = {Izhikevich, E. M.},
 year = {2003},
 title = {Simple model of spiking neurons},
 pages = {1569--1572},
 volume = {14},
 number = {6},
 issn = {1045-9227},
 journal = {IEEE transactions on neural networks},
 doi = {10.1109/TNN.2003.820440}
}

@article{Izhikevich.2004,
 author = {Izhikevich, Eugene M.},
 year = {2004},
 title = {Which model to use for cortical spiking neurons?},
 pages = {1063--1070},
 volume = {15},
 number = {5},
 issn = {1045-9227},
 journal = {IEEE transactions on neural networks},
 doi = {10.1109/TNN.2004.832719}
}

@article{Izhikevich.2006,
 author = {Izhikevich, Eugene M.},
 year = {2006},
 title = {Polychronization: computation with spikes},
 pages = {245--282},
 volume = {18},
 number = {2},
 issn = {0899-7667},
 journal = {Neural computation},
 doi = {10.1162/089976606775093882}
}

@article{Izhikevich.2006b,
 author = {Izhikevich, Eugene M. and FitzHugh, Richard},
 year = {2006},
 title = {FitzHugh-Nagumo model},
 pages = {1349},
 volume = {1},
 number = {9},
 issn = {1941-6016},
 journal = {Scholarpedia},
 doi = {10.4249/scholarpedia.1349}
}

@misc{Jarosz.2009,
 author = {{Jarosz Quasar}},
 title = {Creative Commons --- Attribution-ShareAlike 3.0 Unported --- CC BY-SA 3.0},
 url = {https://creativecommons.org/licenses/by-sa/3.0/deed.en},
 urldate = {09/22/2017}
}

@article{Jimbo.1999,
 author = {Jimbo, Y. and Tateno, T. and Robinson, H.P.C.},
 year = {1999},
 title = {Simultaneous Induction of Pathway-Specific Potentiation and Depression in Networks of Cortical Neurons},
 pages = {670--678},
 volume = {76},
 number = {2},
 issn = {00063495},
 journal = {Biophysical Journal},
 doi = {10.1016/S0006-3495(99)77234-6}
}

@article{Juergens.1997,
 author = {Juergens, E. and Eckhorn, R.},
 year = {1997},
 title = {Parallel processing by a homogeneous group of coupled model neurons can enhance, reduce and generate signal correlations},
 pages = {217--227},
 volume = {76},
 number = {3},
 issn = {0340-1200},
 journal = {Biological cybernetics},
 doi = {10.1007/s004220050334}
}

@article{Kiemel.1998,
 author = {Kiemel, T. and Cohen, A. H.},
 year = {1998},
 title = {Estimation of coupling strength in regenerated lamprey spinal cords based on a stochastic phase model},
 pages = {267--284},
 volume = {5},
 number = {3},
 issn = {1573-6873},
 journal = {Journal of computational neuroscience}
}

@misc{Kohler.2016,
 author = {K{\"o}hler, Tim and W{\"o}lfel, Maximilian and Bochtler, Ulrich and Thielemann, Christiane},
 year = {2016},
 title = {TETRA specific long-term exposure of neuronal in vitro networks}
}

@article{Lewicki.1998,
 author = {Lewicki, Michael S.},
 year = {1998},
 title = {A review of methods for spike sorting: The detection and classification of neural action potentials},
 pages = {R53-R78},
 volume = {9},
 number = {4},
 issn = {0954-898X},
 journal = {Network: Computation in Neural Systems},
 doi = {10.1088/0954-898X_9_4_001}
}

@article{Lieb.2017,
 author = {Lieb, Florian and Stark, Hans-Georg and Thielemann, Christiane},
 year = {2017},
 title = {A stationary wavelet transform and a time-frequency based spike detection algorithm for extracellular recorded data},
 volume = {14},
 number = {3},
 issn = {1741-2552},
 journal = {Journal of neural engineering},
 doi = {10.1088/1741-2552/aa654b}
}

@article{Lungarella.2006,
 author = {Lungarella, Max and Sporns, Olaf},
 year = {2006},
 title = {Mapping information flow in sensorimotor networks},
 pages = {e144},
 volume = {2},
 number = {10},
 issn = {1553-7358},
 journal = {PLoS computational biology},
 doi = {10.1371/journal.pcbi.0020144}
}

@article{Luo.2014,
 author = {Luo, Yvonne H-L and {da Cruz}, Lyndon},
 year = {2014},
 title = {A review and update on the current status of retinal prostheses (bionic eye)},
 pages = {31--44},
 volume = {109},
 issn = {1471-8391},
 journal = {British medical bulletin},
 doi = {10.1093/bmb/ldu002}
}

@article{Maccione.2009b,
 author = {Maccione, Alessandro and Gandolfo, Mauro and Massobrio, Paolo and Novellino, Antonio and Martinoia, Sergio and Chiappalone, Michela},
 year = {2009},
 title = {A novel algorithm for precise identification of spikes in extracellularly recorded neuronal signals},
 pages = {241--249},
 volume = {177},
 number = {1},
 issn = {0165-0270},
 journal = {Journal of neuroscience methods},
 doi = {10.1016/j.jneumeth.2008.09.026}
}

@article{Maccione.2012,
 author = {Maccione, Alessandro and Garofalo, Matteo and Nieus, Thierry and Tedesco, Mariateresa and Berdondini, Luca and Martinoia, Sergio},
 year = {2012},
 title = {Multiscale functional connectivity estimation on low-density neuronal cultures recorded by high-density CMOS Micro Electrode Arrays},
 pages = {161--171},
 volume = {207},
 number = {2},
 issn = {0165-0270},
 journal = {Journal of neuroscience methods},
 doi = {10.1016/j.jneumeth.2012.04.002}
}

@article{Mason.1991,
 author = {Mason, A. and Nicoll, A. and Stratford, K.},
 year = {1991},
 title = {Synaptic transmission between individual pyramidal neurons of the rat visual cortex in vitro},
 pages = {72--84},
 volume = {11},
 number = {1},
 issn = {1529-2401},
 journal = {The Journal of neuroscience : the official journal of the Society for Neuroscience}
}

@article{Masud.2017,
 author = {Masud, Mohammad Shahed and Borisyuk, Roman and Stuart, Liz},
 year = {2017},
 title = {Advanced correlation grid: Analysis and visualisation of functional connectivity among multiple spike trains},
 pages = {78--101},
 volume = {286},
 issn = {0165-0270},
 journal = {Journal of neuroscience methods},
 doi = {10.1016/j.jneumeth.2017.05.016}
}

@article{Mayer.2016,
 author = {Mayer, Margot and Arrizabalaga, Onetsine and Ritter, Sylvia and Thielemann, Christiane},
 year = {2016},
 title = {Human Embryonic Stem Cell Derived Neurospheres - A Novel Three Dimensional Model For Neurotoxicological Studies},
 volume = {10},
 issn = {1662-453X},
 journal = {Frontiers in Neuroscience},
 doi = {10.3389/conf.fnins.2016.93.00081}
}

@article{Muthmann.2015,
 author = {Muthmann, Jens-Oliver and Amin, Hayder and Sernagor, Evelyne and Maccione, Alessandro and Panas, Dagmara and Berdondini, Luca and Bhalla, Upinder S. and Hennig, Matthias H.},
 year = {2015},
 title = {Spike Detection for Large Neural Populations Using High Density Multielectrode Arrays},
 pages = {28},
 volume = {9},
 issn = {1662-5196},
 journal = {Frontiers in neuroinformatics},
 doi = {10.3389/fninf.2015.00028}
}

@article{Nakhnikian.2014,
 author = {Nakhnikian, Alexander and Rebec, George V. and Grasse, Leslie M. and Dwiel, Lucas L. and Shimono, Masanori and Beggs, John M.},
 year = {2014},
 title = {Behavior modulates effective connectivity between cortex and striatum},
 pages = {e89443},
 volume = {9},
 number = {3},
 issn = {1932-6203},
 journal = {PloS one},
 doi = {10.1371/journal.pone.0089443}
}

@article{Nick.2013,
 author = {Nick, Christoph and Goldhammer, Michael and Bestel, Robert and Steger, Frederik and {W. Daus}, Andreas and Thielemann, Christiane},
 year = {2013},
 title = {DrCell -- A Software Tool for the Analysis of Cell Signals Recorded with Extracellular Microelectrodes},
 pages = {96--109},
 volume = {7},
 journal = {Signal Processing: An International Journal}
}

@article{Obeid.2004,
 author = {Obeid, Iyad and Wolf, Patrick D.},
 year = {2004},
 title = {Evaluation of spike-detection algorithms for a brain-machine interface application},
 pages = {905--911},
 volume = {51},
 number = {6},
 issn = {0018-9294},
 journal = {IEEE transactions on bio-medical engineering},
 doi = {10.1109/TBME.2004.826683}
}

@article{Olesen.2003,
 author = {Olesen, J. and Leonardi, M.},
 year = {2003},
 title = {The burden of brain diseases in Europe},
 pages = {471--477},
 volume = {10},
 number = {5},
 issn = {1351-5101},
 journal = {European Journal of Neurology},
 doi = {10.1046/j.1468-1331.2003.00682.x}
}

@article{Overbey.2009,
 author = {Overbey, L. A. and Todd, M. D.},
 year = {2009},
 title = {Dynamic system change detection using a modification of the transfer entropy},
 pages = {438--453},
 volume = {322},
 number = {1-2},
 issn = {0022460X},
 journal = {Journal of Sound and Vibration},
 doi = {10.1016/j.jsv.2008.11.025}
}

@article{Pan.2015,
	
	AUTHOR={Pan, Liangbin and Alagapan, Sankaraleengam and Franca, Eric and Leondopulos, Stathis S. and DeMarse, Thomas B. and Brewer, Gregory J. and Wheeler, Bruce C.},    
	TITLE={An in vitro method to manipulate the direction and functional strength between neural populations},      
	JOURNAL={Frontiers in Neural Circuits},      
	VOLUME={9},     
	PAGES={32},     
	YEAR={2015},      
	URL={https://www.frontiersin.org/article/10.3389/fncir.2015.00032},       
	DOI={10.3389/fncir.2015.00032},      
	
	ISSN={1662-5110} 
	
	
}

@article{Pasquale.2008,
 author = {Pasquale, V. and Massobrio, P. and Bologna, L. L. and Chiappalone, M. and Martinoia, S.},
 year = {2008},
 title = {Self-organization and neuronal avalanches in networks of dissociated cortical neurons},
 pages = {1354--1369},
 volume = {153},
 number = {4},
 issn = {0306-4522},
 journal = {Neuroscience},
 doi = {10.1016/j.neuroscience.2008.03.050}
}

@article{Pastore.2016,
 author = {Pastore, Vito Paolo and Poli, Daniele and Godjoski, Aleksandar and Martinoia, Sergio and Massobrio, Paolo},
 year = {2016},
 title = {ToolConnect: A Functional Connectivity Toolbox for In vitro Networks},
 pages = {13},
 volume = {10},
 issn = {1662-5196},
 journal = {Frontiers in neuroinformatics},
 doi = {10.3389/fninf.2016.00013}
}

@article{Perkel.1967,
 author = {Perkel, Donald H. and Gerstein, George L. and Moore, George P.},
 year = {1967},
 title = {Neuronal Spike Trains and Stochastic Point Processes},
 pages = {391--418},
 volume = {7},
 number = {4},
 issn = {00063495},
 journal = {Biophysical Journal},
 doi = {10.1016/S0006-3495(67)86596-2}
}

@article{Pinyon.2014,
 author = {Pinyon, Jeremy L. and Tadros, Sherif F. and Froud, Kristina E. and {Y Wong}, Ann C. and Tompson, Isabella T. and Crawford, Edward N. and Ko, Myungseo and Morris, Ren{\'e}e and Klugmann, Matthias and Housley, Gary D.},
 year = {2014},
 title = {Close-field electroporation gene delivery using the cochlear implant electrode array enhances the bionic ear},
 pages = {233ra54},
 volume = {6},
 number = {233},
 issn = {1946-6242},
 journal = {Science translational medicine},
 doi = {10.1126/scitranslmed.3008177}
}

@article{Poli.2015,
 author = {Poli, Daniele and Pastore, Vito P. and Massobrio, Paolo},
 year = {2015},
 title = {Functional connectivity in in vitro neuronal assemblies},
 pages = {57},
 volume = {9},
 issn = {1662-5110},
 journal = {Frontiers in neural circuits},
 doi = {10.3389/fncir.2015.00057}
}

@article{Poli.2016,
 author = {Poli, Daniele and Pastore, Vito Paolo and Martinoia, Sergio and Massobrio, Paolo},
 year = {2016},
 title = {From functional to structural connectivity using partial correlation in neuronal assemblies},
 pages = {026023},
 volume = {13},
 number = {2},
 issn = {1741-2552},
 journal = {Journal of neural engineering},
 doi = {10.1088/1741-2560/13/2/026023}
}

@article{Renault.2015,
 author = {Renault, Renaud and Sukenik, Nirit and Descroix, St{\'e}phanie and Malaquin, Laurent and Viovy, Jean-Louis and Peyrin, Jean-Michel and Bottani, Samuel and Monceau, Pascal and Moses, Elisha and Vignes, Ma{\'e}va},
 year = {2015},
 title = {Combining microfluidics, optogenetics and calcium imaging to study neuronal communication in vitro},
 pages = {e0120680},
 volume = {10},
 number = {4},
 issn = {1932-6203},
 journal = {PloS one},
 doi = {10.1371/journal.pone.0120680}
}

@article{Saalmann.2012,
 author = {Saalmann, Yuri B. and Pinsk, Mark A. and Wang, Liang and Li, Xin and Kastner, Sabine},
 year = {2012},
 title = {The pulvinar regulates information transmission between cortical areas based on attention demands},
 pages = {753--756},
 volume = {337},
 number = {6095},
 issn = {1095-9203},
 journal = {Science (New York, N.Y.)},
 doi = {10.1126/science.1223082}
}

@article{Saneyoshi.2010,
 author = {Saneyoshi, Takeo and Fortin, Dale A. and Soderling, Thomas R.},
 year = {2010},
 title = {Regulation of spine and synapse formation by activity-dependent intracellular signaling pathways},
 pages = {108--115},
 volume = {20},
 number = {1},
 issn = {1873-6882},
 journal = {Current opinion in neurobiology},
 doi = {10.1016/j.conb.2009.09.013}
}

@article{Humphries.2008,
	title={Network ‘small-world-ness’: a quantitative method for determining canonical network equivalence},
	author={Humphries, Mark D and Gurney, Kevin},
	journal={PloS one},
	volume={3},
	number={4},
	pages={e0002051},
	year={2008},
	publisher={Public Library of Science}
}

@article{Schreiber.2000,
 author = {Schreiber, Thomas},
 year = {2000},
 title = {Measuring information transfer},
 pages = {461--464},
 volume = {85},
 number = {2},
 issn = {1079-7114},
 journal = {Physical review letters},
 doi = {10.1103/PhysRevLett.85.461}
}

@article{Seth.2010,
 author = {Seth, Anil K.},
 year = {2010},
 title = {A MATLAB toolbox for Granger causal connectivity analysis},
 pages = {262--273},
 volume = {186},
 number = {2},
 issn = {0165-0270},
 journal = {Journal of neuroscience methods},
 doi = {10.1016/j.jneumeth.2009.11.020}
}

@article{Shepherd.1998,
 author = {Shepherd, G. M. and Harris, K. M.},
 year = {1998},
 title = {Three-dimensional structure and composition of CA3--CA1 axons in rat hippocampal slices: implications for presynaptic connectivity and compartmentalization},
 pages = {8300--8310},
 volume = {18},
 number = {20},
 issn = {1529-2401},
 journal = {The Journal of neuroscience : the official journal of the Society for Neuroscience}
}

@article{Song.2005,
 author = {Song, Pengcheng and Wang, Xiao-Jing},
 year = {2005},
 title = {Angular path integration by moving ''hill of activity'': a spiking neuron model without recurrent excitation of the head-direction system},
 pages = {1002--1014},
 volume = {25},
 number = {4},
 issn = {1529-2401},
 journal = {The Journal of neuroscience : the official journal of the Society for Neuroscience},
 doi = {10.1523/JNEUROSCI.4172-04.2005}
}

@article{Sporns.2002,
 author = {Sporns, Olaf},
 year = {2002},
 title = {Network analysis, complexity, and brain function},
 pages = {56--60},
 volume = {8},
 number = {1},
 issn = {1076-2787},
 journal = {Complexity},
 doi = {10.1002/cplx.10047}
}

@article{Sporns.2005,
 author = {Sporns, Olaf and Tononi, Giulio and K{\"o}tter, Rolf},
 year = {2005},
 title = {The human connectome: A structural description of the human brain},
 pages = {e42},
 volume = {1},
 number = {4},
 issn = {1553-7358},
 journal = {PLoS computational biology},
 doi = {10.1371/journal.pcbi.0010042}
}

@article{Sporns.2007,
 author = {Sporns, Olaf and Honey, Christopher J. and K{\"o}tter, Rolf},
 year = {2007},
 title = {Identification and classification of hubs in brain networks},
 pages = {e1049},
 volume = {2},
 number = {10},
 issn = {1932-6203},
 journal = {PloS one},
 doi = {10.1371/journal.pone.0001049}
}

@article{Stetter.2012,
 author = {Stetter, Olav and Battaglia, Demian and Soriano, Jordi and Geisel, Theo},
 year = {2012},
 title = {Model-free reconstruction of excitatory neuronal connectivity from calcium imaging signals},
 pages = {e1002653},
 volume = {8},
 number = {8},
 issn = {1553-7358},
 journal = {PLoS computational biology},
 doi = {10.1371/journal.pcbi.1002653}
}

@article{Svendsen.2005,
 author = {Svendsen, Ove},
 year = {2005},
 title = {Ethics and animal welfare related to in vivo pharmacology and toxicology in laboratory animals},
 pages = {197-199; author reply 200-1},
 volume = {97},
 number = {4},
 issn = {1742-7835},
 journal = {Basic {\&} clinical pharmacology {\&} toxicology},
 doi = {10.1111/j.1742-7843.2005.pto_letter_974.x}
}

@article{Swadlow.1994,
 author = {Swadlow, H. A.},
 year = {1994},
 title = {Efferent neurons and suspected interneurons in motor cortex of the awake rabbit: axonal properties, sensory receptive fields, and subthreshold synaptic inputs},
 pages = {437--453},
 volume = {71},
 number = {2},
 issn = {0022-3077},
 journal = {Journal of neurophysiology}
}

@article{Timme.2014,
 author = {Timme, Nicholas and Ito, Shinya and Myroshnychenko, Maxym and Yeh, Fang-Chin and Hiolski, Emma and Hottowy, Pawel and Beggs, John M.},
 year = {2014},
 title = {Multiplex networks of cortical and hippocampal neurons revealed at different timescales},
 pages = {e115764},
 volume = {9},
 number = {12},
 issn = {1932-6203},
 journal = {PloS one},
 doi = {10.1371/journal.pone.0115764}
}

@article{Vitay.2015,
 author = {Vitay, Julien and Dinkelbach, Helge and Hamker, Fred},
 year = {2015},
 title = {ANNarchy: a code generation approach to neural simulations on parallel hardware},
 pages = {19},
 volume = {9},
 issn = {1662-5196},
 journal = {Frontiers in neuroinformatics},
 doi = {10.3389/fninf.2015.00019}
}

@article{Vogels.2005,
 author = {Vogels, Tim P. and Abbott, L. F.},
 year = {2005},
 title = {Signal propagation and logic gating in networks of integrate-and-fire neurons},
 pages = {10786--10795},
 volume = {25},
 number = {46},
 journal = {The Journal of neuroscience: the official journal of the Society for Neuroscience},
 doi = {10.1523/JNEUROSCI.3508-05.2005}
}

@article{Watts.1998,
 author = {Watts, D. J. and Strogatz, S. H.},
 year = {1998},
 title = {Collective dynamics of 'small-world' networks},
 pages = {440--442},
 volume = {393},
 number = {6684},
 journal = {Nature.},
 doi = {10.1038/30918}
}

@article{Xu.1997,
 author = {Xu, Jinghua and Liu, Zeng-rong and Liu, Ren and Yang, Qing-fei},
 year = {1997},
 title = {Information transmission in human cerebral cortex},
 pages = {363--374},
 volume = {106},
 number = {3-4},
 issn = {01672789},
 journal = {Physica D: Nonlinear Phenomena},
 doi = {10.1016/S0167-2789(97)00042-0}
}


\end{document}